\documentclass[11pt,a4paper]{article}
\pdfoutput=1

\usepackage{jheppub}
\usepackage{amsmath,amsfonts,mathrsfs,graphicx,bbm}

\usepackage{kantlipsum}
\allowdisplaybreaks

\usepackage{fixltx2e}
\usepackage{cancel}

\usepackage{upgreek}
\usepackage{t4phonet}
\usepackage{pifont}

\usepackage[font=small]{caption}

\usepackage{lmodern}

\usepackage{graphicx}
\usepackage{booktabs}
\usepackage{dsfont}

\usepackage{mathrsfs}
\usepackage{amsmath,amssymb}
\usepackage{mathtools}
\usepackage{bbm}
\usepackage{slashed}
\usepackage{braket}

\usepackage{hyperref}

\usepackage{xspace}

\usepackage{color}

\definecolor{grey}{rgb}{0.4,0.4,0.5}
\definecolor{darkgreen}{rgb}{0,0.5,0}
\definecolor{darkred}{rgb}{0.6,0.0,0}
\definecolor{lightbrown}{rgb}{1,0.9,0.8}
\definecolor{brown}{rgb}{0.6,0.3,0.3}
\definecolor{darkblue}{rgb}{0,0,0.8}
\definecolor{darkmagenta}{rgb}{0.5,0,0.5}
\def\db{\color{darkblue}}

\def\cT{{\mathcal T}}

\def\de{\delta }

\def\da{{\dot a}}
\def\dal{{\dot \a}}
\def\db{{\dot b}}
\def\dbe{{\dot \b}}

\numberwithin{equation}{section}

\makeatletter
 \let\old@startsection=\@startsection
 \let\oldl@section=\l@section
 \renewcommand{\@startsection}[6]{\old@startsection{#1}{#2}{#3}{#4}{#5}{#6\mathversion{bold}}}
 \renewcommand{\l@section}[2]{\oldl@section{\mathversion{bold}#1}{#2}}
\makeatother

\newcommand{\lagr}{{\mathscr L}}

\DeclareMathOperator{\Str}{str}
\DeclareMathOperator{\tr}{tr}

\def\XXint#1#2#3{{\setbox0=\hbox{$#1{#2#3}{\int}$}
    \vcenter{\hbox{$#2#3$}}\kern-.5\wd0}}

\newcommand{\alg}[1]{\mathfrak{#1}}

\def\be{\begin{equation}}
\def\ee{\end{equation}}

\newcommand{\bea}{\begin{eqnarray}}
\newcommand{\eea}{\end{eqnarray}}
\newcommand{\bei}{\begin{itemize}}
\newcommand{\eei}{\end{itemize}}
\newcommand{\bee}{\begin{enumerate}}
\newcommand{\eee}{\end{enumerate}}
\newcommand{\bal}{\begin{equation}\begin{aligned}}
\newcommand{\eal}{\end{aligned}\end{equation}}

\newcommand{\ads}{${\rm  AdS}_5\times {\rm S}^5\ $}

\def\ov{\over}

\def\la{\label}

\def\a {\alpha}
\def\b {\beta}
\def\s {\sigma}
\def\g {\gamma}
\def\om {\omega}

\def\p{\phi}

\def\vk{\varkappa}

\def\T{\Theta}
\def\z{\zeta}
\def\eps{\epsilon}
\def\r {\rho}
\def\G{\Gamma}

\def\pa {\partial}

\newcommand{\su}{\alg{su}}
\newcommand{\so}{\alg{so}}

\newcommand{\as}{\alg{s}}
\newcommand{\psu}{\alg{psu}}

\newcommand{\ag}{\alg{g}}
\newcommand{\op}{\mathcal{O}}
\newcommand{\opinv}{\op^{\text{inv}}}
\newcommand{\gb}{\alg{g_b}}
\newcommand{\gf}{\alg{g_f}}
\newcommand{\gx}{\alg{g}_x}

\newcommand{\Ab}{A^\alg{b}}
\newcommand{\Af}{A^\alg{f}}
\newcommand{\gen}{\mathbf}

\newcommand{\ul}[1]{\underline{#1}}
\newcommand{\genQ}[3]{\gen{Q}^{#1 \, \ul{#2}}_{\ \, \ul{#3}}}
\newcommand{\ferm}[4]{#1_{#2 \, \ul{#3}}^{\ \, \ul{#4}}}
\def\bg{\boldsymbol{\gamma}}

\def\L{\mathscr L}
\def\H{{\mathcal H}}
\def\mI{\mathbbm{1}}

\newcommand{\sfrac}[2]{{\textstyle\frac{#1}{#2}}}

\def\dpnu{\nu}

\def\tg{\widetilde{g}}

\newcommand{\optilde}{\widetilde{\mathcal{O}}}

\newcommand{\genQind}[3]{\gen{Q}^{#1 \, \ul{#2}}_{\ \, \ul{#3}}}

\title{Puzzles of $\eta$-deformed ${\rm AdS}_5\times {\rm S}^5 $ }

\author[a,1]{Gleb Arutyunov,}
\note{Correspondent fellow at Steklov
Mathematical Institute, Moscow.}
\author[b]{Riccardo Borsato}
\author[c,1]{and Sergey Frolov}
\affiliation[a]{II. Institut f\"ur Theoretische Physik, Universit\"at Hamburg, Luruper Chaussee 149, 22761 Hamburg, Germany\\
Zentrum f\"ur Mathematische Physik, Universit\"at Hamburg, Bundesstrasse 55, 20146 Hamburg, Germany
}
\affiliation[b]{Institute for Theoretical
Physics and Spinoza Institute, Utrecht University, \\ Leuvenlaan \ 4, 3584 CE
Utrecht, The Netherlands}
\affiliation[c]{Hamilton Mathematics Institute and School of Mathematics, \\
~~Trinity College, Dublin 2, Ireland}
\emailAdd{gleb.arutyunov@desy.de} 
\emailAdd{R.Borsato@uu.nl}
\emailAdd{frolovs@maths.tcd.ie}

\abstract{
We derive the part of the  Lagrangian  for the sigma model on the $\eta$-deformed ${\rm AdS}_5\times {\rm S}^5 $ space  which is quadratic in fermions and has the full dependence on bosons. 
We  then show that there exists a field redefinition which brings the corresponding Lagrangian to the standard form of type IIB Green-Schwarz superstring. 
Reading off the corresponding RR couplings, we observe that they fail to satisfy the supergravity equations of motion, despite the presence of 
$\kappa$-symmetry. 
However, 
in a special scaling limit our solution reproduces the supergravity background found by Maldacena and Russo.   
Further, using the fermionic Lagrangian, we compute a number of new matrix elements of the tree level world-sheet scattering matrix.
We then show that after a unitary transformation  on the basis of two-particle states which is \mbox{\emph{not one-particle factorisable}}, the corresponding T-matrix factorises into two equivalent parts. 
Each part satisfies the classical Yang-Baxter equation  and coincides 
with the large tension limit of the $q$-deformed S-matrix.

}

\begin{document}

\begin{flushright}\scriptsize{ ITP-UU-15-10 \\ TCD-MATH-15-05 \\ ZMP-HH-15-19}\end{flushright}

\maketitle
\flushbottom

\section{Introduction and summary}
In many instances  a better understanding of a physical system or theory takes place once this system or theory is put under deformation.  
Recently there was an interesting proposal on how to deform the sigma model for strings on \ads  while keeping its classical
integrability \cite{Delduc:2013qra}. Deformations of this type constitute
a general class of the so-called Yang-Baxter deformations \cite{Klimcik:2002zj, Klimcik:2008eq}, which in modern parlance comprise 
$\eta$- \cite{Delduc:2013qra},  \cite{Arutyunov:2013ega}-\cite{Engelund:2014pla} and $\lambda$-deformations   \cite{Sfetsos:2013wia}-\cite{Demulder:2015lva}, as well as deformations related to solutions of the classical Yang-Baxter equation \cite{Kawaguchi:2014qwa}-\cite{vanTongeren:2015uha}.
Our primary interest in studying these deformations is that they typically break (super)symmetries of the original string model,
yet allowing for a possibility to solve them exactly.

Here we continue the studies of the $\eta$-deformed \ads sigma model based on a solution of the modified classical Yang-Baxter equation 
corresponding to the standard Dynkin diagram of the $\psu(2,2|4)$ superalgebra.   Recall that for this model the metric and the $B$-field are explicitly known  \cite{Arutyunov:2013ega}.
 At the classical level the model exhibits a local fermionic $\kappa$-symmetry and a hidden ${\rm PSU}_q(2,2|4)$ symmetry \cite{Delduc:2014kha}.
It was shown \cite{Arutyunov:2013ega} that its world-sheet bosonic tree-level scattering matrix factorises into two copies, each of which coincides under proper identification of the parameters 
with the large tension limit of the $q$-deformed S-matrix found from quantum group symmetries, unitarity
and crossing \cite{Beisert:2008tw,Hoare:2011wr}.

The aim of the present  paper is to  clarify an important question of whether or not the $\eta$-deformed model is type IIB string sigma model. 
As we will show, under certain assumptions the answer turns out to be negative. 

One way to approach this question would be to try to find an embedding of the given NSNS background  into a full solution of type IIB supergravity. 
Given complexity of the NSNS background, this appears however a rather difficult task. First of all the equation for the dilaton has many solutions 
and also many components of the RR forms seem to be switched on.  Surprisingly, $\lambda$-deformations and deformations based on solutions of 
the classical Yang-Baxter equation behave better in this respect, and some
of the metrics could be completed to a full supergravity solution. Even if successful, this approach does not however guarantee that the string sigma model in the corresponding 
supergravity background will actually coincide with a deformed model.

Another way to proceed is to note that
the Green-Schwarz action restricted to quadratic order in fermions contains all the information about the background fields.
The corresponding Lagrangian has the form, see {\it e.g.} \cite{Tseytlin:1996hs, Cvetic:1999zs, Wulff:2013kga}, 
\be\nonumber
\begin{aligned}
{\mathscr L}_{\Theta^2}= - \frac{g}{2}  \, i \, \bar{\Theta}_I \,(\g^{\a\b} \delta^{IJ}+\epsilon^{\a\b} \sigma_3^{IJ}) e^m_\a \G_m \, D^{JK}_\b \Theta_K  , 
\end{aligned}
\ee
where $\Theta_I$ are two Majorana-Weyl fermions of the same chirality.
The operator $D^{IJ}_\a$ acting on fermions has the following expression
\be
\nonumber
\begin{aligned}
D^{IJ}_\a  = &
\delta^{IJ} \left( \pa_\a  -\frac{1}{4} \omega^{mn}_\a \G_{mn}  \right)
+\frac{1}{8} \sigma_3^{IJ} e^m_\a H_{mnp} \G^{np}
\\
&-\frac{1}{8} e^{\varphi} \left( \epsilon^{IJ} \G^p F^{(1)}_p + \frac{1}{3!}\sigma_1^{IJ} \G^{pqr} F^{(3)}_{pqr} + \frac{1}{2\cdot5!}\epsilon^{IJ} \G^{pqrst} F^{(5)}_{pqrst}   \right) 
e^m_\a \G_m 
\, ,
\end{aligned}
\ee
where $(e,\omega,H)$ constitute a vielbein, the spin connection and the field strength of a $B$-field, while $F$'s are RR forms and $\varphi$ is a dilaton. Note that the dilaton and RR forms appear only through the combination $e^{\varphi}F$.
An approach we undertake in this paper will be therefore to work out the quadratic fermionic action starting from the $\eta$-deformed action of \cite{Delduc:2013qra} and some conveniently chosen representative of the  
coset  ${\rm PSU}(2,2|4)/{\rm SO}(1,4)\times{\rm SO}(5)$. Then we need to find a field redefinition which brings this action into the Green-Schwarz canonical form above.
This would allow us to identify the background fields and further check if they satisfy the equations of motion of type IIB supergravity and, in particular, to find a solution for the dilaton.
Such a strategy works perfectly, for instance, for the \ads sigma model \cite{Metsaev:1998it}.

We succeeded in constructing a field redefinition which brings the quadratic fermionic Lagrangian of the $\eta$-deformed theory to the canonical form. 
However, reading off the corresponding RR couplings\footnote{Throughout the  paper we loosely refer to $F$-forms as to RR couplings although as found a posteriori they are not a part of a supergravity background.} in section \ref{RRresults}, 
we find that they fail to satisfy the supergravity equations! The next surprising observation is  that these couplings do not meet the necessary conditions of the mirror duality \cite{Arutyunov:2014jfa}, and, as the consequence, 
the mirror background \cite{Arutyunov:2014cra} is not reproduced in the expected limit $\eta\to 1$. Although this duality is a symmetry of the exact S-matrix, it involves rescaling of the string tension and therefore 
its absence in the classical Lagrangian might be explained by the order of limits problem.

Another interesting observation, which supports the correctness of our result, concerns a reproduction of a known string background. As was previously noted by one of us \cite{SF},
there is a special scaling limit under which the $\eta$-deformed metric and $B$-field reproduce the NSNS part of the Maldacena-Russo background \cite{Maldacena:1999mh} dual to a non-commutative Yang-Mills theory.\footnote{The NSNS part of the Maldacena-Russo background also appears in the context of deformations related to solutions of the classical Yang-Baxter equation \cite{Matsumoto:2014gwa}.}  
Now we observe that in this limit the RR couplings we found precisely reproduce the rest of the Maldacena-Russo background which is a genuine solution of type IIB supergravity.

In view of these surprising results it is time to ask how our findings are compatible with $\kappa$-symmetry, especially in view of the work  \cite{Grisaru:1985fv,Bergshoeff:1985su}, where 
it was shown that the fulfilment of the supergravity constraints is sufficient for the Green-Schwarz action to be invariant under $\kappa$-symmetry.
To answer this question, we have explicitly developed the $\kappa$-symmetry transformations of the $\eta$-deformed model \cite{Delduc:2013qra}  to the leading order in fermions. 
We then find that the {\it same} field redefinition which brings the original Lagrangian to the canonical Green-Schwarz form also brings the $\kappa$-symmetry variations 
of the target-space coordinates to the standard form in type IIB theory. Then the variation of the world-sheet metric automatically acquires the standard form as well and 
contains RR couplings, allowing therefore for their independent determination. The RR couplings we read off from the $\kappa$-variations of the world-sheet metric coincide 
with what we found from the canonical Lagrangian. Clearly, at the level of the quadratic Lagrangian $\kappa$-symmetry cannot say anything about equations of motion for 
RR couplings. Indeed, the latter couple to fermion bilinears and their leading order $\kappa$-symmetry variations should be combined with variations of the 
quartic fermionic terms to produce differential constraints on $F$'s which guarantee invariance of the action. The failure of the RR couplings 
to satisfy the supergravity equations including the Bianchi identities suggests  that $\kappa$-symmetry transformations in the $\eta$-deformed theory 
will deviate  from that of the Green-Schwarz superstring beyond the leading order.

Now we comment on the issue of field redefinitions. How can one be sure that no other field redefinitions exist which produce better results for RR couplings?
Note that we already brought our Lagrangian to the canonical form where NSNS fields $(e,\omega,H)$  appear automatically to be the same as determined from the 
bosonic action. Thus, if we want to perform further field redefinitions we have to require that they keep the NSNS part of the fermionic action untouched and
change exclusively the RR content. Moreover, in the limit $\eta\to 0$ such redefinitions should either trivialize  or become a symmetry transformation of the 
undeformed model and the same must be true for the scaling limit to the Maldacena-Russo background.  By performing an infinitesimal analysis we then show that 
there is no smooth $\eta$-dependent transformation of fields which reduces to the identity in the limit $\eta\to 0$ and does not modify the NSNS part of the action.
An existence of discrete, {\it i.e.} $\eta$-independent transformations is much more difficult to rule out and, therefore, our result on  non-existence 
of the supergravity background is only applied if no such transformation  exists.

Since inclusion of fermions leads to a variety of puzzling results, we find it interesting to extend our earlier computation of the 
bosonic tree-level two-particle S-matrix  \cite{Arutyunov:2013ega} to include fermions. What we are computing is in fact T-matrix. In the purely bosonic case 
this T-matrix factorises into two parts, each  satisfies the classical Yang-Baxter equation ({\it i.e.} it is a classical $r$-matrix) and coincides with the leading term 
of the large tension expansion of the known $q$-deformed S-matrix. In other words, this T-matrix has precisely the same properties as its undeformed counterpart.
We then use our quadratic fermionic Lagrangian  to compute new elements in the scattering matrix and discover that this time it does not factorise on two copies. 
This nice property is spoiled by  Boson+Fermion $\to$ Boson+Fermion scattering elements.
However, there exists 
a  unitary momentum-independent transformation of the basis of two-particle states which brings our T-matrix to a factorisable form.
Each factor coincides with the large tension limit of the $\psu_q(2|2)$-invariant S-matrix. 
The  transformation of the two-particle basis we found
does not however admit a factorisation on transformations of one-particle states. A similar situation has been observed at the one- and two-loop level where 
integrability of the corresponding S-matrix obtained through unitarity-based methods also required a (momentum-dependent) one-particle-unfactorisable rotation
on the basis of two-particle states \cite{Engelund:2014pla}.\footnote{One important difference, though, is
that in \cite{Engelund:2014pla} factorisation of the T-matrix could be also achieved by performing a one-particle transformation which made however the spin and dimension of single-particle states complex. }   
We note that one can think about our unitary transformation as acting on the 
Hamiltonian which then becomes highly non-local.  Moreover, this transformation is $\eta$-independent and is therefore a symmetry of the undeformed S-matrix. 

The light-cone Hamiltonian has an important feature. Although the theory has only $q$-deformed supersymmetry, the masses of bosons and fermions in the light-cone 
Hamiltonian appear to be the same and they both have a mild dependence of the deformation parameter. Thus, the BMN vacuum is supersymmetric just as it was in the undeformed case.  

The paper is organised as follows. In the next section we recall the basic facts about the $\eta$-deformed \ads sigma model, describe the main steps in the derivation of the 
fermionic quadratic Lagrangian, present and discuss our main result on the RR couplings.  Section \ref{sec:kappa} contains an alternative derivation of the 
RR couplings from $\kappa$-symmetry. Section \ref{sec:red} is devoted to the discussion of residual field redefinitions. In section \ref{sec:Tmatrix} we present the T-matrix and discuss
how to achieve its factorisation and fulfilment of the Yang-Baxter equation. Definitions and technical derivations are relegated to three appendices. For the reader's convenience 
we also attach appendix \ref{app:IIBsugra} with the equations of motion of type IIB supergravity.

\section{Quadratic fermionic Lagrangian and RR couplings}

\subsection{$\eta$-deformed model}\label{subsec:def_model}

Let us recall that  the Lagrangian density of the  $\eta$-deformed model is given by  \cite{Delduc:2013qra}
\bea
\label{defLag}
{\mathscr L}=-\frac{g}{4}(1+\eta^2)\big(\gamma^{\a\b}-\eps^{\a\b}\big)\, {\rm str}\Big[\tilde{d}(A_{\a})\frac{1}{1-\eta R_\ag \circ d}(A_{\beta}) \Big]\,,
\eea
and the action $S$ is normalised as $S=\int {\rm d\sigma}{\rm d\tau} \mathscr{L}$.
We use the notations and conventions from  \cite{Arutyunov:2009ga}: $\eps^{\tau\sigma}=1$; $\gamma^{\a\b}=h^{\a\b}\sqrt{-h}$ , $\gamma^{\tau\tau}<0$; $g$ is the effective string tension. 
The current $A_{\a}=-\ag^{-1}\pa_{\a}\ag$, where $\ag\equiv \ag(\tau,\sigma)$ is a  coset representative from ${\rm PSU}(2,2|4)/{\rm SO}(4,1)\times {\rm SO}(5)$. The operators
$d$ and $\tilde{d}$ acting on the currents $A_{\alpha}$ are defined as 
\bea
\nonumber
d&=&P_1+\frac{2}{1-\eta^2}P_2-P_3,\qquad\qquad
\tilde{d}=-P_1+\frac{2}{1-\eta^2}P_2+P_3\, ,
\eea
where $P_i$, $i=0,1,2,3$, are projections on the corresponding components of the ${\mathbb Z}_4$-graded decomposition of the superalgebra $\psu(2,2|4)$,
%$
%\mathscr{G}=\mathscr{G}^{(0)}\oplus \mathscr{G}^{(1)}\oplus \mathscr{G}^{(2)}\oplus \mathscr{G}^{(3)}$, and 
%$\mathscr{G}^{(0)}$ coincides with $\so(4,1)\times \so(5)$, 
see appendix \ref{sec:algebra-basis}.

The operator $R_\ag$ acts on $M\in {\mathscr G}$ as follows
\be\label{Rgop}
R_\ag(M) = \ag^{-1}R(\ag M\ag^{-1})\ag\, ,
\ee
where $R$ is a linear operator on $\mathscr{G}$ which in this paper we  define as \be\label{Rop}
R(M)_{ij} = -i\, \eps_{ij} M_{ij}\,,\quad \eps_{ij} = \left\{\begin{array}{ccc} 1& \rm if & i<j \\
0&\rm if& i=j \\
-1 &\rm if& i>j \end{array} \right.\,,
\ee
where $M$ is an arbitrary  $8\times 8$ matrix. This choice of $R$ corresponds to the standard Dynkin diagram of $\psu(2,2|4)$.
 
\begin{figure}[t]
\begin{center}
\includegraphics[width=1.0\textwidth]{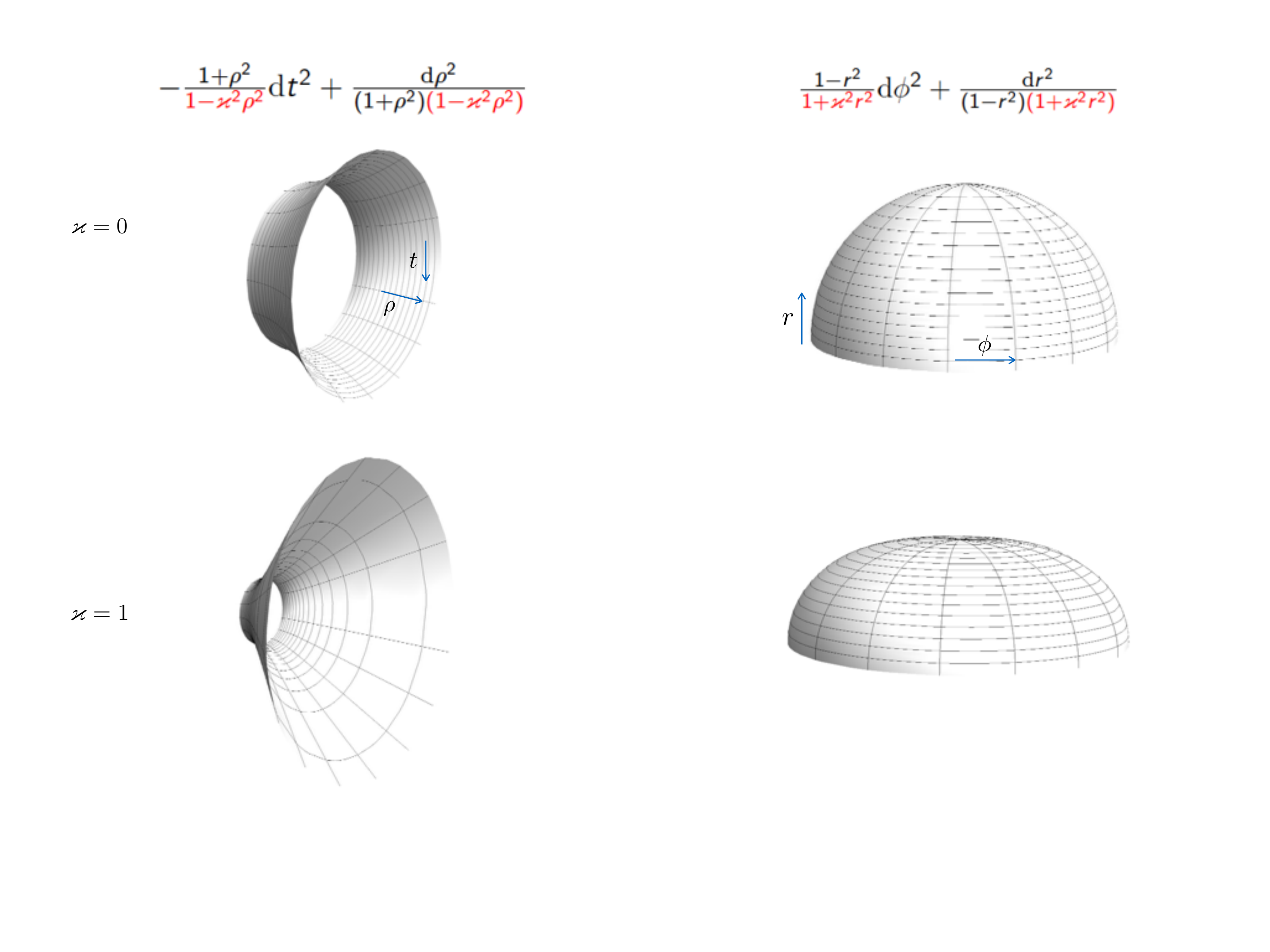}
\vspace{-2cm}
\caption{Geometry of the $\eta$-deformed background depicted via embeddings of two-dimensional surfaces  $(t,\rho)$ and $(\phi,r)$ and  into three-dimensional
pseudo-euclidean and euclidean spaces, respectively.}
\label{MH}
\end{center}
\end{figure}

In our previous paper \cite{Arutyunov:2013ega} the fermions were switched off, a particular choice of the bosonic coset element $\gb$ was made, and  the operator $1/(1-\eta R_\gb \circ d)$ was found and used to determine the 
bosonic part of the $\eta$-deformed action. Introducing the convenient deformation parameter $\varkappa=\frac{2\eta}{1-\eta^2}$ and 
 $\tilde g=g\sqrt{1+\vk^2}$,
the $\eta$-deformed metric and the $B$-field can be written in the form
  \bal\la{dma}
{1\ov \tilde g}ds^2_{\alg{a}}=&
-\frac{dt^2\left(1+\rho ^2\right)}{ 1-\varkappa ^2 \rho ^2}
+\frac{d\r^2}{ \left(1+\rho ^2\right) \left(1-\varkappa ^2 \rho ^2\right)}
\\&+\frac{d\z^2 \rho ^2}{1+ \varkappa ^2 \rho ^4 \sin ^2\z }
+\frac{d\psi_1^2\rho ^2 \cos
   ^2\z}{ 1+\varkappa ^2 \rho ^4 \sin ^2\z}+d\psi_2^2 \rho ^2 \sin ^2\z \,,
\eal
\bal\la{dms}
{1\ov \tilde g}ds^2_{\alg{s}}=&\frac{d\p^2
  \left(1-r^2\right)}{1+\varkappa^2 r^2}+\frac{dr^2
  }{ \left(1-r^2\right) \left(1+\varkappa ^2 r^2\right)}
 \\& +\frac{d\xi^2  r^2}{1+ \varkappa ^2 r^4 \sin ^2\xi}
+\frac{d\p_1^2  r^2 \cos ^2\xi }{1+ \varkappa ^2
   r^4 \sin ^2\xi } +d\p_2^2  r^2 \sin^2\xi \,,
\eal
\bal\la{Ba}
B_{\psi_1\z} ={\tilde g\ov2} \varkappa\frac{ \rho ^4 \sin 2 \zeta}{1+ \varkappa ^2 \rho ^4 \sin ^2\z}\,,\quad
B_{\phi_1\xi} =-{\tilde g\ov2} \varkappa\frac{ r^4 \sin 2 \xi }{1+ \varkappa ^2 r^4 \sin^2\xi}\,  .
\eal
The effect of the deformation on the shape of AdS$_2$ and ${\rm S}^2$ is shown on Figure  \ref{MH}.

In what follows for convenience we are enumerating the coordinates as 
\be\label{param}
\begin{aligned}
& X^0 = t, & \quad X^1= \psi^2,& \quad X^2= \psi^1, &\quad X^3=\zeta, &\quad X^4= \rho,\\
& X^5 = \phi, & \quad X^6= \phi^2,& \quad X^7= \phi^1, &\quad X^8=\xi, &\quad X^9= r,
\end{aligned}
\ee
so that the non-vanishing components of the $B$-field are $B_{23}$ and $B_{78}$ while the non-vanishing components of the field strength $H_{KLM}$ are $H_{234}$ and $H_{789}$.
\medskip

To find the part of the $\eta$-deformed action quadratic in fermions we use the following coset element
\bea
\label{original_coset}
\alg{g}=\gb \gf\,,
\eea
where the bosonic element  $\gb=\Lambda \gx$ is the same as in \cite{Arutyunov:2013ega}. The element $\gf$ which comprises fermionic degrees of freedom 
can be defined through the exponential map $\gf= \exp\chi$, or as $\gf= \chi + \sqrt{1+\chi^2}$. The two choices produce the same expression if we stop at quadratic order.  The Lie algebra element $\chi$ is a linear combination of odd generators of the $\psu(2,2|4)$ algebra\footnote{See appendix \ref{sec:algebra-basis} for the definition of the  $\psu(2,2|4)$ generators we use.} $\chi \equiv \genQ{I}{\a a}{} \ferm{\theta}{I}{\a a}{}$.

The current $A=-\alg{g}^{-1}{\rm d}\alg{g}$ can be decomposed in terms of linear combinations of the generators of the $\psu(2,2|4)$ algebra 
\be
A= L^m \gen{P}_m + \frac{1}{2} L^{mn} \gen{J}_{mn} + \ferm{L}{I}{\a a}{} \genQ{I}{\a a}\,.
\ee
It is useful to look at the purely bosonic and purely fermionic currents separately, that are found by switching off  fermions and  bosons respectively. The purely bosonic current is a combination of even generators $\gen{P}_m$ and $\gen{J}_{mn} $ 
\be
\Ab= -\gb^{-1} {\rm d} \gb = e^m \gen{P}_m + \frac{1}{2} \omega^{mn} \gen{J}_{mn} \,,
\ee
where $e^m = e^m_M dX^M$ is the \ads  vielbein  and $\omega^{mn} = \omega^{mn}_M dX^M$ is  the corresponding spin connection whose explicit expressions can be found in appendix \ref{app:CD}.

The purely fermionic current is decomposed in terms of even and odd generators
\be
\Af= -\gf^{-1} {\rm d} \gf = \Omega^m \gen{P}_m + \frac{1}{2} \Omega^{mn} \gen{J}_{mn} + \ferm{\Omega}{I}{\a a}{} \genQind{I}{\a a}{}
\ee
where we have defined the yet to-be-determined quantities $\Omega^m, \Omega^{mn}, \ferm{\Omega}{I}{\a a}{} $.
Expanding $\gf$ in powers of $\theta$ up to quadratic order in fermions we find
\be
\begin{aligned}
\Af=& -\gf^{-1} {\rm d} \gf \\
=& - \gen{Q}^{I} \, d \theta_I 
+ \frac{i}{2} \delta^{IJ} \bar{\theta}_I  \bg^m  d \theta_J \ \gen{P}_m 
- \frac{1}{4} \epsilon^{IJ} \bar{\theta}_I  \bg^{mn}  d \theta_J \ \check{\gen{J}}_{mn} 
+ \frac{1}{4} \epsilon^{IJ} \bar{\theta}_I   \bg^{mn}  d \theta_J \ \hat{\gen{J}}_{mn} \, ,
\end{aligned}
\ee
where $\check{}$ and $\hat{}$~ refers to the quantities related to ${\rm AdS}_5$ and ${\rm S}^5$, respectively, and  
the matrices $\bg_n$ are defined in (\ref{eq:def16x16-gamma}).
The computation of the full current is similar and one gets
\bea\label{current}
%\begin{aligned}
A= -\alg{g}^{-1}{\rm d}\alg{g}=\Af+ \gf^{-1}\Ab\gf 
&=&\left( e^m +\frac{i}{2}  \bar{\theta}_I  \bg^m  D^{IJ} \theta_J \right) \gen{P}_m 
 - \gen{Q}^{I} \, D^{IJ} \theta_J \\
& +&\frac{1}{2} \omega^{mn} {\gen{J}}_{mn} - \frac{1}{4} \epsilon^{IJ} \bar{\theta}_I \left( \bg^{mn} \check{\gen{J}}_{mn} - \bg^{mn} \hat{\gen{J}}_{mn}  \right)  D^{JK} \theta_K \, ,
\nonumber
%\end{aligned}
\eea
where the operator $D^{IJ}$ acting on fermions $\theta$ is given by
\be\label{eq:op-DIJ-psu224-curr}
D^{IJ} = \delta^{IJ} \left( {\rm d} - \frac{1}{4} \omega^{mn} \bg_{mn}  \right)
+ \frac{i}{2} \epsilon^{IJ} e^m  \bg_m  . 
\ee
%We then check that it is enough to compute the fermionc current $\Af$ and use the ``covariantisation rule'' ${\rm d}\theta_I\to D^{IJ}\theta_J$ to get the full current~\cite{Metsaev:1998it}.
Sometimes it is useful to split this operator as
\be
D^{IJ} = \mathcal{D}^{IJ}
+ \frac{i}{2} \epsilon^{IJ} e^m  \bg_m  ,
\qquad
\mathcal{D}^{IJ} \equiv\delta^{IJ}\mathcal{D},
\ee
where $\mathcal{D}=  {\rm d} - \frac{1}{4} \omega^{mn} \bg_{mn} $ is the covariant derivative acting on fermions.

%Imposing on the current the flatness condition
%\be
%\epsilon^{\a\b} \Big(\pa_\a A_\b -\frac{1}{2} [A_\a,A_\b]\Big)=0
%\ee
%and projecting on the bosonic generators, we find the following equations for the vielbein and the spin connection
%\be\label{eq:d-veilbein}
%\begin{aligned}
%\epsilon^{\a\b} (\pa_\a e^m_\b - \omega^{mq}_\a e_{q\b}) &= 0,
%\end{aligned}
%\ee
%\be\label{eq:d-spin-conn}
%\begin{aligned}
%\epsilon^{\a\b} (\pa_\a \check{\omega}^{mn}_\b - \check{\omega}^{m}_{\ p\a} \check{\omega}^{pn}_\b - \check{e}^m_\a \check{e}^n_\b) &= 0,
%\qquad
%\epsilon^{\a\b} (\pa_\a \hat{\omega}^{mn}_\b - \hat{\omega}^{m}_{\ p\a} \hat{\omega}^{pn}_\b + \hat{e}^m_\a \hat{e}^n_\b) &= 0.
%\end{aligned}
%\ee

The action of the projections $d$ and $\tilde{d}$ on the current $A$ are found from the formulae
{\small
\be
\begin{aligned}
d(\gen{J}_{mn}) &=\tilde{d}(\gen{J}_{mn})= \gen{0}, ~~
d(\gen{P}_m) =\tilde{d}(\gen{P}_m)= \frac{2}{1-\eta^2} \gen{P}_m, ~~
d(\gen{Q}^{I}) =-\tilde{d}(\gen{Q}^{I}) = (\sigma_3)^{IJ} \gen{Q}^{J}\,.
\end{aligned}
\ee
}

\normalsize
\vskip -0.5cm
\noindent
In  particular, the $\gen{J}$-part of the current is irrelevant for the computation of the Lagrangian, since it is projected out when defining the coset.

The next step consists in constructing the inverse of the operator 
\be\label{eq:defin-op-def-supercoset}
\op=1-\eta R_\ag \circ d\, .
\ee
To this end, we find convenient to expand it in powers of fermions $\theta$ as
\be\la{eq:pertop}
\op=\op_{(0)}+\op_{(1)}+\op_{(2)}+\cdots\, ,
\ee
where $\op_{(k)}$ is the contribution at order $\theta^k$. 
On generators $\gen{J}$ of degree 0 the inverse operator $\op$ acts  as the identity, at any order in fermions.
To find its action  on the other generators, we invert it perturbatively in powers of fermions:
\be
\op^{-1}=\opinv_{(0)}+\opinv_{(1)}+\opinv_{(2)}+\cdots\,,
\ee
where $\opinv_{(k)}$ is the contribution at order $\theta^k$. The leading contribution $\opinv_{(0)}$ was already derived in 
\cite{Arutyunov:2013ega}.
Demanding that $\op\cdot\op^{-1}=\op^{-1}\cdot\op=1$ we find
\begin{equation}\label{eq:expans-ferm-inv-op}
\begin{aligned}
\opinv_{(1)} & = - \opinv_{(0)} \circ \op_{(1)} \circ \opinv_{(0)} , \\
\opinv_{(2)} & = - \opinv_{(0)} \circ \op_{(2)} \circ \opinv_{(0)} - \opinv_{(1)} \circ \op_{(1)} \circ \opinv_{(0)}.
\end{aligned}
\end{equation}
We will not need higher order contributions. To keep the discussion transparent, for an explicit construction of $\op^{-1}$ up to quadratic order in fermions we refer the reader to appendix \ref{sec:inverse-op}.

\subsection{Quadratic fermionic Lagrangian}\label{sec:QFL}

\newcommand{\1}{{\bf -}}
\newcommand{\2}{{ \bf +}}

Substituting now all the ingredients, that is the current (\ref{current}) and $\op^{-1}$ into the Lagrangian (\ref{defLag}), we expand it up to quadratic order in fermions.
At leading order we find the already known  \cite{Arutyunov:2013ega} bosonic Lagrangian 
\be
{\mathscr L}_{(0)} = - \frac{\tilde{g}}{2}  (\gamma^{\a\b} - \epsilon^{\a\b}) \ e^m_\a e^n_\b {k_n}^p \eta_{mp},
\ee
where $\ e^m_\a=e^m_M \pa_{\a}X^M$ is the vielbein of \ads and the coefficients ${k_n}^p$ are presented in appendix  \ref{sec:inverse-op}, see eqs.(\ref{eq:k-res1}) and (\ref{eq:k-res2}).
We can rewrite this result in the standard sigma model form recovering the deformed metric (\ref{dma}), (\ref{dms}) and the $B$-field (\ref{Ba}), which happens due to the identities 
\bea
\ e^m_{(M }e^n_{N)} {k_n}^p \eta_{mp} =\widetilde{e}^m_M\widetilde{e}^n_N \eta_{mn}\, , ~~~~~~~ e^m_{[M }e^n_{N]} {k_n}^p \eta_{mp} =B_{MN}\, ,
\eea
where $\widetilde{e}^m_M$ is a vielbein for the deformed metric which we present in appendix  \ref{app:CD} and $\eta$ is the Minkowski metric (\ref{eq:min_metric}).
\smallskip

In the expansion of the Lagrangian (\ref{defLag}) in powers of fermions, contributions to a given power come from three sources: from the current $A_\a$, from the operator $\op^{-1}$ and from the current $A_\b$. 
Thus, the quadratic fermionic Lagrangian is a sum of six terms 
\be\la{eq:qadr_lagr}
\begin{aligned}
{\mathscr L}_{(2)}&=\lagr_{\{002\}}+\lagr_{\{200\}}+\lagr_{\{101\}}\\
&+\lagr_{\{011\}}+\lagr_{\{110\}}+\lagr_{\{020\}}   \, ,
\end{aligned}
\ee
where three numbers in the brackets indicate powers of fermions coming from $A_\a$, $\op^{-1}$ and $A_\b$, respectively.
For the first  two contributions we find
\be
\label{L002}
\begin{aligned}
\lagr_{\{002\}} & = - \frac{\tilde{g}}{2} (\gamma^{\a\b} - \epsilon^{\a\b}) \, \frac{i}{2}\bar{\theta}_I (e^m_\a {k^n}_{m}\bg_n ) D^{IJ}_\b \theta_J , \\
\lagr_{\{200\}} & = - \frac{\tilde{g}}{2} (\gamma^{\a\b} - \epsilon^{\a\b}) \, \frac{i}{2}\bar{\theta}_I (e^m_\b {k_{m}}^{n}\bg_n ) D^{IJ}_\a \theta_J ,
\end{aligned}
\ee
where ${k^n}_{m} = \eta^{nq}{k_q}^p \eta_{pm}$. Note that the sum of $\lagr_{\{002\}}+\lagr_{\{200\}}$ gives a non-trivial contribution also to the Wess-Zumino term, since the matrix $k_{mn}$ has a non-vanishing anti-symmetric part. 

Concerning the contribution $\{101\}$, 
in appendix~{\color{red}\ref{app:der-Lagr-101}} we manipulate the initial result~\eqref{eq:orig-lagr-101} to bring it to the form most close to the canonical one  
\bea
\label{eq:Lagr-101}
%\begin{aligned}
\lagr_{\{101\}} &=& - \frac{\tilde{g}}{2}  \epsilon^{\a\b} \bar{\theta}_L \, i \, e^m_\a \bg_m \left( \sigma_3^{LK}  D^{KJ}_\b \theta_J
-\frac{\vk}{1+\sqrt{1+\vk^2}} \ \epsilon^{LK}  \mathcal{D}^{KJ}_\b \theta_J \right) \, ,\\
%&=& - \frac{\tilde{g}}{2}  \epsilon^{\a\b} \bar{\theta}_I \left( \sigma_3^{IJ}  
 %-\frac{\vk}{1+\sqrt{1+\vk^2}} \ \epsilon^{IJ}  \right) \, i \, e^m_\a \bg_m \mathcal{D}_\b \theta_J  + \frac{\tilde{g}}{4}  \epsilon^{\a\b} \bar{\theta}_I  \sigma_1^{IJ}  e^m_\a \bg_m e^n_\b \bg_n  \theta_J ,
 \nonumber
%\end{aligned}
\eea
which holds up to a total derivative.

Now we spell out the contributions stemming from the inverse operator taken at first order in the $\theta$ expansion.
The two contributions $\{011\},\{110\}$ can be naturally considered together\footnote{The result can be put in this form thanks to the properties~\eqref{eq:swap-lambda}.}
\bea\nonumber
%\begin{aligned}
\lagr_{\{011\}+\{110\}}  %= & - \frac{\tilde{g}}{4}  (\gamma^{\a\b} - \epsilon^{\a\b}) \\
%& \bar{\theta}_K \Bigg[  (-\vk \sigma_1^{KI}+(-1+\sqrt{1+\vk^2})\delta^{KI}) \bar{\Delta}^1_n 
%+ (\vk \sigma_3^{KI} - (-1+\sqrt{1+\vk^2})\epsilon^{KI}) \bar{\Delta}^3_n  \Bigg] \\
%& (k^{nm}e_{m\a} D^{IJ}_\b +k^{mn}e_{m\b} D^{IJ}_\a )\theta_J \\
&= & - \frac{\tilde{g}}{4}  (\gamma^{\a\b} - \epsilon^{\a\b}) 
 \bar{\theta}_K 
\Bigg[ - (\vk \sigma_1^{KI}-(-1+\sqrt{1+\vk^2})\delta^{KI}) \left( i\bg_p +\frac{1}{2}\bg_{mn}  \lambda_{p}^{mn} \right) \\
& +& (\vk \sigma_3^{KI} - (-1+\sqrt{1+\vk^2})\epsilon^{KI}) \ i\bg_n {\lambda_p}^n  \Bigg] 
 (k^p_{\ q}e^q_{\a} D^{IJ}_\b +{k_{q}}^{p}e^q_{\b} D^{IJ}_\a )\theta_J.
%\end{aligned}
\eea
Finally, the last contribution to the Lagrangian is delivered by the term where the inverse operator is taken at order $\theta^2$. We find
\bea
%\begin{aligned}
\lagr_{\{020\}}  &= & - \frac{\tilde{g}}{2} (\gamma^{\a\b} - \epsilon^{\a\b}) \  \frac{\vk}{4} e^v_\a e^m_\b \, {k^{u}}_v {k_m}^n \, \bar{\theta}_K  \nonumber \\
\Bigg[  
&-& 2 \delta^{KI} \left(  \bg_u \left(\bg_n +\frac{i}{4} \lambda_n^{pq} \bg_{pq} \right)
- \frac{i}{4} \bg_{pq}  \bg_n \lambda^{pq}_{\ u}\right) 
  - \epsilon^{KI} \left(  \bg_u {\lambda_n}^p \bg_p 
-  \bg_p \bg_n {\lambda_u}^p \right) \nonumber \\
& -&(-1+\sqrt{1+\vk^2}) \delta^{KI} \bigg( \left( \bg_u -\frac{i}{2} \bg_{pq}  \lambda_u^{pq} \right) \left(\bg_n +\frac{i}{2} \lambda_n^{rs} \bg_{rs} \right) 
 + \bg_p  {\lambda_u}^p {\lambda_n}^r \bg_r \bigg) \nonumber \\
& -&(-1+\sqrt{1+\vk^2}) \epsilon^{KI} \bigg(- \bg_p {\lambda_u}^p \left(\bg_n +\frac{i}{2} \lambda_n^{rs} \bg_{rs}\right) 
 + \left( \bg_u -\frac{i}{2} \bg_{pq}  \lambda_u^{pq} \right) {\lambda_n}^r \bg_r \bigg) \nonumber \\
& + &\vk \sigma_1^{KI} \bigg( \left( \bg_u -\frac{i}{2} \bg_{pq}  \lambda_u^{pq} \right) \left(\bg_n +\frac{i}{2} \lambda_n^{rs} \bg_{rs}\right) 
 - \bg_p  {\lambda_u}^p {\lambda_n}^r \bg_r \bigg) \nonumber \\
& - & \vk \sigma_3^{KI} \bigg( \bg_p{\lambda_u}^p \left(\bg_n +\frac{i}{2} \lambda_n^{rs} \bg_{rs}\right) 
 + \left( \bg_u -\frac{i}{2} \bg_{pq}  \lambda_u^{pq} \right) {\lambda_n}^r \bg_r \bigg)
\Bigg] \theta_I .
%\end{aligned}
\eea
The last two expressions involve the coefficients ${\lambda_m}^n, \lambda_m^{np}, \lambda_{mn}^{p}, \lambda_{mn}^{pq}$ which are collected in appendix \ref{sec:inverse-op}.

Summing up all the above contributions, we discover that the result is \emph{not}  the standard Green-Schwarz Lagrangian, see the Introduction. 
Yet, in the undeformed limit it reduces to that one. Indeed, when $\varkappa\to 0$, the contributions $\lagr_{\{011\}+\{110\}}$ and $\lagr_{\{020\}}$ vanish,
while $k_{mn}$ in eq.(\ref{L002}) becomes $\eta_{mn}$, so that  (\ref{L002}) transforms into the standard kinetic term, while (\ref{eq:Lagr-101}) provides its
Wess-Zumino completion.\footnote{In particular, the self-dual five-form of the \ads background arises from the term with $\eps^{IJ}$ in the definition (\ref{eq:op-DIJ-psu224-curr}) of the operator $D^{IJ}$ 
\cite{Metsaev:1998it}. } This is of course expected because the canonical form of  the undeformed Lagrangian is intrinsically built in our construction based on global symmetries 
and the choice  (\ref{original_coset}) of the coset representative.  
On the other hand in the deformed model the \ads coset plays an auxiliary role because only six commuting isometries remain unbroken. 
It is thus clear that the Lagrangian we got describes couplings of bosons with fermion bilinears written with a more or less arbitrary choice of coordinates and that 
field redefinitions will in general modify its form.
Our next task is therefore to find a field redefinition that will cast (\ref{eq:qadr_lagr}) in the desired canonical form.

To search for necessary field redefinitions we need a guidance principle. All terms in $\lagr_{(2)}$ can be split into two parts: the kinetic part $\lagr^\pa$, which contains all couplings of the form  $\theta\pa\theta$,
and the mass part  $\theta\theta$, which constitutes the rest of the Lagrangian.
The idea is to concentrate just on the kinetic part
and find field redefinitions which bring it to the canonical form. In the process new mass terms will be generated  and we look at all of them at the very end. 
Clearly, two types of field redefinitions are possible: rotations of fermions $\theta^I\to U^{IJ}\theta^J$ with 
coefficients $U^{IJ}$ depending on bosons and shifts of bosons by fermion bilinears. In the second case, the bosonic Lagrangian $\lagr_{(0)}$ will generate contributions to $\lagr_{(2)}$ and if 
we do not want to create higher derivatives of $\theta$, the corresponding shifts should be of the form 
\bea\label{eq:shift_bosons}
X^M \to X^M +\bar{\theta}^I f^M_{IJ} (X)\, \theta^J\, 
\eea
with boson-dependent coefficients $f^M_{IJ} (X)$.

Next, all the terms in  $\lagr^\pa$ are naturally divided according to their symmetry properties into two categories 
$\lagr_\2^\pa$ and $\lagr_\1^\pa$.
Given an expression of the form $\theta_I M^{IJ} \pa\theta_J$, we classify it according to
\be\la{eq:sym_prop}
\begin{aligned}
\theta_I M^{IJ} \pa\theta_J = +\pa\theta_I M^{IJ} \theta_J &\quad\implies\quad \lagr_\2^\pa\, ,\\
\theta_I M^{IJ} \pa\theta_J = -\pa\theta_I M^{IJ} \theta_J &\quad\implies\quad \lagr_\1^\pa\, .
\end{aligned}
\ee
The symmetry properties are manifested through purely algebraic manipulations, not by integrating by parts. They are inherited from symmetries of gamma matrices contained in $M^{IJ}$
and from the behaviour of $M^{IJ}$ under the exchange of $I,J$. We then show in appendix \ref{app:CGS}  that there exists a choice of the coefficients in (\ref{eq:shift_bosons})
such that the corresponding shift completely removes $\lagr_\2^\pa$, leaving behind a bunch of new mass terms. As to $\lagr_\1^\pa$, it remains untouched under this shift because 
the symmetry properties of the derivative couplings generated by (\ref{eq:shift_bosons}) are opposite to that of $\lagr_\1^\pa$. The only manipulations we are left with at this point are 
boson-dependent rotations of fermions. Since $\lagr_\1^\pa$ and the canonical kinetic term share the same symmetry (\ref{eq:sym_prop}), a rotation which transforms one into the other 
always exists and we find its explicit form in appendix \ref{app:CGS}. 

\smallskip

Through the shift of bosons and the rotation of fermions we generated quite a lot of new mass terms. It is now time to sum them up 
and group together according to their tensorial structures. Quite remarkably, after this is done, the mass part turns out to automatically fit the canonical 
arrangement. In terms of a $32$-dimensional Majorana fermion $\T$ of positive chirality \eqref{eq:def-32-dim-Theta}, our Lagrangian is therefore
\be\label{eq:lagr-quad-ferm}
\begin{aligned}
\lagr_{(2)}&= - \frac{\tilde{g}}{2}  \, i \, \bar{\Theta}_I \,(\g^{\a\b} \delta^{IJ}+\epsilon^{\a\b} \sigma_3^{IJ}) \widetilde{e}^m_\a \G_m \, \widetilde{D}^{JK}_\b \Theta_K  , 
\end{aligned}
\ee
where the operator $\widetilde{D}^{IJ}_\a$  enjoys the canonical form\footnote{The $10$-dimensional $\G$-matrices are given in~\eqref{eq:def-10-dim-Gamma}.} 
\be
\label{eq:deform-D-op}
\begin{aligned}
\widetilde{D}^{IJ}_\a  = &
\delta^{IJ} \left( \pa_\a  -\frac{1}{4} \widetilde{\omega}^{mn}_\a \G_{mn}  \right)
+\frac{1}{8} \sigma_3^{IJ} \widetilde{e}^m_\a H_{mnp} \G^{np}
\\
&-\frac{1}{8} e^{\varphi} \left( \epsilon^{IJ} \G^p F^{(1)}_p + \frac{1}{3!}\sigma_1^{IJ} \G^{pqr} F^{(3)}_{pqr} + \frac{1}{2\cdot5!}\epsilon^{IJ} \G^{pqrst} F^{(5)}_{pqrst}   \right) 
\widetilde{e}^m_\a \G_m 
\, ,
\end{aligned}
\ee
 In the last equation $\widetilde{e}^m_M$ is the same vielbein of the $\eta$-deformed metric, {\it c.f.} appendix \ref{app:CD}, that features in the bosonic Lagrangian, 
 while $ \widetilde{\omega}^{mn}_{M}$ is the spin connection that is related to $\widetilde{e}^m_M$ by the standard formula (\ref{eq:spin_con_deformed}). Finally, for the 3-form $H_{mnp}$
we find the following two non-vanishing components
\be\la{eq:H}
H_{234} = - 4 \vk \rho \frac{\sqrt{1+\rho^2}\sqrt{1-\vk^2\rho^2}\sin \zeta}{1+\vk^2 \rho^4 \sin^2 \zeta},
\quad
H_{789} = + 4 \vk r \frac{\sqrt{1-r^2}\sqrt{1+\vk^2r^2}\sin \xi}{1+\vk^2 r^4 \sin^2 \xi} \, .
\ee
These are precisely the field strength components of the $B$-field  (\ref{Ba}) written with the flat space indices. Thus, we completely restore the NSNS background of the $\eta$-deformed theory 
at the level of the quadratic fermionic action, which is rather non-trivial by itself and provides a strong validity check of our computation.  

Postponing the discussion of the RR couplings till the next section, we conclude by pointing out that the field redefinitions of $(X,\theta)$ we used do not involve world-sheet derivatives and,
as such, they can be viewed as a certain $\varkappa$-dependent reparametrisation of the original coset representative (\ref{original_coset}) of ${\rm AdS}_5\times {\rm S}^5$. 

%We also use the ``Majorana-flip''relations of Eq.~\eqref{eq:symm-gamma-otimes-gamma}. 

%starting from more or less arbitrary choice of coordinates on the coset spaceAt the Lagrangian level changes of coordinates correspond to redefinitions of fields which do not involve world-sheet derivatives.  We are thus %prompted to look for appropriate field redefinitions of this type.
%In the undeformed case this happens because the choice (\ref{original_coset}) has an underlying geometric meaning: 
%Bosonic isometries of \ads act on fermions by Lorentz transformations and this is what enforces 
 %fermions to couple to vielbein and spin connection in the standard fashion as to maintain the invariance of the fermionic action under global symmetries of the bosonic background.}.

\subsection{RR couplings}\label{RRresults}
Here we present our main result -- the RR couplings of the $\eta$-deformed theory, and then discuss some of their features.
From eq.(\ref{eq:deform-D-op}) we find the following non-vanishing RR forms written with {\it flat} indices of the tangent space 
%% F1
\be\label{eq:flat-comp-F1}
\begin{aligned}
&\hspace{-2.2cm}e^{\varphi} F_1 =-4 \vk ^2  \ c_{F}^{-1} \ \rho ^3 \sin \zeta , \qquad
&&~~~e^{\varphi} F_6&= +4 \vk ^2  \ c_{F}^{-1} \ r^3 \sin\xi ,
\end{aligned}
\ee
%% F3
\be\label{eq:flat-comp-F3}
\begin{aligned}
&e^{\varphi} F_{014} = + 4 \vk   \ c_{F}^{-1} \ \rho ^2 \sin\zeta, \qquad
&&e^{\varphi} F_{123} = -4 \vk   \ c_{F}^{-1} \ \rho  , \\
&e^{\varphi} F_{569}= + 4 \vk   \ c_{F}^{-1} \ r^2 \sin\xi, \qquad
&&e^{\varphi} F_{678} = -4 \vk   \ c_{F}^{-1} \ r, \\
&e^{\varphi} F_{046} = +4 \vk^3   \ c_{F}^{-1} \ \rho  r^3 \sin \xi, \qquad
&&e^{\varphi} F_{236} = -4 \vk^3   \ c_{F}^{-1} \ \rho ^2 r^3 \sin \zeta   \sin\xi, \\
&e^{\varphi} F_{159} = - 4 \vk^3   \ c_{F}^{-1} \ \rho ^3 r \sin\zeta, \qquad
&&e^{\varphi} F_{178} = -4 \vk^3   \ c_{F}^{-1} \  \rho ^3 r^2 \sin\zeta \sin \xi, \\
\end{aligned}
\ee
%% F5
\be\label{eq:flat-comp-F5}
\begin{aligned}
&e^{\varphi} F_{01234} = + 4    \ c_{F}^{-1} , \qquad
&&e^{\varphi} F_{02346} = -4 \vk ^4   \ c_{F}^{-1}\rho ^3 r^3 \sin \zeta  \sin\xi , 
\\
&e^{\varphi} F_{01459} = +4 \vk ^2 \ c_{F}^{-1} \rho ^2 r \sin\zeta, \qquad
&&e^{\varphi} F_{01478} = +4 \vk ^2 \ c_{F}^{-1} \rho ^2 r^2   \sin\zeta \sin \xi , 
\\
&e^{\varphi} F_{04569}= +4 \vk ^2 \ c_{F}^{-1} \rho  r^2 \sin \xi, \qquad
&&e^{\varphi} F_{04678} = -4 \vk ^2 \ c_{F}^{-1} \rho  r.
\end{aligned}
\ee
%%%%%%%%%%%%%%%%%%%%%%%%%%%%%%%%%%%%%%%%%%
For simplicity we have defined the common coefficient 
\be
c_{F} = \frac{1}{\sqrt{1+\vk ^2}}\sqrt{1-\vk ^2 \rho^2} \sqrt{1+\vk ^2 \rho ^4 \sin ^2\zeta} \sqrt{1+\vk ^2 r^2} \sqrt{1+\vk ^2 r^4 \sin ^2\xi}.
\ee
For the five-form we presented here only half of all its non-vanishing components, namely those which involve the index $0$. The other half is obtained from the self-duality equation for the five-form.
The answer appears to be rather simple and in the limit $\kappa\to 0$ all the components vanish except $F_{01234}$ which reduces to the constant five-form flux of the \ads background. 
\smallskip

\noindent
For the reader's convenience we present the same results written in \emph{curved} indices

{\footnotesize
%% F1
\be\label{eq:curved-comp-F1}
\begin{aligned}
&\hspace{-2.5cm}e^{\varphi} F_{\psi_2} &=4 \vk ^2 \ c_{F}^{-1} \ \rho ^4 \sin ^2 \zeta , ~~~\qquad
e^{\varphi} F_{\phi_2} &= -4 \vk ^2  \ c_{F}^{-1} \ r^4 \sin ^2\xi ,
\end{aligned}
\ee
%% F3
\be\label{eq:curved-comp-F3}
\begin{aligned}
e^{\varphi} F_{t\psi_2 \rho } &= + 4 \vk   \ c_{F}^{-1} \ \frac{ \rho ^3 \sin ^2\zeta}{1-\vk ^2 \rho ^2}, \qquad
&e^{\varphi} F_{\psi_2 \psi_1 \zeta } &= +4 \vk   \ c_{F}^{-1} \ \frac{ \rho ^4 \sin \zeta  \cos \zeta }{1+\vk ^2 \rho ^4 \sin^2\zeta }, \\
e^{\varphi} F_{\phi \phi_2 r } &= + 4 \vk  \ c_{F}^{-1} \ \frac{ r^3 \sin ^2\xi}{1+\vk ^2 r^2}, \qquad
&e^{\varphi} F_{\phi_2 \phi_1 \xi } &= +4 \vk  \ c_{F}^{-1} \ \frac{ r^4 \sin \xi \cos\xi}{1+\vk ^2 r^4 \sin ^2\xi}, \\
e^{\varphi} F_{t \rho \phi_2 } &= + 4 \vk^3  \ c_{F}^{-1} \ \frac{\rho  r^4 \sin ^2\xi }{1-\vk ^2 \rho ^2}, \qquad
&e^{\varphi} F_{\psi_1 \zeta \phi_2 } &= +4 \vk^3   \ c_{F}^{-1} \ \frac{ \rho ^4 r^4 \sin \zeta  \cos \zeta  \sin ^2\xi}{1+\vk ^2 \rho ^4 \sin ^2\zeta}, \\
e^{\varphi} F_{\psi_2 \phi r } &= - 4 \vk^3   \ c_{F}^{-1} \ \frac{ \rho ^4 r \sin ^2\zeta}{1+\vk ^2 r^2}, \qquad
&e^{\varphi} F_{\psi_2 \phi_1 \xi } &= +4 \vk^3   \ c_{F}^{-1} \ \frac{ \rho ^4 r^4 \sin^2\zeta \sin \xi \cos \xi}{1+\vk ^2 r^4 \sin ^2\xi}, \\
\end{aligned}
\ee
%% F5
\be\label{eq:curved-comp-F5}
\begin{aligned}
\hspace{-3cm}e^{\varphi} F_{t\psi_2\psi_1\zeta\rho } &= + \ \frac{ 4    \ c_{F}^{-1}  \rho ^3 \sin\zeta \cos\zeta}{\left(1-\vk ^2 \rho ^2\right) \left(1+\vk ^2 \rho ^4 \sin ^2\zeta\right)}, 
&e^{\varphi} F_{t\psi_1\zeta\rho\phi_2 } &= -\frac{4 \vk ^4   \ c_{F}^{-1}\rho ^5 r^4 \sin \zeta \cos \zeta \sin ^2\xi }{\left(1-\vk ^2 \rho ^2\right) \left(1+\vk ^2 \rho ^4 \sin ^2\zeta\right)}, 
\\
e^{\varphi} F_{t\psi_2\rho\phi r } &= -\frac{4 \vk ^2 \ c_{F}^{-1} \rho ^3 r \sin ^2\zeta}{\left(1-\vk ^2 \rho ^2\right) \left(1+\vk ^2 r^2\right)}, 
&e^{\varphi} F_{t\psi_2\rho\phi_1\xi } &= +\frac{4 \vk ^2 \ c_{F}^{-1} \rho ^3 r^4   \sin ^2\zeta \sin \xi \cos\xi}{\left(1-\vk ^2 \rho ^2\right) \left(1+\vk ^2 r^4 \sin ^2\xi\right)}, 
\\
e^{\varphi} F_{t\rho\phi\phi_2 r } &= -\frac{4 \vk ^2 \ c_{F}^{-1} \rho  r^3 \sin ^2\xi}{\left(1-\vk ^2 \rho ^2\right) \left(1+\vk ^2 r^2\right)}, 
&e^{\varphi} F_{t\rho \phi_2 \phi_1\xi } &= -\frac{4 \vk ^2 \ c_{F}^{-1} \rho  r^4 \sin   \xi \cos\xi}{\left(1-\vk ^2 \rho ^2\right) \left(1+\vk ^2 r^4 \sin ^2\xi\right)}.
\end{aligned}
\ee
}

\normalsize
Inspection of the found RR couplings reveals that contrary to the natural expectations they do not obey  equations of motion of type IIB supergravity.

First of all for the Bianchi identities this is already obvious from the expression (\ref{eq:curved-comp-F1}) for the 1-form. To fit the supergravity content this form must be exact $F^{(1)}={\rm d}\chi$, where 
$\chi$ is axion.  
One can verify that there is no way to split off an integrating factor $e^{\varphi}$  in  (\ref{eq:curved-comp-F1}), such that the corresponding $F^{(1)}$ becomes exact.

Concerning other equations of motion, consider, for instance, the Einstein equations (\ref{Einstein}) which involve an unknown dilaton. One can show that to achieve vanishing of the off-diagonal 
components of the Einstein equations the dilaton $\varphi$ \emph{must} be of the form 
\bea\label{dil}
\varphi=\Phi_{\alg{a}}(\rho,\zeta)+\Phi_{\alg{s}}(r,\xi)\, ,
\eea
where $\Phi_{\alg{a}}$ and $\Phi_{\alg{s}}$ are some functions. However, analysis of the diagonal components of the Einstein equations shows that a solution for 
 $\Phi_{\alg{a}}$ and $\Phi_{\alg{s}}$ does not exist.

Now we will attempt to make contact of our findings with known supergravity solutions by considering special limits.

\paragraph{\sl Mirror background}
We first analyse  a special limit $\vk \to \infty$. Rescaling the bosonic coordinates of the $\eta$-deformed metric as
\be\la{eq:rescaling_mirror}
t \to \frac{t}{\vk},
\qquad
\rho \to \frac{\rho}{\vk},
\qquad
\phi \to \frac{\phi}{\vk},
\qquad
r \to \frac{r}{\vk},
\ee
and then sending $\vk\to \infty$,  yields upon an overall rescaling the metric for the \ads mirror model~\cite{Arutyunov:2014cra}. The $B$-field vanishes in this limit. 
The resulting metric can be then embedded into a full solution of type IIB supergravity by supplementing it with a dilaton and a five-form flux~\cite{Arutyunov:2014cra}.

Now we look at how the actual RR couplings behave in this limit.  Upon rescaling (\ref{eq:rescaling_mirror}) it is enough to keep only those components with tangent indices that are of order $\mathcal{O}(\vk)$ at large $\varkappa$ 
to compensate the power $1/\vk$ coming from the vielbein that multiplies the RR couplings in eq.\eqref{eq:deform-D-op}. The surviving components are thus
\be
\begin{aligned}
e^\varphi \, F_{123} = - \frac{4 \rho}{\sqrt{1-\rho^2}\sqrt{1+r^2}}\, ,
\qquad
e^\varphi \, F_{678} = - \frac{4 r}{\sqrt{1-\rho^2}\sqrt{1+r^2}}\, ,
\\
e^\varphi \, F_{01234} = +\frac{4 }{\sqrt{1-\rho^2}\sqrt{1+r^2}}\, ,
\qquad
e^\varphi \, F_{04678} = - \frac{4 \rho r}{\sqrt{1-\rho^2}\sqrt{1+r^2}}\, .
\end{aligned}
\ee
This result does not match the proposed mirror background~\cite{Arutyunov:2014cra}, and the limiting couplings continue to displease the supergravity equations.

\paragraph{\sl  Maldacena-Russo background}
Here we look at a special $\vk\to 0$ limit and show that the solution we found reproduces in this limit the Maldacena-Russo (MR) background \cite{Maldacena:1999mh}
which is a genuine solution of supergravity equations.

To achieve this limit, we first rescale the coordinates parameterising the deformed AdS space as 
\be
t\to \sqrt{\vk} \, t\,,
\quad
\psi_2\to \frac{\sqrt{\vk}}{\sin \zeta_0} \, \psi_2\,,
\quad
\psi_1\to \frac{\sqrt{\vk}}{\cos \zeta_0} \, \psi_1\,,
\quad
\zeta\to\zeta_0+ \sqrt{\vk} \, \zeta\,,
\quad
\rho\to \frac{\rho}{\sqrt{\vk}}\,,
\ee
where $\zeta_0$ is a parameter, and then send $\vk\to 0$. 
Because  the coordinates of the deformed S$^5$ do not undergo any rescaling, the corresponding part of the metric just reduces in this limit to the underformed  metric on S$^5$, and the components of the $B$-field in those directions vanish. The AdS part of the metric and the $B$-field remain non-trivial and we find
\be\label{eq:Malda-Russo-sph-coord}
\begin{aligned}
{\rm d}s^2_{(\text{MR})}&=\rho^2\left(-{\rm d}t^2+ {\rm d}\psi_2^2\right)
  + \frac{\rho^2}{1+\rho^4\sin^2\zeta_0}\left( {\rm d}\psi_1^2+ {\rm d}\zeta^2\right) 
  +\frac{{\rm d}\rho^2}{\rho^2}
+{\rm d}s^2_{\text{S}^5}\,,
\\
{B}_{(\text{MR})} &= +  \frac{\rho^4 \sin \zeta_0}{1+ \rho^4\sin^2 \zeta_0} {\rm d}\psi_1\wedge{\rm d}\zeta ,
\end{aligned}
\ee
which is precisely the NSNS content of the MR background.

Now we apply the same limiting procedure to the components of the RR couplings \eqref{eq:flat-comp-F1},~\eqref{eq:flat-comp-F3} and~\eqref{eq:flat-comp-F5} and find that the axion vanishes, 
and only one component of $F^{(3)}$ and one of $F^{(5)}$ (plus its dual) survive
\be
e^{\varphi} F_{014} = \frac{4\rho^2\sin \zeta_0}{\sqrt{1+ \rho^4\sin^2 \zeta_0}}\,,
\qquad
e^{\varphi} F_{01234} = \frac{4}{\sqrt{1+ \rho^4\sin^2 \zeta_0}}\,.
\ee
If we identify the dilaton as 
\be
\varphi=\varphi_0-\frac{1}{2}\log ( 1+ \rho^4\sin^2 \zeta_0)\,,
\ee
where $\varphi_0$ is a constant, we then find that the non-vanishing components for the RR fields, written both with tangent and curved indices, are
\be
\begin{aligned}
& F_{014} = e^{-\varphi_0}\, 4\rho^2\sin \zeta_0\,,
\qquad
&& F_{01234} = e^{-\varphi_0}\,4\,,
\\
& F_{t\psi_2\rho} = e^{-\varphi_0}\, 4\rho^3\sin \zeta_0\,,
\qquad
&& F_{t\psi_2 \psi_1\zeta\rho} = e^{-\varphi_0}\, \frac{4\rho^3}{1+\rho^4\sin^2\zeta_0}\,.
\end{aligned}
\ee
These are precisely the dilaton and the RR fields of the MR background~\cite{Maldacena:1999mh}.
It is very interesting that despite  incompatibility with supergravity equations for generic values of the deformation parameter, there exists a certain limit, different from ${\rm AdS}_5\times {\rm S}^5$, where this 
compatibility is retrieved.

\section{RR couplings from $\kappa$-symmetry}\label{sec:kappa}
As was shown in \cite{Delduc:2013qra,Delduc:2014kha}, the Lagrangian of the deformed model is invariant under $\kappa$-symmetry transformations.
Recall that in the undeformed case $\kappa$-transformations are implemented by multiplying a group representative of a coset element 
from the right:
\be\label{eq:right-ferm-action}
\alg{g}\cdot \text{exp}(\varepsilon) = \alg{g}'\cdot \alg{h}\,,
\ee
where $\varepsilon$ is a local fermionic parameter which takes values in $\psu(2,2|4)$. Here on the right hand side
 $\alg{g}'$ is a new coset representative and $\alg{h}$ is a compensating transformation from  $\text{SO}(4,1)\times \text{SO}(5)$.
For generic $\varepsilon$ this transformation is not a symmetry of the action, but for a special choice
\be\label{eq:undef-eps-kappa-tr}
\begin{aligned}
\varepsilon &=  \frac{1}{2} (\gamma^{\a\b} \delta^{IJ}- \epsilon^{\a\b}\sigma_3^{IJ}) \left( \gen{Q}^I\kappa_{J\a}  A_\b^{(2)}  + A_\b^{(2)} \gen{Q}^I \kappa_{J\a} \right),
\end{aligned}
\ee
one can show that this is indeed the case \cite{Arutyunov:2009ga}. The spinors $\kappa_{1\a}$ and $\kappa_{2\a}$ are local transformation parameters which under ${\mathbb Z}_4$-decomposition have degree $1$ and $3$, respectively.

In the deformed case one can still prove the existence of a local fermionic symmetry of the form~\eqref{eq:right-ferm-action}.
%meaning that the parameter $\varepsilon$ is related to the \emph{infinitesimal} variation of the coset representative as
%\be
%\delta_\kappa \alg{g}=\alg{g}\cdot \varepsilon\,.
%\ee
However, to achieve the invariance of the action the definition~\eqref{eq:undef-eps-kappa-tr}  has to be modified, in particular $\varepsilon$ will no longer lie just in the odd part of the algebra, 
but will have a non-trivial overlap with the even part. Precisely, $\varepsilon$
is written in terms of an odd element $\varrho$ as~\cite{Delduc:2013qra}
\be\label{eq:eps-op-rho-kappa}
\varepsilon = \op \varrho, \qquad \varrho= \varrho^{(1)} +\varrho^{(3)} .
\ee
where $\op$ is the operator defined in (\ref{eq:defin-op-def-supercoset}) and the two projections $\varrho^{(k)}$ are\footnote{Comparing to~\cite{Delduc:2013qra} we have dropped the factor of $i$ because we use ``anti-hermitian'' generators.}
\be\label{eq:def-varrho-kappa-def}
\begin{aligned}
\varrho^{(1)} &= \frac{1}{2} (\gamma^{\a\b} - \epsilon^{\a\b}) \left( \gen{Q}^1\kappa_{1\a}  \left(\op^{-1} A_\b \right)^{(2)}  + \left(\op^{-1} A_\b \right)^{(2)} \gen{Q}^1\kappa_{1\a} \right),\\
\varrho^{(3)} &=  \frac{1}{2} (\gamma^{\a\b} + \epsilon^{\a\b}) \left( \gen{Q}^2\kappa_{2\a}  \left(\optilde^{-1} A_\b \right)^{(2)}  + \left(\optilde^{-1} A_\b \right)^{(2)} \gen{Q}^2\kappa_{2\a} \right),
\end{aligned}
\ee
where we defined
\be
\optilde=\mathbf{1} + \eta R_{\alg{g}} \circ \widetilde{d}\,.
\ee
In appendix~\ref{app:kappa} we  explicitly derive the variations of bosonic and fermionic fields implied by the above definitions, and observe that they do not have the usual form of the $\kappa$-variations 
of type IIB superstring. However, after implementing the field redefinitions of  appendix~\ref{app:CGS}, which were needed to put the Lagrangian in the canonical Green-Schwarz form,
we find that also the kappa-variations become indeed standard
\be\label{eq:kappa-var-32}
\begin{aligned}
 \delta_{\kappa}X^M &= - \frac{i}{2} \ \bar{\T}_I \delta^{IJ} \widetilde{e}^{Mm}  \G_m  \delta_{\kappa} \T_J + \mathcal{O}(\T^3),
\\
\delta_{\kappa} \T_I &= -\frac{1}{4} (\delta^{IJ} \gamma^{\a\b} - \sigma_3^{IJ} \epsilon^{\a\b})  \widetilde{e}_{\b}^m  \G_m  \widetilde{K}_{\a J}+ \mathcal{O}(\T^2),
\end{aligned}
\ee
where
\be
\widetilde{K} \equiv \left( \begin{array}{c} 0 \\ 1 \end{array} \right) \otimes \widetilde{\kappa},
\ee
and $\tilde{\kappa}$ is related to $\kappa$ as in~\eqref{eq:def-kappa-tilde-k-symm}.
It is now instructive to also look at the kappa-variation for the world-sheet metric, as this provides an independent way to derive the couplings of the fermions to the background fields.
The variation is given by~\cite{Delduc:2013qra}
\be\label{eq:defin-kappa-var-ws-metric}
\delta_{\kappa}\g^{\a\b}=\frac{1-\eta^2}{2} \Str\left( \Upsilon \left[\gen{Q}^1\kappa^{\a}_{1+},P^{(1)}\circ \widetilde{\op}^{-1}( A^{\b}_+ ) \right]
+\Upsilon \left[\gen{Q}^2\kappa^{\a}_{2-},P^{(3)}\circ {\op}^{-1}( A^{\b}_- ) \right] \right)\,,
\ee
where $\Upsilon=\text{diag}(\mathbf{1}_4,-\mathbf{1}_4)$ and the projections of a vector $V_\a$ are defined as
\be\label{defpm}
V^\a_{\pm}= \frac{\g^{\a\b}\pm \epsilon^{\a\b}}{2} V_\b\,.
\ee
As we show in appendix~\ref{app:kappa}, after taking into account the field redefinitions performed to get the canonical action, we find a standard kappa-variation also for the world-sheet metric
\be\label{eq:standardkappametric}
\begin{aligned}
\delta_\kappa \g^{\a\b}&=2i\Bigg[ 
\bar{\widetilde{K}}^\a_{1+} \widetilde{D}^{\b 1J}_+\Theta_J+\bar{\widetilde{K}}^\a_{2-} \widetilde{D}^{\b 2J}_-\Theta_J
 \Bigg]+ \mathcal{O}(\T^3) \\
&= 2i\ \Pi^{IJ\, \a\a'}\Pi^{JK\, \b\b'}
\ \bar{\widetilde{K}}_{I\a'}\widetilde{D}^{KL}_{\b'}\Theta_{L}+ \mathcal{O}(\T^3),
\end{aligned}
\ee
where we have defined
\be
\Pi^{IJ\, \a\a'}\equiv\frac{\delta^{IJ}\g^{\a\a'}+\sigma_3^{IJ}\epsilon^{\a\a'}}{2}\,.
\ee
The operator $\widetilde{D}^{IJ}_{\a}$ turns out to be the same as obtained earlier in the Lagrangian approach. It is given by eq.\eqref{eq:deform-D-op}, and, in particular, it contains the same RR couplings 
as found in section \ref{RRresults}.

We point out that at the level of the quadratic fermionic action, the requirement of $\kappa$-symmetry is unable to produce differential constraints on the RR fields, in particular, the Bianchi identities. 
Constraints will start to emerge from the quartic  action, because to check its invariance, one has to vary the RR couplings entering the quadratic part of the fermionic action, which will lead to the appearance 
of their derivatives. Thus, if our result for the RR couplings is an ultimate one, {\it i.e.} if there are no further field redefinitions changing the RR couplings only, one could expect 
that at higher orders in fermions both $\kappa$-symmetry transformations and the corresponding Lagrangian start to deviate from the standard form in the theory of IIB Green-Schwarz superstring,
and this  could explain why our results are compatible with the work  \cite{Grisaru:1985fv,Bergshoeff:1985su}. It is also worth stressing that in  \cite{Grisaru:1985fv,Bergshoeff:1985su} it was shown that 
the supergravity constrains are sufficient for  $\kappa$-symmetry of the Green-Schwarz action, whether they are also necessary is unknown to us.

\section{On field redefinitions}\label{sec:red}
In the previous section we were able to transform the original Lagrangian into the canonical form and further observed  that the RR couplings derived from the latter
do not satisfy the supergravity equations. On the other hand, the NSNS couplings in the quadratic fermionic action are properly reproduced
and they are the same as found earlier from the bosonic Lagrangian. Therefore we are motivated to ask whether further field redefinitions could be performed 
which exclusively change  the RR content of the theory. It appears to be rather difficult to answer this question in full generality. We will argue however 
that  no field redefinition of this type, continuous in the deformation parameter exists. 

We will work in the formulation with 32-dimensional fermions $\T_I$ obeying the Majorana and Weyl conditions, see appendix \ref{sec:10-dim-gamma}. We start with considering 
a generic rotation of fermions\footnote{One could imagine more complicated redefinitions like $\Theta_I \to F_{IJ} \Theta_J+G_{IJ}^{\a}\pa_{\a}\T_J$, etc. 
They were not needed to bring the original Lagrangian to the canonical form and we do not consider them here. These redefinitions will generate higher derivative terms in the action,
whose cancellation would imply further stringent constraints on their possible form.} 
\be
\label{FIJ}
\begin{aligned}
\Theta_I &\to F_{IJ} \Theta_J,\quad \bar\Theta_I \to  \bar\Theta_J\bar F_{IJ}\, , \quad  \bar F_{IJ} =- \G_0 F_{IJ}^\dagger\G_0\, , 
\end{aligned}
\ee
where $F_{IJ}$ are rotation matrices which depend on bosonic fields. We write $F_{IJ}$ as an expansion over a complete basis in the space of $2\times 2$-matrices 
\be
\label{FIJ1}
\begin{aligned}
F_{IJ} &\equiv \delta^{IJ} F_\delta + \sigma_1^{IJ} F_{\sigma_1} + \eps^{IJ} F_{\epsilon} + \sigma_3^{IJ} F_{\sigma_3} = \sum_{a=0}^3\as_a^{IJ}F_a\,,\\
\bar F_{IJ} &= \delta^{IJ} \bar F_\delta + \sigma_1^{IJ} \bar F_{\sigma_1} + \eps^{IJ} \bar F_{\epsilon} + \sigma_3^{IJ} \bar F_{\sigma_3}= \sum_{a=0}^3\as_a^{IJ}\bar F_a\, ,
\end{aligned}
\ee
where we have introduced 
$$
 \as_0^{IJ}=\delta^{IJ}\,,\quad \as_1^{IJ}=\s_1^{IJ}\,,\quad \as_2^{IJ}=\eps^{IJ}\,,\quad \as_3^{IJ}=\s_3^{IJ}\, .
 $$
Next, the coefficients $F_{a}$ and $\bar{F}_{a}$ are $32\times 32$-matrices and they can be expanded over the complete basis generated by $\Gamma^{(r)}$ and identity, see appendix \ref{sec:10-dim-gamma}
for the definition and properties of $\Gamma^{(r)}$. Further, we  require that the transformation $F_{IJ}$ preserves chirality and the Majorana condition. Conservation of chirality implies that 
the $\Gamma$-matrices appearing in the expansion of $F_{IJ}$ must commute with $\Gamma_{11}$, {\it i.e.} the expansion involves $\Gamma^{(r)}$ of even rank only
\bal\label{eq:def-redefinition}
F_a &=  f_a\,{\mathbb I}_{32} + {1\ov2}f^{mn}_a\G_{mn} + {1\ov24}f^{klmn}_a\G_{klmn} \,,\\
\bar F_a &= \bar f_a\,{\mathbb I}_{32} + {1\ov2}\bar f^{mn}_a\G_{mn} + {1\ov24}\bar f^{klmn}_a\G_{klmn} \,.
\eal
In this expansion there are no matrices of higher rank, because those by virtue of duality relations are re-expressed via matrices of lower rank. 
The Majorana condition imposes the requirement 
\bea\label{Maj}
\G_0F^\dagger_{IJ}\G_0=\mathcal{C}F^t_{IJ}\mathcal{C}\, 
\eea
which implies that the coefficients $f$ are real. Coefficients of $\bar{F}_a$ are then given by 
\bea
\bar f_{a} = f_a\,,\quad  \bar f_{a}^{mn} =- f_a^{mn}\,,\quad  \bar f_{a}^{klmn} = f_a^{klmn}\,.
\eea
Thus, the total number of degrees of freedom in the rotation matrix is 
$$
4\cdot\Big(1+\frac{10\cdot 9}{2}+\frac{10\cdot 9\cdot 8\cdot 7}{4!}\Big)=2^{10}=(16+16)^2 \, ,
$$
which is precisely the dimension of ${\rm GL}(32,{\mathbb R})$. This correctly reflects the freedom to perform general linear transformations on 
32 real fermions of type IIB. 

\smallskip

Under these rotations the kinetic part of the fermionic Lagrangian transforms into
\bea
&&(\gamma^{\a\b}\de^{IJ}+\eps^{\a\b}\sigma_3^{IJ})\bar{\T}_I \tilde{e}_{\a}^m\G_m\pa_{\beta}\T_{I} \to \\
\nonumber
&&~~~~~~~~~~\to (\gamma^{\a\b}\de^{IJ}+\eps^{\a\b}\sigma_3^{IJ})\Big( \bar{\Theta}_K \, \bar{F}_{IK} \, \widetilde{e}^m_\a  \G_m F_{JL} \, \pa_\b \Theta_L +   \bar{\Theta}_K \, \bar{F}_{IK} \, \widetilde{e}^m_\a \G_m  (\pa_\b F_{JL})\Theta_L  \Big)\, .
\eea 
The requirement that under rotations the kinetic part remains unchanged  can be formulated as the following conditions on $F_{IJ}$:
\be\label{eq:redef_cond}
\begin{aligned}
\de^{IJ}  \bar{F}_{IK} \,  \G_m F_{JL} &=\delta_{KL}\G_m+{\rm removable~terms}\, , \\
\sigma_3^{IJ}\bar{F}_{IK} \,  \G_m F_{JL}&=\sigma_3^{KL}\G_m+{\rm removable~terms}\, ,
\end{aligned}
\ee
where ``removable terms" means terms which can be removed by shifting bosons in the bosonic action by fermion bilinears. 
The equations (\ref{eq:redef_cond}) should hold on chiral fermions, 
that is as sandwiched between two chirality projectors. To make the discussion simple, we will not indicate these projectors explicitly till the very end.
In the following it is enough to analyse the first equation in (\ref{eq:redef_cond}) and, thus,
we are led to understand the structure of $  \bar{F}_{IK} \,  \G_m F_{IL} $, which in general has an expansion over a basis of odd rank $\G^{(r)}$. The strategy is to determine first the structure of removable terms. 
To this end we need to study the properties of fermion bilinears.

Suppose that $s^{JI}=s \ s^{IJ}$, with $s=\pm 1$. Now we take two sets of Majorana-Weyl fermions, that we call $\Theta_A$ and $\Theta_B$ in order to distinguish them.
We consider odd rank $\G$-matrices (not to get vanishing expressions)
\be\label{eq:symm-TGT}
\begin{aligned}
s^{IJ} \ \bar{\Theta}_{A,I} \G^{(r)} \Theta_{B,J} 
&= s^{IJ} \ \Theta_{A,I\dot{\a}} (\mathcal{C}\G^{(r)})^{\dot{\a}\dot{\b}} \Theta_{B,J\dot{\b}}  
= -s^{IJ} \ \Theta_{B,J\dot{\b}}( \mathcal{C}\G^{(r)})^{\dot{\a}\dot{\b}}  \Theta_{A,I\dot{\a}} \\
&= - \ s^{IJ} \ \bar{\Theta}_{B,J}\,  \mathcal{C}(\mathcal{C}\G^{(r)})^t  \Theta_{A,I} = s \cdot t_r^{\G} \ s^{IJ} \ \bar{\Theta}_{B,I} \G^{(r)}  \Theta_{A,J}  ,
\end{aligned}
\ee
see appendix \ref{sec:10-dim-gamma} for the definition of the numbers $t_r^\Gamma$.
The kinetic term for bosons under the shift, which can be schematically represented as 
\bea\la{redbos}
X^M\to X^M+\,w_{(r)}^{M, a} s^{IJ}_a\bar{\T}_I\G^{(r)}\Theta_J\, ,
\eea
will generate the fermionic terms containing the terms 
\bea
s^{IJ}_a\pa_{\a}(\bar{\T}_I\G^{(r)}\Theta_J)=s^{IJ}_a\pa_{\a}\bar{\T}_I\G^{(r)}\Theta_J+s^{IJ}_a\bar{\T}_I\G^{(r)}\pa_{\a}\Theta_J\, \la{2terms}
\eea
Clearly, for this expression to fit the structure of the fermionic kinetic term, the two terms in the right hand side of \eqref{2terms} must be equal.
Identifying $\pa_{\a}\T$ with $\T_A$ and $\T$ with $\T_B$ in eq.(\ref{eq:symm-TGT}) shows that removable structures in the fermionic action are those for which $s\cdot t_r^\Gamma=+1$.
Indeed, the structures with $s \cdot t_r^G=-1$ entering in the shift (\ref{redbos}) simply vanish because of the same equation (\ref{eq:symm-TGT}) considered for $A=B$. 
Using the results of appendix \ref{sec:10-dim-gamma} one can determine $s\cdot t_r^\Gamma$ for various $r$ and $s$
and the  corresponding values are collected in Table 1. 
\begin{table}[h]
\begin{center}\begin{tabular}{r|ccc}
  & $r=1$ & $r=3$ & $r=5$ 
\\
\hline
$s=+1$ & $-1$ & $+1$ & $-1$
\\
\hline
$s=-1$ & $+1$ & $-1$ & $+1$
\end{tabular}
\caption{ Values of $s \cdot t_r^G$ for  different  $r$ and  $s$.}\end{center}\end{table}

\noindent
According to this table, the condition that the kinetic term is invariant up to the terms removable by a shift of bosons can be now written as 
\be\label{FGF}
\begin{aligned}
\bar{F}_{IK} \G_m F_{IL} =& \delta_{KL}\G_m 
+  \eps_{KL}\left[  (h_{\epsilon})^n_m \G_n +  (h_{\epsilon})^{npqrs}_m \G_{npqrs} \right]
\\
&+ \left[\delta_{KL}  (h_{\delta})^{npq}_m + \sigma_{1KL}  (h_{\sigma_1})^{npq}_m + \sigma_{3KL}  (h_{\sigma_3})^{npq}_m \right] \G_{npq}\,.
\end{aligned}
\ee

Here $h$-tensors are arbitrary coefficients which parametrise the structures which can be removed from the action by shifting bosons. Obviously, putting a generic $F$ satisfying the Majorana-Weyl conditions 
in the left hand side of \eqref{FGF}, one would expect an appearance
on the right hand side of all these $h$-tensors. But do they actually appear? As we show in a moment the answer is negative.    

Combining equations  (\ref{FIJ}) and (\ref{Maj}), we get 
\be\la{symFIJ}
\mathcal{C}\bar{F}^t_{IJ}\mathcal{C}= -F_{IJ}\,, \quad \text{ and } \quad \mathcal{C}F^t_{IJ}\mathcal{C}= -\bar{F}_{IJ}\, .
\ee
Now collect all terms on the right hand side of~\eqref{FGF} that are removable by shifting bosons into a tensor $M_{KL,m}$. This tensor has the following symmetry property\footnote{Notice that to exhibit  this symmetry property, one has to transpose also the indices $K,L$, on top of transposition in the $32\times32$ space.}
\be\la{sym1}
\mathcal{C}\left(M_{KL,m}\right)^t\mathcal{C}=- M_{LK,m}\,.
\ee
Note that the tensor in the canonical kinetic term has exactly the opposite symmetry property
\be\la{sym2}
\mathcal{C}(\delta_{KL} \G_m^t)\mathcal{C}=\delta_{LK} \G_m\, .
\ee
Putting this information together, let us consider~\eqref{FGF} written as
\be
\bar{F}_{IK}\G_mF_{IL}= \delta_{KL}\G_m+M_{KL,m}\,.
\ee
We take transposition and we multiply by $\mathcal{C}$ from the left and from the right
\be
\begin{aligned}
\mathcal{C}\left(\bar{F}_{IK}\G_mF_{IL}\right)^t \mathcal{C}
&= \delta_{KL}\mathcal{C}\left(\G_m\right)^t\mathcal{C}+\mathcal{C}\left(M_{KL,m}\right)^t\mathcal{C}
\end{aligned}
\ee
and further manipulate as
\be
\begin{aligned}
\mathcal{C}\left(F_{IL}\right)^t \mathcal{C}\cdot  \mathcal{C}\left(\G_m\right)^t \mathcal{C}\cdot \mathcal{C}\left(\bar{F}_{IK}\right)^t \mathcal{C}
&= \delta_{KL}\mathcal{C}\left(\G_m\right)^t\mathcal{C}+\mathcal{C}\left(M_{KL,m}\right)^t\mathcal{C}\, .
\end{aligned}
\ee
With the help of eqs.(\ref{symFIJ}) , (\ref{sym1}) and (\ref{sym2}) and relabelling the indices $K$ and $L$, we get 
\be
\begin{aligned}
\bar{F}_{IK}\G_mF_{IL}= \delta_{KL}\G_m-M_{KL,m}\, ,
\end{aligned}
\ee
which shows that $M_{KL,m}=0$, that is this structure cannot appear because it is incompatible with the symmetry properties of the rotated kinetic term.
It is clear that the same considerations are also applied to the second equation in (\ref{eq:redef_cond}), where $\sigma_3^{IJ}$ replaces $\delta^{IJ}$.
Thus, to keep the kinetic term invariant, the rotation matrix $F$ must satisfy the following system of equations\footnote{Would not be there indices $I,J$, we would immediately conclude that the first equation in (\ref{eq:redef_cond1}) has only a trivial solution $F=1$, because $\G_m$ form an irreducible representation of the Clifford algebra.   }
\be\label{eq:redef_cond1}
\begin{aligned}
\de^{IJ}  \bar{F}_{IK} \,  \G_m F_{JL} &=\delta_{KL}\G_m\, , \\
\sigma_3^{IJ}\bar{F}_{IK} \,  \G_m F_{JL}&=\sigma_3^{KL}\G_m\, ,
\end{aligned}
\ee
We have also learned that we cannot shift bosons anymore, any shift would spoil the kinetic term in a way that cannot be fixed by rotations.
In order not to deal with indices of the $2\times 2$ space, we can introduce $64\times 64$-matrices 
\be
U\equiv \sum_{a=0}^3 s_a \otimes F_a\,,
\qquad
\bar{U}\equiv \sum_{a=0}^3 s_a^t \otimes \bar{F}_a=-\sum_{a=0}^3 s_a^t \otimes \mathcal{C}F_a^t\mathcal{C}\, ,
\ee
which allow us to rewrite the equations above in the form 
\be
\begin{aligned}\la{stringent1}
\Uppi_-\Big(\bar{U} \,(\mathbf{1}_2 \otimes \G_m)U &- \mathbf{1}_2 \otimes \G_m\Big)\Uppi_+=0\,,
\\
\Uppi_-\Big(\bar{U} \,( \sigma_3 \otimes \G_m)U &- \sigma_3 \otimes \G_m\Big)\Uppi_+=0\, .
\end{aligned}
\ee
Here we reinstated the two chirality projectors $\Uppi_{\pm}={\mathbf 1}_2\otimes \frac{1}{2}({\mathbf 1}_{32}\pm \Gamma_{11})$.
\smallskip

Finally, we assume that $U$ is a smooth function of $\eta$:
\bea
U=\mathbf{1}_{64}+\eta u+\mathcal{O}(\eta^2)\, .
\eea
At first order in $\eta$ we get a system of linear equations for $u$:
\be
\begin{aligned}\la{stringent2}
\Uppi_-\Big(\bar{u} \,(\mathbf{1}_2 \otimes \G_m) &+(\mathbf{1}_2 \otimes \G_m)u\Big)\Uppi_+=0\,,
\\
\Uppi_-\Big(\bar{u} \,( \sigma_3 \otimes \G_m) &+ ( \sigma_3 \otimes \G_m)u\Big)\Uppi_+=0\,.
\end{aligned}
\ee
This system appears to have no solution which acts non-trivially on chiral fermions. Thus, non-trivial field redefinitions of the type we considered here 
do not exists.  Whether equation (\ref{stringent1}) has solutions which do not depend on $\eta$ is unclear to us.  Finally, let us mention that similar considerations of field redefinitions can be done
for $\kappa$-symmetry transformations with the same conclusion.

\section{T-matrix and factorisation}\label{sec:Tmatrix}

To find the Lagrangian quadratic in fermions we used  a coset element of the  form 
 \begin{equation} \alg{g} = \Lambda(t,\phi) \cdot\gx\cdot \alg{g_f}\,, \end{equation} 
where $\gx$ depends on the transverse bosons and $\alg{g_f}$ on the fermions. With this particular choice fermions are uncharged under bosonic isometries. On the other hand to impose a uniform l.c. gauge, one uses a coset element of the  form  $\alg{g} = \Lambda(t,\phi) \cdot \alg{g_f}' \cdot \gx$. In the undeformed case this guarantees that the $\kappa$-gauge-fixed bosons and fermions transform in a bi-fundamental irreducible representation of the centrally-extended  $\psu(2|2)\oplus \psu(2|2)$  which is the symmetry algebra of the l.c. \ads Lagrangian, and this also allows one to develop a perturbative expansion of the l.c. Lagrangian in inverse  powers of string tension.
It is clear that the two  fermionic group elements are related as follows 
\begin{equation} \alg{g_f} = \gx^{-1} \alg{g_f}' \gx\,. \end{equation} 
This redefinition of the fermionic group element is obviously equivalent to the corresponding redefinition of the fermionic coordinates
\begin{equation} \chi = \gx^{-1} \chi'\gx \end{equation} 
The new version for the coset representative $\alg{g}$ has the same form as the one used in the review \cite{Arutyunov:2009ga} to construct the l.c. Lagrangian and develop the perturbative expansion, but it is \emph{not} exactly the same choice. The reason is that the bosonic coset element $\alg{g_b}$ used here and the one of the review $\alg{g_b}^F$ differ by the action of a local Lorentz transformation $h$ 
\begin{equation} \alg{g_b} = \alg{g_b}^F \cdot h. \end{equation}
In the limit $\vk\to 0$ one does not get the same Lagrangian as in  \cite{Arutyunov:2009ga}. The Lagrangians are related by a nontrivial redefinition of bosons. This however does not change the physical quantities, and in particular both Lagrangians would give the same T-matrix.

%%%%%%%%%%%%%%%%%%%%%%%%
\subsection{T-matrix}\la{app:T}
%%%%%%%%%%%%%%%%%%%%%%%%
\newcommand{\cpp}{p\,\om'  - p'\,\om }
\newcommand{\ppcpp}{\tfrac{pp'}{\cpp}}
\newcommand{\lrbrk}[1]{\left(#1\right)}
\def \nn {\nonumber}
\newcommand{\half}{\sfrac{1}{2}}
\def\bT{{\mathbb T}}
\def\bU{{\mathbb U}}
\def\bI{{\mathbb I}}
\def\bL{{\mathbb L}}
\def\bH{{\mathbb H}}

%%%%%%%%%%%%%%%%%%%%%%%%%%%%%

Here we list the action of the T-matrix on two-particle states in the uniform $a=1/2$  light-cone
gauge. Since we do not know the quartic fermionic Lagrangian the terms quadratic in fermions are missed in the scattering processes  Fermion-Fermion  $\to$ Boson-Boson $+$ Fermion-Fermion. However if the T-matrix factorises then the missing matrix elements are fixed unambiguously.  The derivation of the l.c. Hamiltonian and its quantisation is sketched in Appendix \ref{lcHam}. We follow the same notations and conventions as in  \cite{Arutyunov:2009ga}
\bea\nonumber
&&a^\dagger_{a\da}(p)\to Y_{a\da}\,,\quad a^\dagger_{a\da}(p')\to Y'_{a\da}\,,\quad
a^\dagger_{\a\dal}(p)\to Z_{\a\dal}\,,\quad a^\dagger_{\a\dal}(p')\to Z'_{\a\dal}\,,\\\nonumber
&&a^\dagger_{\a\da}(p)\to \eta_{\a\da}\,,\quad a^\dagger_{\a\da}(p')\to \eta'_{\a\da}\,,\quad
a^\dagger_{a\dal}(p)\to \theta_{a\dal}\,,\quad a^\dagger_{a\dal}(p')\to \theta'_{a\dal}\,,
\eea
so that we have, in particular
$$|Y_{a\da}\eta'_{\b\db}\rangle\equiv |a^\dagger_{a\da}(p)a^\dagger_{\b\db}(p')\rangle\,,\quad
|\theta_{a\dal}Z'_{\b\dbe}\rangle\equiv |a^\dagger_{a\dal}(p)a^\dagger_{\b\dbe}(p')\rangle\,.
$$
Then we introduce the rapidity $\theta$ related to the momentum $p$ and energy $\om$ as follows
$$
p=\sinh\theta\,,\quad \om=\sqrt{1+\vk^2}\,\cosh\theta\,.
$$

%%%%%%%%%%%%%%%%%%%%%%%%%%%%%%%%%%%%%%%%%%%%%%%%%%%

\subsubsection*{Boson-Boson $\longrightarrow$ Boson-Boson $+$ Fermion-Fermion}
\[
\begin{split}
\bT\cdot \ket{Y_{{a}{\da}} Y'_{{b}{\db}}} = \ &
  2A
  \ket{Y_{{a}{\da}} Y'_{{b}{\db}}}
  + (B+W\eps_{\dot{a}\dot{b}})\ket{Y_{{a}{\db}} Y'_{{b}{\da}}} +
  (B+W\eps_{{a}{b}})\ket{Y_{{b}{\da}} Y'_{{a}{\db}}}  \\ &
+ C
  \eps_{{\da}{\db}} \eps^{{\dal}{\dbe}} \ket{\theta_{{a}{\dal}}\theta'_{{b}{\dbe}}} +
  C\eps_{{a}{b}} \eps^{{\a}{\b}} \ket{\eta_{{\a}{\da}}\eta'_{{\b}{\db}}}  \\[1mm]
\bT\cdot  \ket{Z_{{\a}{\dal}} Z'_{{\b}{\dbe}}} = \ &
  - 2A
  \ket{Z_{{\a}{\dal}} Z'_{{\b}{\dbe}}}
  + (-B+W\eps_{\dot{\a}\dot{\b}})
  \ket{Z_{{\a}{\dbe}} Z'_{{\b}{\dal}}} + (-B+W\eps_{{\a}{\b}})
  \ket{Z_{{\b}{\dal}} Z'_{{\a}{\dbe}}}  \\ &
  -C
  \eps_{{\dal}{\dbe}} \eps^{{\da}{\db}} \ket{\eta_{{\a}{\da}}\eta'_{{\b}{\db}}} -C
  \eps_{{\a}{\b}} \eps^{{a}{b}} \ket{\theta_{{a}{\dal}}\theta'_{{b}{\dbe}}}  \\[1mm]
\bT\cdot  \ket{Y_{{a}{\da}} Z'_{{\a}{\dal}}} = \ &
  2G \ket{Y_{{a}{\da}} Z'_{{\a}{\dal}}}
  + H \ket{\eta_{{\a}{\da}}\theta'_{{a}{\dal}}}
        -H \ket{\theta_{{a}{\dal}}\eta'_{{\a}{\da}}}  \\[1mm]
\bT\cdot  \ket{Z_{{\a}{\dal}} Y'_{{a}{\da}}} = \ &
 -2G\ket{Z_{{\a}{\dal}} Y'_{{a}{\da}}}
 +H \ket{\eta_{{\a}{\da}}\theta'_{{a}{\dal}}}
        - H\ket{\theta_{{a}{\dal}}\eta'_{{\a}{\da}}} 
\end{split}
\]
% \end{document}

%%%%%%%%%%%%%%%%%%%%%%%%%%%%%%%%%%%%%%%%%%%%%%%%%%%%%%%%%%%%%%%%%%%%%%%%%%%%%%%%
\subsubsection*{Fermion-Fermion  $\longrightarrow$ Boson-Boson}
\[
\begin{split}
\bT\cdot   \ket{\theta_{{a}{\dal}}\theta'_{{b}{\dbe}}} =\ & C 
  \eps_{{\dal}{\dbe}} \eps^{{\da}{\db}} \ket{Y_{{a}{\da}} Y'_{{b}{\db}}} -C
  \eps_{{a}{b}} \eps^{{\a}{\b}} \ket{Z_{{\a}{\dal}} Z'_{{\b}{\dbe}}}  \\[1mm]
\bT\cdot   \ket{\eta_{{\a}{\da}}\eta'_{{\b}{\db}}} = \ &
 -C
  \eps_{{\da}{\db}} \eps^{{\dal}{\dbe}} \ket{Z_{{\a}{\dal}} Z'_{{\b}{\dbe}}} +C
  \eps_{{\a}{\b}} \eps^{{a}{b}} \ket{Y_{{a}{\da}} Y'_{{b}{\db}}}  \\[1mm]
\bT\cdot   \ket{\theta_{{a}{\dal}}\eta'_{{\b}{\db}}} = \ &
  -H  \ket{Y_{{a}{\db}} Z'_{{\b}{\dal}}} -H
  \ket{Z_{{\b}{\dal}} Y'_{{a}{\db}}} \\[1mm]
\bT\cdot   \ket{\eta_{{\a}{\da}}\theta'_{{b}{\dbe}}} = \ &
  H
  \ket{Z_{{\a}{\dbe}} Y'_{{b}{\da}}} +H
  \ket{Y_{{b}{\da}} Z'_{{\a}{\dbe}}} 
\end{split}
\]

\vspace{-0.3cm}
%%%%%%%%%%%%%%%
%%%%%%%%%%%%%%%%%%%%%%%%%%%%%%%%%%%%%%%%%%%%%%%%%%%%%%%%%%%%%%%%%

\subsubsection*{Boson-Fermion $\longrightarrow$ Boson-Fermion}
\[
\begin{split}
\bT\cdot   \ket{Y_{{a}{\da}}\theta'_{{b}{\dbe}}} = \ &
  (A+G) \ket{Y_{{a}{\da}}\theta'_{{b}{\dbe}}}
  + (B-W\eps_{ab}) \ket{Y_{{b}{\da}}\theta'_{{a}{\dbe}}} \\ &
  +H\ket{\theta_{{a}{\dbe}} Y'_{{b}{\da}}}
 +C\eps_{{a}{b}} \eps^{{\a}{\b}} \ket{\eta_{{\a}{\da}} Z'_{{\b}{\dbe}}} \\[1mm]
\bT\cdot   \ket{Y_{{a}{\da}}\eta'_{{\b}{\db}}} = \ &
 (A+G) \ket{Y_{{a}{\da}}\eta'_{{\b}{\db}}}
 + (B-W\eps_{\da\db})  \ket{Y_{{a}{\db}}\eta'_{{\b}{\da}}} \\ &
  +H \ket{\eta_{{\b}{\da}} Y'_{{a}{\db}}}
  -C\eps_{{\da}{\db}} \eps^{{\dal}{\dbe}} \ket{\theta_{{a}{\dal}} Z'_{{\b}{\dbe}}} \\[1mm]
\bT\cdot   \ket{\theta_{{a}{\dal}} Y'_{{b}{\db}}} = \ &
 (A-G)\ket{\theta_{{a}{\dal}} Y'_{{b}{\db}}}
  + (B-W\eps_{ab}) \ket{\theta_{{b}{\dal}} Y'_{{a}{\db}}} \\ &
  + H \ket{Y_{{a}{\db}}\theta'_{{b}{\dal}}}
 -C\eps_{{a}{b}} \eps^{{\a}{\b}} \ket{Z_{{\a}{\dal}}\eta'_{{\b}{\db}}} \\[1mm]
\bT\cdot   \ket{\eta_{{\a}{\da}} Y'_{{b}{\db}}} = \ &
  (A-G)\ket{\eta_{{\a}{\da}} Y'_{{b}{\db}}}
  + (B-W\eps_{\da\db})  \ket{\eta_{{\a}{\db}} Y'_{{b}{\da}}} \\ &
  +H \ket{Y_{{b}{\da}}\eta'_{{\a}{\db}}}
 +C\eps_{{\da}{\db}} \eps^{{\dal}{\dbe}} \ket{Z_{{\a}{\dal}}\theta'_{{b}{\dbe}}}
\end{split}
\]

\[
\begin{split}
\bT\cdot   \ket{Z_{{\a}{\dal}}\theta'_{{b}{\dbe}}} = \ &
  -(A+G)\ket{Z_{{\a}{\dal}}\theta'_{{b}{\dbe}}}
  - (B+W\eps_{\dal\dbe}) \ket{Z_{{\a}{\dbe}}\theta'_{{b}{\dal}}} \\ &
  - H\ket{\theta_{{b}{\dal}} Z'_{{\a}{\dbe}}}
+C \eps_{{\dal}{\dbe}} \eps^{{\da}{\db}} \ket{\eta_{{\a}{\da}} Y'_{{b}{\db}}} \\[3mm]
\bT\cdot   \ket{Z_{{\a}{\dal}}\eta'_{{\b}{\db}}} = \ &
 -(A+G)\ket{Z_{{\a}{\dal}}\eta'_{{\b}{\db}}}
  -  (B+W\eps_{\a\b}) \ket{Z_{{\b}{\dal}}\eta'_{{\a}{\db}}} \\ &
  - H\ket{\eta_{{\a}{\db}} Z'_{{\b}{\dal}}}
 -C \eps_{{\a}{\b}} \eps^{{a}{b}} \ket{\theta_{{a}{\dal}} Y'_{{b}{\db}}} \\[3mm]
\bT\cdot   \ket{\theta_{{a}{\dal}} Z'_{{\b}{\dbe}}} = \ &
  - (A-G) \ket{\theta_{{a}{\dal}} Z'_{{\b}{\dbe}}}
  -  (B+W\eps_{\dal\dbe}) \ket{\theta_{{a}{\dbe}} Z'_{{\b}{\dal}}} \\ &
  -H \ket{Z_{{\b}{\dal}}\theta'_{{a}{\dbe}}}
 -C\eps_{{\dal}{\dbe}} \eps^{{\da}{\db}} \ket{Y_{{a}{\da}}\eta'_{{\b}{\db}}} \\[3mm]
\bT\cdot   \ket{\eta_{{\a}{\da}} Z'_{{\b}{\dbe}}} = \ &
  - (A-G) \ket{\eta_{{\a}{\da}} Z'_{{\b}{\dbe}}}
  -  (B+W\eps_{\a\b}) \ket{\eta_{{\b}{\da}} Z'_{{\a}{\dbe}}} \\ &
  -H\ket{Z_{{\a}{\dbe}}\eta'_{{\b}{\da}}}
 +C \eps_{{\a}{\b}} \eps^{{a}{b}} \ket{Y_{{a}{\da}}\theta'_{{b}{\dbe}}}
\end{split}
\]
Here the coefficients are defined as follows\footnote{Note that the coefficients $C(p,p')$ and $H(p,p')$ differ by sign from the ones in \cite{Engelund:2014pla} if the signs of $p$ and $p'$ are opposite.}
\bea
\begin{aligned}
\la{Tmatrcoef}
&A(p,p')= \frac{1}{4} \frac{ (p-p')^2 +\dpnu^2 (\omega-\omega')^2}{p \omega' - p' \omega} \,,\\
&B(p,p')= \frac{p p' + \dpnu^2 \omega \omega'  }{p \omega' - p' \omega} \,, \\
&D(p,p')=-\frac{1}{4} \frac{ (p-p')^2 +\dpnu^2 (\omega-\omega')^2  }{p \omega' - p' \omega} \,,\\
&G(p,p')=-\frac{1}{4} \frac{\om^2-\om'^2}{p \omega' - p' \omega} \,,\\
&W(p,p')= i\dpnu  \,,\\
&C(p,p')=-(1+\vk^2)p p' \sqrt{1+{\nu^2\ov p^2}}\sqrt{1+{\nu^2\ov p'^2}}\,\frac{\sinh{\theta-\theta'\ov2}}{p \omega' - p' \omega} \,,\\
&H(p,p')=(1+\vk^2)p p' \sqrt{1+{\nu^2\ov p^2}}\sqrt{1+{\nu^2\ov p'^2}}\,\frac{\cosh{\theta-\theta'\ov2}}{p \omega' - p' \omega} \, .
\end{aligned}
\eea

\subsection{Factorisation}

Let us recall that in the undeformed case, as a consequence of invariance of ${\mathbb S}$ with respect to two copies of the centrally extended superalgebra $\psu(2|2)$, there is a basis of two-particle states such that the ${\mathbb T}$-matrix elements with respect to this basis admit a factorisation
\bea
{\mathbb T}^{P\dot{P},Q\dot{Q}}_{M\dot{M},N\dot{N}}=(-1)^{\eps_{\dot M}(\eps_{N}+\eps_{Q})}{\cal T}_{MN}^{PQ}\delta_{\dot{M}}^{\dot{P}}\delta_{\dot{N}}^{\dot{Q}}
+(-1)^{\eps_Q(\eps_{\dot{M}}+\eps_{\dot{P}})}\delta_{M}^{P}\delta_{N}^{Q} {\cal T}_{\dot{M}\dot{N}}^{\dot{P}\dot{Q}}\, .
\eea 
Here $M=(a, \alpha)$ and $\dot{M}=(\dot{a},\dot{\alpha})$, and  dotted and undotted indices are referred to two copies of $\psu(2|2)$, respectively, while 
$\eps_{M}$ and $\eps_{\dot{M}}$ describe statistics of the corresponding indices, {\it i.e.} they  are zero for bosonic (Latin) indices and equal to one for fermionic (Greek) ones.
The factor ${\mathcal T}$ can be regarded as $16\times 16$ matrix.

\smallskip

As was shown in \cite{Arutyunov:2013ega}, in the deformed model the bosonic  T-matrix elements Boson-Boson $\to$ Boson-Boson  enjoy the same type of factorisation.  It is not difficult to see that the  T-matrix elements Boson-Boson $\to$ Boson-Boson $+$ Fermion-Fermion, and Fermion-Fermion $\to$ Boson-Boson  also admit the same factorisation. In fact these T-matrix elements determine all the coefficients \eqref{Tmatrcoef}, and  the  elements of the ${\mathcal T}$-matrix
\bea\la{cTmatr}
\begin{aligned}
&\cT_{ab}^{cd}=
A\,\de_a^c\de_b^d+(B+W\, \eps_{ab})\de_a^d\de_b^c\, ,   \\ 
&\cT_{\a\b}^{\g\de}=
-A\,\de_\a^\g\de_\b^\de+(-B+W\, \eps_{\a\b}) \de_{\a}^{\delta}\de_{\b}^{\gamma}\,,  \\
&\cT_{a\b}^{c\de}= G\,\de_a^c\de_\b^\de\,,\qquad
~\cT_{\a b}^{\g d}= -G\,\de_\a^\g\de_b^d\,,
 \\
&\cT_{ab}^{\g \de}= C\,\eps_{ab}\eps^{\g\de}\,,\qquad
~\cT_{\a \b}^{c d}= C\,\eps_{\a\b}\eps^{cd}\,,
 \\
&\cT_{a\b}^{\g d}= H\,\de_a^d\de_\b^\de\,,\qquad
~\cT_{\a b}^{c \de}= H\,\de_\a^\de\de_b^c\,.
\end{aligned}
\eea 
It is straightforward to check that this  ${\mathcal T}$-matrix coincides with the first nontrivial term in the large $g$ expansion of the properly normalised q-deformed $\psu_q(2|2)$ invariant S-matrix, i.e.  with the corresponding classical $r$-matrix.

Despite this promising agreement, the full  T-matrix does not factorise. Indeed, by using  \eqref{cTmatr}, it is not difficult to see that the scattering elements Boson-Fermion $\to$ Boson-Fermion  listed in the previous subsection cannot be written in the same factorised form  because they have wrong signs in front of $W$. One can also check that there is no unitary transformation of the basis of one-particle states which would restore the factorisability.  Nevertheless, there exists a change of the basis of 2-particle states which brings these T-matrix elements to the factorised form.\footnote{Obviously, the resulting factorised $\bT$ satisfies the cYBE, while the original $\bT$-matrix does not  for some scattering processes. To be precise, those are the processes which involve Boson-Fermion to Boson-Fermion transmission amplitudes.}  Let us consider a 2-particle state made of one boson and one fermion. In any such a state there is exactly one pair of indices of the same type, e.g. $(a,b)$ or $(\dal,\dbe)$, for example
\be\la{Yathb}
\ket{Y_{a\dot{b}}\theta_{b\dot\a}} 
\ee
has the pair $(a,b)$. For any such a state we perform the transformation which exchanges the indices $1\leftrightarrow 2$ or $3\leftrightarrow 4$, or the corresponding dotted indices, and in addition multiplies each of these states by $i$. This changes the sign  in front of $W$, and restores the factorisability.  The existence of this transformation means that the T-matrix can be written in the form 
\be
\bT = \bU\cdot \bT_{\rm q}\cdot \bU^\dagger\,,\quad \bU^\dagger\cdot \bU = \bI\,,
\ee
 where $\bU$ is a unitary operator which realises the transformation just described, and $\bT_{\rm q}$ is the T-matrix which factorises in the standard way with the $q$-deformed $\cT$-matrix as its building block. It is clear that the restriction of the operator onto the space of one- and two-particle states satisfies the condition $\bU^2= -\mI$. The Hamiltonian $ \bH_{\rm q}$ which leads in a natural way to the $q$-deformed scattering T-matrix is obviously given by
 \be
\bH_{\rm q} = \bU^\dagger\cdot \bH\cdot \bU.
\ee
 It is easy to construct an operator $\bU$ which satisfies the necessary properties. For example the operator $\bU_{12}$ which exchanges the indices 1 and 2 of two-particle states \eqref{Yathb}  while acting trivially on all the other two-particle states is given by
\be\la{U12}
\bU_{12} = e^{i{\pi\ov 2}(\bL_{1}^{\rm b}{}^2+\bL_{2}^{\rm b}{}^1)(\bL_{1}^{\rm f}{}^2+\bL_{2}^{\rm f}{}^1)}\,,
\ee
 where the operators $\bL_{a}^{\rm b}{}^b$ and $\bL_{a}^{\rm f}{}^b$ are bosonic and fermionic parts of the $\su(2)$ generators
  \bal
\bL_{a}{}^{b}&=\bL_{a}^{\rm b}{}^b+\bL_{a}^{\rm f}{}^b =\int {\rm d}p\, \sum_{{\dot M}}\, {1\ov 2} \Big(\,  a^\dagger_{a {\dot M}} a^{b {\dot M}} - \eps_{ad} \eps^{bc}\, a^\dagger_{c {\dot M}} a^{d {\dot M}}\, \Big) \,,\\
\bL_{a}^{\rm b}{}^b &=\int {\rm d}p\, \sum_{{\dot c}}a^\dagger_{a {\dot c}} a^{b {\dot c}} \,,\quad \bL_{a}^{\rm f}{}^b =\int {\rm d}p\, \sum_{{\dot \g}}a^\dagger_{a {\dot \g}} a^{b {\dot \g}} \,,\quad a\neq b\,.
\eal
The full operator $\bU$ is obviously given by the product
\be\label{opU}
\bU = \bU_{12}\cdot  \bU_{34}\cdot  \bU_{\dot1\dot2}\cdot  \bU_{\dot3\dot4}\,.
\ee
Since the exponential of $\bU$ is a linear combination of products of integrals, the Hamiltonian $\bH_{\rm q}$ is seemingly highly nonlocal.  
 
We conclude this section by pointing out that while we have found 16 non-vanishing RR couplings, 
the quartic Lagrangian we used to compute the T-matrix depends only on six of them
$$
F_{014}, \,   F_{123}, \,  F_{569}, \,  F_{678}, \,  F_{01234}, \,  F_{04678}\, .
$$
Other couplings will apparently contribute beyond the quartic order.

%%%%%%%%%%%%%%%%%%%%%%%%%%%%%%
\section{Conclusions}

The main result of this paper is the calculation of the part of the  Lagrangian of the $\eta$-deformed model which is quadratic in fermions and has the full dependence on the bosonic fields. 

We have shown that a field redefinition is necessary to cast the original Lagrangian in the standard form and that 
the simplest and natural redefinition leads to RR couplings which do not satisfy the equations of motion of type IIB supergravity. Moreover, the wide class of transformations considered in section \ref{sec:red}  does not allow one to change the RR couplings while keeping the NSNS couplings untouched. We have not however analysed more involved changes of fields which depend for example on fermions and their derivatives, or even are nonlocal.
One cannot also exclude the existence of a discrete transformation (maybe of the type considered in section \ref{sec:Tmatrix}?). 

 Assuming however that this is the final answer for the RR couplings the question is whether the $\eta$-deformed sigma model can be considered as a string theory sigma model. The usual ways to address this question (the vanishing of conformal anomaly or the modular invariance of the partition function) are difficult to implement for a $\kappa$-symmetric Green-Schwarz sigma model. A related and more general question is -- given a sigma model with 8  bosonic and  8  fermionic physical degrees of freedom how to determine whether it is a string theory.

The Lagrangian we found can be used to address many interesting questions. Let us list some of them.

There are many different choices of a light-cone gauge 
because there are three isometry directions on the $\eta$-deformed sphere. We have shown that the standard choice leads to a vacuum which does not receive quantum corrections at least at one-loop level. It would be interesting to see what happens with other simple choices where one chooses the angle $\p_1$ or $\p_2$ as the light-cone gauge space isometry direction. 

There are many explicit classical solutions for the $\eta$-deformed model,   see {\it e.g.}  \cite{Arutyunov:2014cda}-\cite{Banerjee:2015nha},  which reduce to known  \ads string solutions. It would be interesting to compute one-loop corrections to the deformed solutions and compare them with the undeformed results.  This may shed some light on the structure of the mysterious dual ``field'' theory. Since in the scaling limit discussed in section \ref{sec:kappa} the $\eta$-deformed background reduces to the Maldacena-Russo background,  the dual model should be a deformation of the non-commutative ${\cal N}=4$ SYM.

We have mentioned that in the limit $\vk\to\infty$ the RR couplings we found do not reduce to those of the \ads mirror background. It would be interesting to compute and compare T-matrices for the $\vk=\infty$ background obtained from our Lagrangian, and from the mirror Lagrangian. 

We found that to get a factorisable two-body S-matrix we have to perform the transformation (\ref{opU}). It would be interesting to investigate the scattering of 3 particles into 3 particles and find out if the same 
transformation would bring  the three-body S-matrix to a factorisable form.

In this paper we considered the $R$-operator corresponding to the standard Dynkin diagram of $\psu(2,2|4)$. 
It is believed  however that in the undeformed case the so-called ``all-loop" Dynkin diagrams \cite{BS05} give the only consistent choice for finite $\lambda$.
It would be very interesting to investigate the $R$-operator corresponding to  these Dynkin diagrams, and determine how it influences the Lagrangian  and T-matrix. 

 Let us finally mention that recently the $\eta$- and $\lambda$-deformations were related by the Poisson-Lie duality   \cite{Vicedo:2015pna,Hoare:2015gda,Sfetsos:2015nya}. It would be interesting to understand how the  Poisson-Lie duality acts on 
the background fields, and hopefully use it to rederive from it the RR couplings we found.

\section*{Acknowledgements}

We would like to thank Niklas Beisert, Ben Hoare, Radu Roiban and Stijn van Tongeren for helpful discussions, and Ben Hoare, Stijn van Tongeren and Arkady Tseytlin for useful comments on the manuscript. 
G.A. is grateful to the organizers of the workshop ``$\eta$ and $\lambda$
Deformations in Integrable Systems and Supergravity" at the ITP Bern where some of the present results were presented, for the stimulating discussions
and kind hospitality.
The work of G.A. is supported by the German Science Foundation (DFG) under the Collaborative
Research Center (SFB) 676 Particles, Strings and the Early Universe.
R.B. acknowledges support by the Netherlands Organization for Scientific Research
(NWO) under the VICI grant 680-47-602. His work is also part of the ERC Advanced
grant research programme No. 246974, ``Supersymmetry: a window to non-perturbative
physics'', and of the D-ITP consortium, a program of the NWO that is funded by the
Dutch Ministry of Education, Culture and Science (OCW). During the first stages of this long-term project S.F. was supported by a DFG grant in the framework of the SFB 647 Raum - Zeit - Materie. Analytische und Geometrische Strukturen
and by the Science Foundation Ireland under Grant 09/RFP/PHY2142.

%\section{Appendices}

\appendix

\section{Conventions}
\subsection{Basis of $\psu(2,2|4)$}\label{sec:algebra-basis}
Here we  introduce a basis of the superalgebra $\psu(2,2|4)$ which we use throughout the paper, present the commutation relations between
the corresponding generators and  recall the ${\mathbb Z}_4$-graded decomposition of $\psu(2,2|4)$.

\smallskip

Recall that the superalgebra $\alg{sl}(4|4)$ is generated by $8\times 8$ matrices 
\be
M=
\left(
\begin{array}{c|c}
 m_{11} & m_{12} 
\\
\hline
m_{21} & m_{22}
\end{array}
\right)\,,
\ee
where each $m_{ij}$ above is a $4\times 4$ block. The matrix $M$ is required to have vanishing supertrace, defined as $\Str M=\tr m_{11}-\tr m_{22}$.
The diagonal blocks $m_{11}, m_{22}$ are even, while the off-diagonal blocks $m_{12},m_{21}$ are odd. 
The algebra $\su(2,2|4)$ is a real form of  $\alg{sl}(4|4)$ which is obtained by demanding the following reality condition
\be\label{eq:real-cond-su224}
M^\dagger H+HM=0\,,
\ee
where the matrix $H$ is
\be
H=\left(
\begin{array}{cc}
 \Sigma & 0  \\
 0 &  {\bf 1}_4  \\
\end{array}
\right)\,,
\ee
and the diagonal matrix $\Sigma$ is defined in (\ref{SKC}). The algebra $\psu(2,2|4)$
is obtained from $\su(2,2|4)$ by projecting out a one-dimensional centre generated by $i {\bf 1}_8$.

\smallskip 

\noindent
\paragraph{Gamma matrices}%\label{sec:gamma-mat}
The bosonic generators of $\psu(2,2|4)$ are constructed with the help of  ${\rm SO}(1,4)$ and ${\rm SO}(5)$ gamma matrices. We introduce the following matrices\footnote{We find it useful to exchange the definition of $\gamma_1, \gamma_4$ in comparison to the one of \cite{Arutyunov:2009ga}.}

\be
\begin{aligned}
&\gamma_0  =
\left(
\begin{array}{cccc}
 1 & 0 & 0 & 0 \\
 0 &  1 & 0 & 0 \\
 0 & 0 & -1 & 0 \\
 0 & 0 & 0 & -1 \\
\end{array}
\right) , \qquad
\gamma_1 =
\left(
\begin{array}{cccc}
 0 & 0 & -i & 0 \\
 0 & 0 & 0 &  i \\
  i & 0 & 0 & 0 \\
 0 & -i & 0 & 0 \\
\end{array}
\right), \qquad
\gamma_2 =
\left(
\begin{array}{cccc}
 0 & 0 & 0 &  i \\
 0 & 0 &  i & 0 \\
 0 & -i & 0 & 0 \\
 -i & 0 & 0 & 0 \\
\end{array}
\right), \\
&\gamma_3 =
\left(
\begin{array}{cccc}
 0\, &\, 0 &\,  1\, & 0\, \\
 0 & 0 & 0 & 1 \\
  1 & 0 & 0 & 0 \\
 0 & 1 & 0 & 0 \\
\end{array}
\right), \qquad~~
\gamma_4 =
\left(
\begin{array}{cccc}
 0 & 0 & 0 & -1 \\
 0 & 0 &  1 & 0 \\
 0 &  1 & 0 & 0 \\
 -1 & 0 & 0 & 0 \\
\end{array}
\right).
\end{aligned}
\ee
\medskip

\noindent
These matrices are hermitian and satisfy the ${\rm SO}(5)$ Clifford algebra $\{ \gamma_m, \gamma_n\} = 2 \delta_{mn}$.
To describe embeddings of the anti-de Sitter space and the five-sphere into the group ${\rm PSU}(2,2|4)$, we introduce
the matrices  $\check{\g}_m$ and $\hat{\g}_m$
\be\label{eq:gamma-AdS5-S5}
\begin{aligned}
\text{AdS}_5:\quad&\check{\gamma}_0 = i \gamma_0, \qquad \check{\gamma}_m = \gamma_m, \quad m=1,\cdots,4, \\
\text{S}^5:\quad&\hat{\gamma}_{m+5} = -\gamma_m, ~\quad\qquad\quad\quad m=0,\cdots,4.
\end{aligned}
\ee
We have chosen to enumerate the gamma matrices for ${\rm AdS}_5$ from 0 to 4 and the ones for ${\rm S}^5$
from 5 to 9 to adopt a smooth transition to the 10-dimensional notation. 
The matrices $\check{\g}_m$ and $\hat{\g}_m$ realise representations of the Clifford algebras ${\rm SO}(4,1)$ and ${\rm SO}(5)$, respectively.
We denote their matrix elements as  ${(\check{\gamma}_m)_{\ul{\a}} }^{\ul{\beta}}$ and ${(\hat{\gamma}_m)_{\ul{a}}}^{\ul{b}}$, where Greek and Latin indices are 
associated with AdS$_5$ and ${\rm S}^5$, respectively.

Further, we introduce the matrices $\Sigma, K,C$  

\be\la{SKC}
\begin{aligned}
\Sigma = 
%\gamma_0= 
\left(
\begin{array}{cccc}
 1 & 0 & 0 & 0 \\
 0 &  1 & 0 & 0 \\
 0 & 0 & -1 & 0 \\
 0 & 0 & 0 & -1 \\
\end{array}
\right), \qquad
K = 
%-\gamma_2 \gamma_1 =
\left(
\begin{array}{cccc}
 0 & -1 & 0 & 0 \\
 1 & 0 & 0 &  0 \\
 0 & 0 & 0 & -1 \\
 0 & 0 & 1 & 0 \\
\end{array}
\right), \qquad
C = 
%\gamma_4 \gamma_3 = 
\left(
\begin{array}{cccc}
  0 & -1 & 0 & 0 \\
 1 & 0 & 0 &  0 \\
 0 & 0 & 0 & 1 \\
 0 & 0 & -1 & 0 \\
\end{array}
\right).
\end{aligned}
\ee
\medskip

\noindent
The matrix elements of these matrices are assumed to carry upper indices
$\Sigma^{\ul{a}\ul{b}},K^{\ul{a}\ul{b}},C^{\ul{a}\ul{b}}$, while  the matrix elements of their inverses are defined with lower indices.
The matrices $\Sigma, K, C$ generate the following automorphisms of the Clifford algebra
\be
\gamma_m^t = K \gamma_m K^{-1},
\ee
\be
\begin{aligned}\label{aut}
\gamma_m^t & = -C \gamma_m C^{-1}, \quad m=1,...,4, \qquad
\gamma_0^t & = C \gamma_0 C^{-1}, \\
\gamma_m^\dagger & = -\Sigma \gamma_m \Sigma^{-1}, \quad m=1,...,4, \qquad
\gamma_0^\dagger & = \Sigma \gamma_0 \Sigma^{-1}.
\end{aligned}
\ee
It follows from the last line that $\check{\gamma}_m^\dagger = -\Sigma \check{\gamma}_m \Sigma^{-1}, \quad m=0,...,4$. Note that 
if we keep the same notations as in \cite{Arutyunov:2009ga}, it is the matrix $K$, not $C$, which plays the role of the charge conjugation matrix.

%For our conventions concerning spinors we refers the reader to appendix \ref{app-spinors}.

\medskip
\noindent
\paragraph{Even generators}%\label{sec:even-gen}
The bosonic subalgebra of $\psu(2,2|4)$ is $\su(2,2)\oplus\su(4)$. 
We spilt the generators of $\su(2,2)$ as $(\check{\gen{P}}_m,\check{\gen{J}}_{mn} )$ with $m,n=0,\ldots , 4$.
Here $\check{\gen{J}}_{mn} $ generate the subalgebra $\so(1,4)\subset \su(2,2)$. Analogously, 
the generators of $\su(4)$ are $(\hat{\gen{P}}_{m}, \hat{\gen{J}}_{mn} )$ with $m,n=5,\ldots, 9$, and 
$\hat{\gen{J}}_{mn} $ generate $\so(5)\subset \su(4)$. Explicitly we choose 
\bea
\check{\gen{P}}_m &=& 
\left(
\begin{array}{cc}
 -\frac{1}{2} \check{\gamma}_m & \mathbf{0}_4  \\
 \mathbf{0}_4 & \mathbf{0}_4 \\
\end{array}
\right), 
\qquad\qquad
\check{\gen{J}}_{mn} = 
\left(
\begin{array}{cc}
 \frac{1}{2} \check{\gamma}_{mn} & \mathbf{0}_4  \\
 \mathbf{0}_4 & \mathbf{0}_4 \\
\end{array}
\right), ~~\quad m,n=0,\ldots4\, ,
\\
\nonumber
\\
\hat{\gen{P}}_{m} &=& 
\left(
\begin{array}{cc}
 \mathbf{0}_4 & \, \mathbf{0}_4  \\
 \mathbf{0}_4 &\,  \frac{i}{2} \hat{\gamma}_m \\
\end{array}
\right), 
~~\qquad\qquad
\hat{\gen{J}}_{mn} = 
\left(
\begin{array}{cc}
 \mathbf{0}_4 & \mathbf{0}_4  \\
 \mathbf{0}_4 & \frac{1}{2} \hat{\gamma}_{mn} \\
\end{array}
\right), ~~\quad m,n=5,\ldots , 9\, .
\eea
Here we defined $\check{\gamma}_{mn} \equiv \frac{1}{2} [\check{\gamma}_m,\check{\gamma}_n]$ and $\hat{\gamma}_{mn} \equiv \frac{1}{2} [\hat{\gamma}_m,\hat{\gamma}_n]$.
%In the $\mathbb{Z}_4$-graded decomposition of $\psu(2,2|4)$ the space of degree zero is spanned by the generators $\gen{J}$, while the space of degree 2 by  $\gen{P}$.

\bigskip
\noindent
\paragraph{Odd generators} %\label{sec:odd-gen}
The 32 odd generators of $\alg{psu}(2,2|4)$ will be represented by $\gen{Q}_{~\ul{a}}^{I~ \ul{\a}}$ where $I=1,2$ and two spinor indices run $\ul{\a},\ul{a}=1,2,3,4$ correspond to fundamental representations of $\su(2,2)$
and $\su(4)$, respectively. Explicitly we choose
\be\label{eq:def-odd-el-psu224}
\begin{aligned}
\gen{Q}_{~\ul{a}}^{I~ \ul{\a}}&=e^{+i\pi/4}
\left(
\begin{array}{cc}
 \mathbf{0}_4 & m^{\ \, \ul{\a}}_{I \, \ul{a}}  \\
 K \left(m^{\ \, \ul{\a}}_{I \, \ul{a}}\right)^\dagger K & \mathbf{0}_4 \\
\end{array}
\right),
\end{aligned}
\ee
where  $4\times 4$ matrices $m^{\ \, \ul{\a}}_{I \, \ul{a}}$ are
\be\label{eq:def-odd-el-psu224(a)}
\begin{aligned}
{\left(m^{\ \, \ul{\a}}_{1 \, \ul{a}}\right)_j}^k &= e^{+i\pi/4+i\phi_{\gen{Q}}} \, \delta^{\ul{\a}}_j \delta_{\ul{a}}^k,
\qquad\qquad
{\left(m^{\ \, \ul{\a}}_{2 \, \ul{a}}\right)_j}^k = -e^{-i\pi/4+i\phi_{\gen{Q}}} \,  \delta^{\ul{\a}}_j \delta_{\ul{a}}^k.
\end{aligned}
\ee
and  $K$ is defined in~\eqref{SKC}. The phase $\phi_{\gen{Q}}$ reflects the ${\rm U}(1)$ external automorphism of $\su(2,2|4)$, and we set $\phi_{\gen{Q}}=0$. 
The supermatrices $\gen{Q}$   do not satisfy the reality condition (\ref{eq:real-cond-su224}) but rather 
\be \gen{Q}^\dagger (i{\mathcal H})+{\mathcal H}\gen{Q}=0\, , \qquad\qquad {\mathcal H}\equiv \left(
\begin{array}{cc}
 K & 0  \\
 0 &  K  \\
\end{array}
\right).
\ee
These matrices can be however related to supermatrices $\mathcal{Q}$ satisfying~\eqref{eq:real-cond-su224} as
\be
\mathcal{Q}= e^{+i\pi/4} \,
\left(
\begin{array}{cc}
 C & 0  \\
 0 &  K  \\
\end{array}
\right)
\gen{Q},
\qquad
\gen{Q}= -e^{-i\pi/4} \,
\left(
\begin{array}{cc}
 C & 0  \\
 0 &  K  \\
\end{array}
\right)
\mathcal{Q}\, .
\ee

\smallskip

\noindent
If we would take the linear combinations of $\mathcal{Q}$'s with Grassmann variables $\vartheta$ and require that $\ferm{\vartheta}{I}{\a}{a} \mathcal{Q}^{I \, \ul{\a}}_{\ \, \ul{a}}$ is in $\alg{su}(2,2|4)$, then we 
 would have real fermions $\ferm{\vartheta}{I}{\a}{a}$, which is one of the possible realisations of the Majorana condition. Instead, in this paper we choose to construct the Grassmann envelope as 
$\ferm{\theta}{I}{\a}{a} \genQind{I}{\a}{a}$. Requiring 
$$(\ferm{\theta}{I}{\a}{a} \genQind{I}{\a}{a})^\dagger= -H(\ferm{\theta}{I}{\a}{a} \genQind{I}{\a}{a})H^{-1}
$$ 
implies that
\be
\ferm{{\theta^\dagger}}{I}{a}{\a} =-i\, \ferm{\theta}{I}{\nu}{b} \ C^{\ul{\nu}\ul{\a}} K_{\ul{b}\ul{a}}\,  .
\ee
Defining the barred version of a fermion we then obtain the following realisation of the Majorana condition
\be
\ferm{\bar{\theta}}{I}{a}{\a} \equiv \ferm{{\theta^\dagger}}{I}{a}{\nu} {(\check{\gamma}^0)_{\ul{\nu}}}^{\ul{\a}} = - \ferm{\theta}{I}{\nu}{b} \ K^{\ul{\nu}\ul{\a}} K_{\ul{b}\ul{a}} .
\ee
Throughout the paper we are using fermions $\ferm{\theta}{I}{\a a}{}$ with both spinor indices lowered and $\ferm{\bar{\theta}}{I}{}{\a a}$ with both spinor indices raised,\footnote{The rules for raising and lowering spinor indices
are given in appendix \ref{app-spinors}, see eq.(\ref{eq:rl}).} so the above equation reads as
\be\label{eq:Majorana-cond-ferm-sp-ind}
\ferm{\bar{\theta}}{I}{}{\a a}  = + \, \ferm{\theta}{I}{\nu b}{} \ K^{\ul{\nu}\ul{\a}} K^{\ul{b}\ul{a}} ,
\ee
matching the conventions of~\cite{Metsaev:1998it}. In the matrix conventions this equation reads as 
\be\label{eqMajorana-cond-compact-not}
\bar{\theta}_I = \theta_I^\dagger \bg^0=+ \theta_I^t \, (K\otimes K)\,,
\ee
where $\bg^0\equiv \check{\g}^0\otimes \gen{1}_4$, and hermitian conjugation and transposition are implemented on the space spanned by the spinor indices, where the matrices $\bg^0$ and $K\otimes K$ are acting.

\paragraph{Commutation relations}%\label{sec:comm-rel}
In our basis the commutation relations involving the bosonic elements only read as
\bea
\begin{aligned}
%\text{AdS}_5 : 
\quad &[ \check{\gen{P}}_m, \check{\gen{P}}_n ] = \check{\gen{J}}_{mn}, \ 
&\ \  
%\text{S}^5 : 
\quad &[ \hat{\gen{P}}_m, \hat{\gen{P}}_n ] = -\hat{\gen{J}}_{mn}, \\
&[ \check{\gen{P}}_{m}, \check{\gen{J}}_{np} ] = \eta_{mn} \check{\gen{P}}_p - \ _{n \leftrightarrow p}, \ 
& &[ \hat{\gen{P}}_{m}, \hat{\gen{J}}_{np} ] = \eta_{mn} \hat{\gen{P}}_p - \ _{n \leftrightarrow p},\\
&[ \check{\gen{J}}_{mn}, \check{\gen{J}}_{pq} ] = (\eta_{np} \check{\gen{J}}_{mq} - \ _{m \leftrightarrow n} ) - \ _{p \leftrightarrow q} \ 
& &[ \hat{\gen{J}}_{mn}, \hat{\gen{J}}_{pq} ] = (\eta_{np} \hat{\gen{J}}_{mq} - \ _{m \leftrightarrow n} ) - \ _{p \leftrightarrow q},
\end{aligned}
\eea
where 
\bea\la{eq:min_metric}
\eta_{mn}= \text{diag}(-1,1,1,1,1,1,1,1,1,1)\, .
\eea
Generators from two different subalgebras commute with each other.
The commutators between odd and even elements with explicit spinor indices read as
\bea
\begin{aligned}
& [\genQind{I}{\a a}{}, \check{\gen{P}}_m] = - \frac{i}{2} \epsilon^{IJ}  \ \genQind{J}{\nu a}{}\ {(\check{\gamma}_m)_{\ul{\nu}}}^{\ul{\a}}, & \qquad
& [\genQind{I}{\a a}{}, \hat{\gen{P}}_m] =  \frac{1}{2} \epsilon^{IJ}  \ \genQind{J}{\a b}{}\ {(\hat{\gamma}_m)_{\ul{b}}}^{\ul{a}}, & \\
& [\genQind{I}{\a a}{}, \check{\gen{J}}_{mn}] = - \frac{1}{2} \delta^{IJ} \ \genQind{J}{\nu a}{}\  {(\check{\gamma}_{mn})_{\ul{\nu}}}^{\ul{\a}}, & \qquad
& [\genQind{I}{\a a}{}, \hat{\gen{J}}_{mn}] =  -\frac{1}{2} \delta^{IJ}  \ \genQind{J}{\a b}{}\ {(\hat{\gamma}_{mn})_{\ul{b}}}^{\ul{a}}. & 
\end{aligned}
\eea
The anti-commutator of two supercharges gives
\bea
\{\genQind{I}{\a a}{}, \genQind{J}{\nu b}{}\} &=& \delta^{IJ} \left( i\, K^{\ul{\a}\ul{\lambda}} K^{\ul{a}\ul{b}} \ {(\check{\gamma}^m)_{\ul{\lambda}}}^{\ul{\nu}} \, \check{\gen{P}}_m - \, K^{\ul{\a}\ul{\nu}} \, K^{\ul{a}\ul{c}} {(\hat{\gamma}^m)_{\ul{c}}}^{\ul{b}}  \, \hat{\gen{P}}_m -\frac{i}{2} K^{\ul{\a}\ul{\nu}} K^{\ul{a}\ul{b}} \mathbf{1}_8 \right) \nonumber \\
&- &  \frac{1}{2} \epsilon^{IJ} \left( K^{\ul{\a}\ul{\lambda}} K^{\ul{a}\ul{b}} \ {(\check{\gamma}^{mn})_{\ul{\lambda}}}^{\ul{\nu}} \, \check{\gen{J}}_{mn}  - \, K^{\ul{\a}\ul{\nu}} \, K^{\ul{a}\ul{c}} {(\hat{\gamma}^{mn})_{\ul{c}}}^{\ul{b}}  \, \hat{\gen{J}}_{mn} \right),
\eea
where the indices $m,n$ are raised with the metric $\eta_{mn}$. For completeness we also keep the term proportional to the identity, since the supermatrices provide a realisation of $\alg{su}(2,2|4)$. To obtain $\alg{psu}(2,2|4)$ one just needs to drop the term proportional to $i\mathbf{1}_8$ in the r.h.s. of the anti-commutator.
\smallskip

It is convenient to rewrite the commutation relations for the Grassmann enveloping algebra. In this way we may suppress the spinor indices to obtain more compact expressions. We define $\gen{Q}^{I} \theta_I\equiv \genQind{I}{\a a}{} \ferm{\theta}{I}{\a a}{}$ and introduce the $16\times 16$ matrices
\be\label{eq:def16x16-gamma}
\begin{aligned}
& \bg_m \equiv \check{\g}_m \otimes \mathbf{1}_4,
\quad m=0, \cdots, 4,
\qquad
& \bg_m \equiv  \mathbf{1}_4 \otimes i\hat{\g}_m,
\quad m=5, \cdots, 9, \\
& \bg_{mn} \equiv \check{\g}_{mn} \otimes \mathbf{1}_4,
\quad m,n=0, \cdots, 4,
\qquad
& \bg_{mn} \equiv  \mathbf{1}_4 \otimes \hat{\g}_{mn},
\quad m,n=5, \cdots, 9. \\
\end{aligned}
\ee
The first space in the tensor product is spanned by the AdS spinor indices, the second by the sphere spinor indices.
To understand the 10-dimensional origin of these objects see appendix~\ref{sec:10-dim-gamma}.
In the context of type IIB, one loosely refers to $\bg_m$ as gamma matrices even though they do not satisfy the Clifford algebra.
With the above definitions, the commutation relations of $\su(2,2|4)$ involving odd generators are\footnote{For commutators of two odd elements we need to multiply by two \emph{different} fermions $\lambda_I,\psi_I$, otherwise the right hand side vanishes.}
\be
\begin{aligned}
& [\gen{Q}^{I} \theta_I, \gen{P}_m] = - \frac{i}{2} \epsilon^{IJ} \gen{Q}^{J} \bg_m  \theta_I, & \qquad
& [\gen{Q}^{I} \theta_I, \gen{J}_{mn}] =  -\frac{1}{2} \delta^{IJ} \gen{Q}^{J} \bg_{mn}  \theta_I, & 
\end{aligned}
\ee
\be\label{eq:comm-rel-QQ-su224}
\begin{aligned}
[ \gen{Q}^{I} \lambda_I, \gen{Q}^{J} \psi_J ] =& \, i \, \delta^{IJ} \bar{\lambda}_I \bg^m  \psi_J \ \gen{P}_m  %
-  \frac{1}{2} \epsilon^{IJ} \bar{\lambda}_I (\bg^{mn} \check{\gen{J}}_{mn} -\bg^{mn}  \hat{\gen{J}}_{mn}) \psi_J \ 
 - \frac{i}{2} \delta^{IJ} \bar{\lambda}_I \psi_J \mathbf{1}_8.
\end{aligned}
\ee
Here we also used the Majorana condition to rewrite the result in terms of the fermions $\bar{\lambda}_I$.

%%%%%%%%%%%%%%%%%%%%%%%%%%%
%%%%%%%%%%%%%%%%%%%%%%%%%%%
\paragraph{Supertraces}%\label{sec:str-gen}
In the computation for the Lagrangian we will need to take the supertrace of products of two generators of the algebra. For the non-vanishing ones we find
\bea
\begin{aligned}
&\Str[\gen{P}_m\gen{P}_n]=\eta_{mn}, \\
&\Str[\check{\gen{J}}_{mn}\check{\gen{J}}_{pq}]= - (\eta_{mp}\eta_{nq}-\eta_{mq}\eta_{np}),  \\
&\Str[\hat{\gen{J}}_{mn}\hat{\gen{J}}_{pq}]= + (\eta_{mp}\eta_{nq}-\eta_{mq}\eta_{np}),  \\
&\Str[\genQind{I}{\a a}{}\genQind{J}{\nu b}{}] = -2 \epsilon^{IJ} K^{\ul{\a}\ul{\nu}} K^{\ul{a}\ul{b}}\, .
\end{aligned}
\eea
The last formula for the supertrace of two elements from the Grassmann envelope reads as
\bea
\Str[\gen{Q}^I \lambda_I \, \gen{Q}^J \psi_J ]= -2 \epsilon^{IJ} \bar{\lambda}_I \psi_J = -2 \epsilon^{JI} \bar{\psi}_J \lambda_I \,.\eea

%%%%%%%%%%%%%%%%%%%%%%%%%%%%

%%%%%%%%%%%%%%%%%%%%%%%%%%
\paragraph{$\mathbb{Z}_4$-decomposition}%\label{sec:Z4-dec}
The $\su(2,2|4)$ algebra admits a $\mathbb{Z}_4$-graded decomposition.
Introducing the following automorphism $\Omega$ of $\alg{sl}(4|4)$, see {\it e.g.} \cite{Arutyunov:2009ga},
\be
\Omega(M) = - \mathcal{K} M^{st} \mathcal{K}^{-1},
\ee
with $\mathcal{K}=\text{diag}(K,K)$ and $^{st}$ denoting the supertranspose
\be
M^{st}\equiv
\left(
\begin{array}{c|c}
 m_{11}^t & -m_{21}^t
\\
\hline
m_{12}^t & m_{22}^t
\end{array}
\right)\, ,
\ee
the real form $\mathscr{G}=\psu(2,2|4)$ can be decomposed with respect to $\Omega$ into a direct sum of four graded vector subspaces
\bea
\mathscr{G}=\mathscr{G}^{(0)}\oplus \mathscr{G}^{(1)}\oplus \mathscr{G}^{(2)}\oplus \mathscr{G}^{(3)}\, , ~~~~~\mathscr{G}^{(k)}=\{M\in\mathscr{G},~\Omega(M)=i^k\, M\}\, .\eea
The bosonic generators  $\gen{J}$ and $\gen{P}$ have degree 0 and 2, respectively,
\be
\Omega(\gen{J})=+\gen{J}\,,
\qquad
\Omega(\gen{P})=-\gen{P}\,.
\ee
In our basis the action of $\Omega$ on odd generators is also very simple
\be
\Omega(\genQind{I}{\a a}{})=(-1)^{I+1} i\, \genQind{I}{\a a}{}\,,
\ee
meaning that odd elements with $I=1$ and $I=2$ have degree 1 and 3, respectively.
We introduce projectors $P^{(k)}$ on each subspace, whose action is
\be\label{eq:def-proj-Z4-grad}
P^{(k)}(M)=\frac{1}{4}\left( M+i^{3k}\Omega(M)+i^{2k}\Omega^2(M) +i^{k}\Omega^{3}(M)\right)\,.
\ee
Then $P^{(0)}$ will project on generators $\gen{J}$, $P^{(2)}$ on generators $\gen{P}$, and $P^{(1)},P^{(3)}$ on odd elements with labels $I=1,2$
\be
P^{(1)}(\genQind{I}{\a a}{}) = \frac{1}{2} (\delta^{IJ}+\sigma_3^{IJ}) \genQind{J}{\a a}{}, \qquad 
P^{(3)}(\genQind{I}{\a a}{}) = \frac{1}{2} (\delta^{IJ}-\sigma_3^{IJ}) \genQind{J}{\a a}{}. 
\ee
The definition  of the \ads coset implies that the generators $\gen{J}$ of degree zero which span the $\alg{so}(4,1) \oplus \alg{so}(5)$ subalgebra are projected out.

\subsection{Spinor rules}\label{app-spinors}

For raising and lowering spinor indices we adopt the conventions of~\cite{Freedman:2012zz}
\be\la{eq:rl}
\lambda^\a = K^{\a\b} \lambda_\b,
\qquad
\lambda_\a = \lambda^\b K_{\b\a},
\ee
where $K^{\a\b}$ are the components of the matrix $K$, that plays the role of charge conjugation matrix.
We also have
\be
K^{\a\b} K_{\g\b} = \delta^\a_\g,
\qquad
K_{\b\a} K^{\b\g} = \delta_\a^\g,
\qquad
\chi^\a \lambda_\a = - \chi_\a \lambda^\a .
\ee

The five-dimensional gamma matrices have  the following symmetry properties 
\be\label{eq:symm-prop-5dim-gamma}
\begin{aligned}
(K\gamma^{(r)})^t &= - t_r^\g \ K\gamma^{(r)}\,,
\\
K(\gamma^{(r)})^t K&= - t_r^\g \ \gamma^{(r)}\,,
\qquad
t_0^\g=t_1^\g=+1,\quad  t_2^\g=t_3^\g=-1\,.
\end{aligned}
\ee
Here $\gamma^{(r)}$ denotes the antisymmetrised product of $r$ gamma matrices and the coefficients. The coefficients  
$t_r^\g$ are the same for both $\check{\g}^{(r)}$ and $\hat{\gamma}^{(r)}$, and we label them with the superscript 
${}^\g$ to distinguish from the corresponding coefficients of 10-dimensional gamma matrices.
For the rules concerning hermitian conjugation we find
\be\label{eq:herm-conj-prop-5dim-gamma}
\begin{aligned}
&\check{\g}_m^\dagger \phantom{{}_n}=+\check{\g}^0\check{\g}_m\check{\g}^0\,,
\qquad
&&\hat{\g}_m^\dagger \phantom{{}_n}=+\hat{\g}_m\,,
\\
&\check{\g}_{mn}^\dagger=+\check{\g}^0\check{\g}_{mn}\check{\g}^0\,,
\qquad
&&\hat{\g}_{mn}^\dagger=-\hat{\g}_{mn}\,,
\end{aligned}
\ee
With these rules we can derive the following 
useful formulae for  dealing with Dirac conjugate spinors
\be\nonumber
\begin{aligned}
& ((\check{\gamma}_m \otimes \mathbf{1}_4) \theta_I)^\dagger (\check{\gamma}_0 \otimes \mathbf{1}_4) = - \bar{\theta}_I (\check{\gamma}_m \otimes \mathbf{1}_4) , &
~
& ((\mathbf{1}_4 \otimes \hat{\gamma}_m) \theta_I)^\dagger (\check{\gamma}_0\otimes \mathbf{1}_4) = + \bar{\theta}_I (\mathbf{1}_4 \otimes \hat{\gamma}_m ),& 
\\
& ((\check{\gamma}_{mn}\otimes \mathbf{1}_4) \theta_I)^\dagger (\check{\gamma}_0\otimes \mathbf{1}_4) = - \bar{\theta}_I (\check{\gamma}_{mn} \otimes \mathbf{1}_4),& 
~
& ((\mathbf{1}_4 \otimes \hat{\gamma}_{mn}) \theta_I)^\dagger (\check{\gamma}_0\otimes \mathbf{1}_4) = - \bar{\theta}_I (\mathbf{1}_4 \otimes \hat{\gamma}_{mn}). & 
\end{aligned}
\ee
Thanks to~\eqref{eq:symm-prop-5dim-gamma} one can also show that given two Grassmann bi-spinors $\psi_{\ul{\a}\ul{a}},\chi_{\ul{\a}\ul{a}}$ the ``Majorana-flip'' relations are
\be\label{eq:symm-gamma-otimes-gamma}
\bar{\chi} \left(\check{\gamma}^{(r)}\otimes \hat{\gamma}^{(s)} \right) \psi = - t_r^\g t_s^\g \ \bar{\psi} \left(\check{\gamma}^{(r)}\otimes \hat{\gamma}^{(s)} \right) \chi.
\ee
%We show for example the first one
%\be
%\bar{\chi} \psi = +i \chi_{\ul{\a}\ul{a}} K^{\ul{\a}\ul{\nu}} K^{\ul{a}\ul{b}} \psi_{\ul{\nu}\ul{b}}
%= -i \psi_{\ul{\nu}\ul{b}} K^{\ul{\a}\ul{\nu}} K^{\ul{a}\ul{b}} \chi_{\ul{\a}\ul{a}}
%= -i \psi_{\ul{\nu}\ul{b}} K^{\ul{\nu}\ul{\a}} K^{\ul{b}\ul{a}} \chi_{\ul{\a}\ul{a}}
%=-\bar{\psi} \chi
%\ee
%
With this formula at hand, it is easy to prove that
\be\label{eq:Maj-flip}
s^{IJ }\bar{\theta}_I \left(\check{\gamma}^{(r)}\otimes \hat{\gamma}^{(s)} \right) \theta_J=0
\qquad \text{ if }~
\left\{\begin{array}{ccc} 
s^{IJ} = + s^{JI}& \text{ and } & t_r^\g t_s^\g=+1 \\
s^{IJ} = - s^{JI}& \text{ and } & t_r^\g t_s^\g=-1 \\
 \end{array} \right.\,.
\ee
Finally, up to a total derivative the following relations hold
\be
\bar{\psi} \mathcal{D} \lambda =  \bar{\lambda} \mathcal{D} \psi,
\qquad
\bar{\psi}_I D^{IJ} \lambda_J =  \bar{\lambda}_J D^{JI} \psi_I\, .
\ee

%%%%%%%%%%%%%%%%%%%

\subsection{10-dimensional $\Gamma$-matrices}\label{sec:10-dim-gamma}
We use the $4\times 4$ gamma matrices $\check{\g}, \hat{\g}$ to define the $32 \times 32$ gamma matrices $\G_m$:
\be\label{eq:def-10-dim-Gamma}
\G_m = \sigma_1 \otimes \check{\g}_m \otimes {\bf 1}_4  , \ \ m=0, \cdots, 4,
\qquad
\G_m =  \sigma_2 \otimes {\bf 1}_4 \otimes \hat{\g}_m , \ \ m=5, \cdots, 9,
\ee
that satisfy $\{\G_m,\G_n\}= 2\eta_{mn}$. We also define $\G_{11} \equiv \G_0 \cdots \G_9 = \sigma_3 \otimes  {\bf 1}_4 \otimes {\bf 1}_4 $. 
Anti-symmetrised products of gamma matrices are
$\G_{m_1\cdots m_r} = \frac{1}{r!} \G_{[m_1} \cdots \G_{m_r]}$.
The charge conjugation matrix is defined as $\mathcal{C} \equiv  i\, \sigma_2 \otimes K \otimes K$, and $\mathcal{C}^2=-\mathbf{1}_{32}$.
In the chosen representation the matrices $\Gamma^{(r)}$ have the symmetry properties
\be
\begin{aligned}
(\mathcal{C} \G^{(r)})^t &= - t_r^\G \ \mathcal{C} \G^{(r)},
\\
\mathcal{C}( \G^{(r)})^t\mathcal{C} &= - t_r^\G \ \G^{(r)},
\qquad
t_0^\G =t_3^\G =+1,
\qquad
 t_1^\G =t_2^\G=-1.
\end{aligned}
\ee
For hermitian conjugation we find
\be
\G^0 (\G^{(r)})^\dagger \G^0 =
\left\{\begin{array}{c}
+ \G^{(r)}, \quad r=1,2 \text{ mod } 4,
\\
- \G^{(r)}, \quad r=0,3 \text{ mod } 4.
\end{array}
\right.
\ee
Given two 4-component spinors $\check{\psi}, \hat{\psi}$ transforming in the fundamental representations of 
 $\su(2,2)$ and $\su(4)$ respectively, a 32-component spinor is constructed as
\be
\Psi_+ = \left( \begin{array}{c} 1 \\ 0 \end{array} \right) \otimes \check{\psi} \otimes \hat{\psi} , 
\qquad
\Psi_- = \left( \begin{array}{c} 0 \\ 1 \end{array} \right) \otimes \check{\psi} \otimes \hat{\psi} , 
\ee
for the case of positive and negative chirality respectively.
In the main text we use 16-component fermions with two spinor indices $\theta_{\ul{\a}\ul{a}}$, and we construct a 32-component Majorana fermion of positive chirality as 
\be\label{eq:def-32-dim-Theta}
\Theta = \left( \begin{array}{c} 1 \\ 0 \end{array} \right) \otimes \theta ,
\qquad
\bar{\Theta} = \Theta^t \mathcal{C} =  \left( \  0 \ , \ 1 \  \right) \otimes\bar{\theta} .
\ee
In \eqref{eq:def16x16-gamma} we have defined the $16\times 16$-matrices $\bg_m$. Now we see that
\be
\bar{\Theta}_1 \G_m \Theta_2 \equiv \bar{\theta}_1 \bg_m \theta_2 
\implies
\left\{
\begin{array}{rll}
\bg_m &= \check{\g}_m \otimes {\bf 1}_4, & \quad m=0,\cdots 4,
\\
\bg_m &= {\bf 1}_4 \otimes i\hat{\g}_m, & \quad m=5,\cdots 9,
\end{array}
\right.
\ee
The above formulae explain the reason for the factor of $i$ in the definition of $\bg_m$ for the sphere. 
In the same way we can explain why there is a $+$ sign and not $-$ in the definition of $\bg_{mn}$ for the sphere, computing\footnote{Considering even rank $\G$-matrices we need to also insert  an odd rank $\G$-matrix to have a non-vanishing result for $\Theta_{1,2}$ of the same chirality.}
\be
\bar{\Theta}_1 \G_{p} \G_{mn} \Theta_2 \equiv \bar{\theta}_1 \bg_{p} \bg_{mn} \theta_2 
\implies
\left\{
\begin{array}{rll}
\bg_{mn} &= \check{\g}_{mn} \otimes {\bf 1}_4, & \quad m,n=0,\cdots 4,
\\
\bg_{mn} &= {\bf 1}_4 \otimes \hat{\g}_{mn}, & \quad m,n=5,\cdots 9,
\\
\bg_{mn} &= -\check{\g}_{m} \otimes i\hat{\g}_{n}, & \quad m=0,\cdots 4, \ n=5,\cdots 9.
\end{array}
\right.
\ee
Similarly, for matrices of rank 3 we would obtain
\be
\bar{\Theta}_1 \G_{mnp} \Theta_2 \equiv \bar{\theta}_1 \bg_{mnp} \theta_2 
\implies
\left\{
\begin{array}{rll}
\bg_{mnp} &= \check{\g}_{mnp} \otimes {\bf 1}_4, &\quad m,n,p=0,\cdots 4,
\\
\bg_{mnp} &= {\bf 1}_4 \otimes i\hat{\g}_{mnp}, &\quad m,n,p=5,\cdots 9,
\\
\bg_{mnp} &= \frac{1}{3} \check{\g}_{mn} \otimes i\hat{\g}_{p},& \quad  m,n=0,\cdots 4, \ p=5,\cdots 9,
\\
\bg_{mnp} &= \frac{1}{3} \check{\g}_{p} \otimes \hat{\g}_{mn}, &\quad  p=0,\cdots 4, \ m,n=5,\cdots 9.
\end{array}
\right.
\ee

\subsection{Vielbein and spin connection for \ads and $({\rm AdS}_5\times {S}^5)_{\eta}$}\label{app:CD}
Here we list the components of the vielbein and spin connection for \ads. 
In our parameterisation (\ref{param}) the vielbein $e^m = e^m_M dX^M$ is diagonal and given by\footnote{To avoid confusion with tangent indices, we write curved indices with the explicit names of the target-space coordinates.}
\be
\begin{aligned}
e^0_t = \sqrt{1+\rho^2}, \quad & e^1_{\psi_2} = -\rho \sin \zeta ,  &e^2_{\psi_1} =-\rho \cos \zeta, \quad &e^3_{\zeta} = -\rho, \quad& e^4_{\rho} = -\frac{1}{\sqrt{1+\rho^2}}, \\
e^5_{\phi} = \sqrt{1-r^2}, \quad & e^6_{\phi_2} = -r \sin \xi ,  &e^7_{\phi_1} =-r \cos \xi, \quad &e^8_\xi = -r, \quad& e^9_r = -\frac{1}{\sqrt{1-r^2}}. 
\end{aligned}
\ee
The non-vanishing components of the spin connection $\omega^{mn} = \omega^{mn}_M {\rm d}X^M$ are
\be
\begin{aligned}
& \omega^{04}_t = \rho,  \quad\quad                      &\omega^{34}_\zeta =  -\sqrt{1+\rho^2},  &\quad\quad\quad                     \omega^{24}_{\psi_1} = -\sqrt{1+\rho^2} \cos \zeta\, , & \\
& \omega^{13}_{\psi_2} = -\cos \zeta, \quad\quad  &\omega^{14}_{\psi_2} = -\sqrt{1+\rho^2} \sin \zeta , &\quad\quad\quad \omega^{23}_{\psi_1} = \sin \zeta, \quad & \\
\\
& \omega^{59}_\phi = -r, \quad\quad                  &\omega^{89}_\xi =  -\sqrt{1-r^2}, \quad & \quad\quad  \omega^{79}_{\phi_1} = -\sqrt{1-r^2} \cos \xi \, ,& \\
& \omega^{68}_{\phi_2} = -\cos \xi, \quad\quad & \omega^{69}_{\phi_2} = -\sqrt{1-r^2} \sin \xi , \quad &\quad\quad \omega^{78}_{\phi_1} = \sin \xi, \quad & 
\end{aligned}
\ee
and it can be checked that $\omega_M^{mn}$ satisfies an equation
\bea
\omega_M^{mn}=-e^{N[m}\Big(\pa_M e_N^{n]}-\pa_Ne_M^{n]}+e^{n]P}e^p_M\pa_P e_{Np}\Big)\, ,
\eea
where anti-symmetrisation of indices $m$ and $n$ is performed with the weight $1/2$.

For the $({\rm AdS}_5\times {S}^5)_{\eta}$ background we will use the following diagonal vielbein
\be\label{eq:def-vielb-comp}
\begin{aligned}
&\widetilde{e}^0_t=\frac{\sqrt{1+\rho ^2}}{\sqrt{1-\vk ^2 \rho ^2}},
\quad\quad
\widetilde{e}^1_{\psi_2}=-\rho  \sin \zeta,
\quad\quad
\widetilde{e}^2_{\psi_1}=-\frac{\rho  \cos \zeta}{\sqrt{1+\vk ^2 \rho ^4 \sin ^2\zeta}},
\\
&\widetilde{e}^3_\zeta=-\frac{\rho }{\sqrt{1+\vk ^2   \rho ^4 \sin ^2\zeta}},
\quad\quad
\widetilde{e}^4_\rho=-\frac{1}{\sqrt{1+\rho ^2} \sqrt{1-\vk ^2 \rho ^2}},
\\
&\widetilde{e}^5_\phi=\frac{\sqrt{1-r^2}}{\sqrt{1+\vk ^2 r^2}},
\quad\quad
\widetilde{e}^6_{\phi_2}=-r \sin \xi ,
\quad
\widetilde{e}^7_{\phi_1}=-\frac{r \cos \xi }{\sqrt{1+\vk ^2   r^4 \sin ^2\xi}},
\\
&\widetilde{e}^8_\xi=-\frac{r}{\sqrt{1+\vk ^2 r^4 \sin ^2\xi}},
\quad
\widetilde{e}^9_r=-\frac{1}{\sqrt{1-r^2} \sqrt{1+\vk ^2 r^2}}.
\end{aligned}
\ee
Finally, the deformed spin connection compatible with the $\eta$-deformed metric is found from the equation
\be\la{eq:spin_con_deformed}
\widetilde{\omega}_M^{mn}=
- \widetilde{e}^{N \, [m} \left( \pa_M \widetilde{e}^{n]}_N - \pa_N \widetilde{e}^{n]}_M + \widetilde{e}^{n] \, P} \widetilde{e}_M^p \pa_P \widetilde{e}_{Np} \right),
\ee
where tangent indices $m,n$ are raised and lowered with the Minkowski metric $\eta_{mn}$, while curved indices $M,N$ with the deformed metric $\widetilde{G}_{MN}$.

\section{Derivation of the fermionic Lagrangian and $\kappa$-symmetry}

\subsection{Construction of the inverse of $\op$}\label{sec:inverse-op}
In this appendix we construct an operator $\op^{-1}$ to several orders in fermions.  
The perturbative expansion (\ref{eq:pertop}) starts from the operator $\op_{(0)}=1-\eta R_\gb \circ d$, {\it i.e.} we first need to know the action of 
$R_{\gb}$ on  the superalgebra $\psu(2,2|4)$.

\paragraph{Action of $\op$ on $\psu(2,2|4)$}\label{sec:useful-results-eta-def} The action of $R_{\gb}$on  the basis of generators of $\psu(2,2|4)$ is found to be 
\be\label{eq:Rgb-action-lambda}
\begin{aligned}
R_{\gb}(\gen{P}_m) &= {\lambda_m}^n \gen{P}_n + \frac{1}{2} \lambda_m^{np} \gen{J}_{np} , \\
R_{\gb}(\gen{J}_{mn}) &=  \lambda_{mn}^{p} \gen{P}_p +\frac{1}{2} \lambda_{mn}^{pq} \gen{J}_{pq}, \\
R_{\gb}(\gen{Q}^{I}) &= R (\gen{Q}^{I}) = -\epsilon^{IJ} \gen{Q}^{J} ,\\
\end{aligned}
\ee
where the coefficients ${\lambda_m}^n, \lambda_m^{np}, \lambda_{mn}^{p}, \lambda_{mn}^{pq}$  are

\bea\label{eq:lambda11}\begin{aligned}
\lambda_0^{\ 4} &= \lambda_4^{\ 0} = \rho, \qquad~~ \lambda_2^{\ 3} = - \lambda_3^{\ 2} = - \rho^2 \sin \zeta, \qquad \\
\lambda_5^{\ 9} &=- \lambda_9^{\ 5} = r, \qquad \lambda_7^{\ 8} = - \lambda_8^{\ 7} =  r^2 \sin \xi,
\end{aligned}\eea

\be\label{eq:lambda12}
\begin{aligned}
& \lambda_1^{01} =\lambda_2^{02} =\lambda_3^{03} =\lambda_4^{04} = \sqrt{1+\rho^2}, \qquad 
&~~~ \lambda_1^{12} =- \lambda_3^{23}= -\rho \cos \zeta, \qquad 
 \\
& \lambda_6^{56} =\lambda_7^{57} =\lambda_8^{58} =\lambda_9^{59} = -\sqrt{1-r^2}, \qquad 
& \lambda_6^{67} =- \lambda_8^{78}= r \cos \xi, \qquad 
 \\
& \lambda_2^{34} = - \lambda_3^{24}= -\rho \sqrt{1+\rho^2} \sin \zeta,
&~~~~~~ \lambda_7^{89} = - \lambda_8^{79}= r \sqrt{1-r^2} \sin \xi, \qquad\end{aligned}
\ee

\be\label{eq:lambda21}
\begin{aligned}
& \lambda_{01}^1 = \lambda_{02}^2 = \lambda_{03}^3 = \lambda_{04}^4 = -\sqrt{1+\rho^2}, \qquad
& \lambda_{12}^1 = - \lambda_{23}^3 = -\rho \cos \zeta, \qquad
\\
& \lambda_{56}^6 = \lambda_{57}^7 = \lambda_{58}^8 = \lambda_{59}^9 = \sqrt{1-r^2}, \qquad
& \lambda_{67}^6 = - \lambda_{78}^8 = -r \cos \xi, \qquad 
\\
& \lambda_{24}^3 = - \lambda_{34}^2 = \rho \sqrt{1+\rho^2} \sin \zeta, \qquad
& \lambda_{79}^8 = - \lambda_{89}^7 = r \sqrt{1-r^2} \sin \xi, \end{aligned}
\ee

\be\label{eq:lambda22a}
\begin{aligned}
& \lambda_{01}^{14} = \lambda_{02}^{24} = \lambda_{03}^{34} = \lambda^{01}_{14} = \lambda^{02}_{24} = \lambda^{03}_{34} = -\rho, \qquad
& \lambda_{12}^{13} = - \lambda_{13}^{12} = \sin \zeta, \qquad \\
& \lambda_{12}^{14} = -\lambda_{14}^{12}= -\lambda_{23}^{34}= \lambda_{34}^{23}= - \sqrt{1+ \rho^2} \cos \zeta, \qquad 
& \lambda_{24}^{34} = - \lambda_{34}^{24} = (1+\rho^2) \sin \zeta
\end{aligned}
\ee

\be\label{eq:lambda22s}
\begin{aligned}
& \lambda_{56}^{69} = \lambda_{57}^{79} = \lambda_{58}^{89} = -\lambda^{56}_{69} = -\lambda^{57}_{79} = -\lambda^{58}_{89} = -r, \qquad
& \lambda_{67}^{68} = - \lambda_{68}^{67} = \sin \xi, \qquad \\
& \lambda_{67}^{69} = -\lambda_{69}^{67}= -\lambda_{78}^{89}= \lambda_{89}^{78}= - \sqrt{1-r^2} \cos \xi, \qquad 
& \lambda_{79}^{89} = - \lambda_{89}^{79} = (1-r^2) \sin \xi
\end{aligned}
\ee
The $\lambda$-coefficients have the following properties
\be\label{eq:swap-lambda}
{\lambda_m}^n = - \, \eta_{mm'} \eta^{nn'} {\lambda_{n'}}^{m'},
\quad
\check{\lambda}_m^{np} = \eta_{mm'} \eta^{nn'} \eta^{pp'} \check{\lambda}^{m'}_{n'p'},
\quad
\hat{\lambda}_m^{np} = -\, \eta_{mm'} \eta^{nn'} \eta^{pp'} \hat{\lambda}^{m'}_{n'p'},
\ee
that will be used to simplify some terms in the Lagrangian.
For completeness we give also the coefficients $w_m^{np}$ defined in~\eqref{eq:action-Oinv0-P} corresponding to the action of $\opinv_{(0)}$
\be\label{eq:w12a}
\begin{aligned}
& w^{04}_0 =  \vk^2 \frac{\rho \sqrt{1+\rho^2}}{1-\vk^2 \rho^2},  \\
& w^{12}_1 = -\vk \rho \cos \zeta, \quad & w^{01}_1 = \vk \sqrt{1+\rho^2},  \\
& w^{02}_2 = \vk \frac{\sqrt{1+\rho^2}}{1+\vk^2 \rho^4 \sin^2 \zeta}, \quad & w^{03}_2 = -\vk^2  \frac{\rho^2 \sqrt{1+\rho^2} \sin \zeta }{1+\vk^2 \rho^4 \sin^2 \zeta}, \\
& w^{23}_2 = -\vk^2  \frac{\rho^3 \sin \zeta \cos \zeta}{1+\vk^2 \rho^4 \sin^2 \zeta}, \quad & w^{24}_2 = -\vk^2  \frac{\rho^3 \sqrt{1+\rho^2} \sin^2 \zeta }{1+\vk^2 \rho^4 \sin^2 \zeta},  \quad & w^{34}_2 = -\vk  \frac{\rho \sqrt{1+\rho^2} \sin \zeta }{1+\vk^2 \rho^4 \sin^2 \zeta}, \quad & \\
& w^{02}_3 = \vk^2  \frac{\rho^2 \sqrt{1+\rho^2} \sin \zeta }{1+\vk^2 \rho^4 \sin^2 \zeta}, \quad & w^{03}_3 = \vk \frac{\sqrt{1+\rho^2}}{1+\vk^2 \rho^4 \sin^2 \zeta}, \\
& w^{23}_3 = \vk  \frac{\rho  \cos \zeta}{1+\vk^2 \rho^4 \sin^2 \zeta}, \quad & w^{24}_3 = \vk  \frac{\rho \sqrt{1+\rho^2} \sin \zeta }{1+\vk^2 \rho^4 \sin^2 \zeta},  \quad & w^{34}_3 = -\vk^2  \frac{\rho^3 \sqrt{1+\rho^2} \sin^2 \zeta }{1+\vk^2 \rho^4 \sin^2 \zeta}, \quad & \\
& w^{04}_4 = \vk \frac{\sqrt{1+\rho^2}}{1-\vk^2 \rho^2}, \\
\end{aligned}
\ee
%%%%%%%%%%%%%%%%%%%
\be\label{eq:w12s}
\begin{aligned}
& w^{59}_5 = - \vk^2 \frac{r \sqrt{1-r^2}}{1+\vk^2 r^2},  \\
& w^{67}_6 = \vk r \cos \xi, \quad & w^{56}_6 = -\vk \sqrt{1-r^2},  \\
& w^{57}_7 = -\vk \frac{\sqrt{1-r^2}}{1+\vk^2 r^4 \sin^2 \xi}, \quad & w^{58}_7 = -\vk^2  \frac{r^2 \sqrt{1-r^2} \sin \xi }{1+\vk^2 r^4 \sin^2 \xi}, \\
& w^{78}_7 = -\vk^2  \frac{r^3 \sin \xi \cos \xi}{1+\vk^2 r^4 \sin^2 \xi}, \quad & w^{79}_7 = -\vk^2  \frac{r^3 \sqrt{1-r^2} \sin^2 \xi }{1+\vk^2 r^4 \sin^2 \xi},  \quad & w^{89}_7 = \vk  \frac{r \sqrt{1-r^2} \sin \xi }{1+\vk^2 r^4 \sin^2 \xi}, \quad & \\
& w^{57}_8 = \vk^2  \frac{r^2 \sqrt{1-r^2} \sin \xi }{1+\vk^2 r^4 \sin^2 \xi}, \quad & w^{58}_8 = -\vk \frac{\sqrt{1-r^2}}{1+\vk^2 r^4 \sin^2 \xi}, \\
& w^{78}_8 = -\vk  \frac{r  \cos \xi}{1+\vk^2 r^4 \sin^2 \xi}, \quad & w^{79}_8 = -\vk  \frac{r \sqrt{1-r^2} \sin \xi }{1+\vk^2 r^4 \sin^2 \xi},  \quad & w^{89}_8 = -\vk^2  \frac{r^3 \sqrt{1-r^2} \sin^2 \xi }{1+\vk^2 r^4 \sin^2 \xi}, \quad & \\
& w^{59}_9 = -\vk \frac{\sqrt{1-r^2}}{1+\vk^2 r^2}, \\
\end{aligned}
\ee
These formulae allow one to determine the action of $\op_{(0)}$ on $\psu(2,2|4)$.
The next two terms in the expansion (\ref{eq:pertop})  read explicitly as
\bea
\op_{(1)} (M) &=& \eta [ \chi,R_{\gb} \circ d (M)] -\eta R_{\gb} ([\chi , d (M)] ), \nonumber\\
\op_{(2)} (M) &=& \eta [\chi , R_{\gb} ([\chi,d(M)])] - \frac{1}{2} \eta  R_{\gb} ( [\chi,[\chi,d(M)]])- \frac{1}{2} \eta  ( [\chi,[\chi,R_{\gb} \circ d(M)]]) \nonumber\\
& = &\frac{1}{2} \eta \left(  [\chi , [\chi , R_{\gb} \circ d(M)]] -R_{\gb} [\chi, [\chi,d(M)]] \right) - [\chi, \op_{(1)}(M)],
\label{eq:expans-ferm-op}
\eea
where we use again the notation $\chi \equiv \genQind{I}{}{}\ferm{\theta}{I}{}{} $.
\paragraph{Perturbative inversion} Now we are ready to invert the operator $\op$. We will do it up to quadratic order in fermions. 

\subsection*{{\sl Order $\theta^0$}} Using the results above we find that on $\gen{P}_m$ the inverse of $\op_{(0)}$ acts as 
\be\label{eq:action-Oinv0-P}
\opinv_{(0)} (\gen{P}_{m})= {k_m}^n \gen{P}_n + \frac{1}{2} w_m^{np} \gen{J}_{np},
\ee
where we have
\be\label{eq:k-res1}
\begin{aligned}
& k_0^{\ 0} = k_4^{\ 4} = \frac{1}{1-\vk^2 \rho^2} , \qquad 
& k_1^{\ 1} = 1, \qquad 
& k_2^{\ 2} = k_3^{\ 3} = \frac{1}{1+\vk^2 \rho^4 \sin^2 \zeta}, \\
& k_5^{\ 5} =k_9^{\ 9}= \frac{1}{1+\vk^2 r^2}, \qquad 
& k_6^{\ 6} = 1, \qquad 
& k_7^{\ 7} = k_8^{\ 8}= \frac{1}{1+\vk^2 r^4 \sin^2 \xi},
\end{aligned}
\ee
%%%%%%%%%%%%%
\be\label{eq:k-res2}
\begin{aligned}
& k_0^{\ 4} = +k_4^{\ 0}= \frac{\vk \rho}{1-\vk^2 \rho^2}, \qquad 
& k_2^{\ 3}=-k_3^{\ 2}=- \frac{\vk \rho^2 \sin \zeta}{1+\vk^2 \rho^4 \sin^2 \zeta},  \\
& k_5^{\ 9} = - k_9^{\ 5}=\frac{\vk r}{1+\vk^2 r^2}, \qquad 
& k_7^{\ 8}= -k_8^{\ 7}=\frac{\vk r^2 \sin \xi}{1+\vk^2 r^4 \sin^2 \xi}.
\end{aligned}
\ee
The coefficients $w_m^{np}$ do not contribute to the Lagrangian because of the coset projection, and their expression is given by~\eqref{eq:w12a}-\eqref{eq:w12s}.

When acting on odd elements, the inverse operator rotates only the labels $I,J$ without modifying the spinor indices
\be
\opinv_{(0)} (\gen{Q}^{I})= \frac{1}{2} (1+\sqrt{1+\vk^2}) \, \gen{Q}^{I}- \frac{\vk}{2} {\sigma_1}^{IJ} \, \gen{Q}^{J}. 
\ee
\smallskip

\subsection*{\sl Order $\theta^1$}
We use the first formula of~\eqref{eq:expans-ferm-op} and~\eqref{eq:expans-ferm-inv-op} to compute the action of $\op_{(1)}$ and $\opinv_{(1)}$ on $\gen{P}_m$ and $\gen{Q}^I$.
First we find
\be
\op_{(1)}(\gen{P}_m) = 
\frac{\vk}{2} \gen{Q}^I \left[ 
 \delta^{IJ} \left(i \bg_m - \frac{1}{2} \lambda_m^{np} \bg_{np} \right) 
+ i \epsilon^{IJ} {\lambda_m}^n \bg_n  
\right] \theta_J\,,
\ee
and we use this result to get
\be
\begin{aligned}
\opinv_{(1)}(e^m\gen{P}_m) = 
-\frac{\vk}{4} \gen{Q}^I \ e^m {k_m}^n \ \Bigg[ 
& \left((1+\sqrt{1+\vk^2})\delta^{IJ} -\vk \sigma_1^{IJ}\right) \left(i \bg_n - \frac{1}{2} \lambda_n^{pq} \bg_{pq} \right) \\
 & + i \left((1+\sqrt{1+\vk^2}) \epsilon^{IJ} + \vk \sigma_3^{IJ}\right) {\lambda_n}^p \bg_p 
\Bigg] \theta_J\,.
\end{aligned}
\ee
For later convenience we rewrite this as
\be
\begin{aligned}
\opinv_{(1)}(e^m\gen{P}_m) = 
-\frac{\vk}{4} \gen{Q}^I \ e^m {k_m}^n \ \Bigg[ 
& \left((1+\sqrt{1+\vk^2})\delta^{IJ} -\vk \sigma_1^{IJ}\right) \Delta^1_n \\
 & + \left((1+\sqrt{1+\vk^2}) \epsilon^{IJ} + \vk \sigma_3^{IJ}\right) \Delta^3_n
\Bigg] \theta_J\,,
\end{aligned}
\ee
where $\Delta^1_n\equiv\left(i \bg_n- \frac{1}{2} \lambda_n^{pq} \bg_{pq} \right)$, $\Delta^3_n\equiv i{\lambda_n}^p \bg_p $.
On odd generators we find
\be
\begin{aligned}
\op_{(1)}(\gen{Q}^I\psi_I) =
\frac{-1+\sqrt{1+\vk^2}}{\vk} \ \bar{\theta}_J  \Bigg[ 
 -\sigma_1^{JI} \left( i\bg_p +\frac{1}{2} \lambda^{mn}_{ p}  \bg_{mn}   \right) 
 + i \, \sigma_3^{JI} {\lambda_p}^n \bg_n 
\Bigg] \psi_I \ \eta^{pq} \gen{P}_q  + \cdots \,
\end{aligned}
\ee
that helps to compute
\be
\begin{aligned}
\opinv_{(1)}(\gen{Q}^I\psi_I) =
- \frac{1}{2} \ \bar{\theta}_K  \Bigg[ &
 (-\vk \sigma_1^{KI} +(-1+\sqrt{1+\vk^2})\delta^{KI}) \left( i\bg_p +\frac{1}{2}  \lambda^{mn}_{ p} \bg_{mn}  \right) \\
& + i \, (\vk \sigma_3^{KI} -(-1+\sqrt{1+\vk^2})\epsilon^{KI})  {\lambda_p}^n \bg_n
\Bigg] \psi_I \ k^{pq} \ \gen{P}_q  + \cdots \,.
\end{aligned}
\ee
In these formulae we have omitted the terms proportional to $\gen{J}_{mn}$ and replaced them by dots, since they do not contribute to the Lagrangian.
It is interesting to note that the last result can be rewritten as
\be
\begin{aligned}
\opinv_{(1)}(\gen{Q}^I\psi_I) =
- \frac{1}{2} \ \bar{\theta}_K  \Bigg[ &
 (-\vk \sigma_1^{KI} +(-1+\sqrt{1+\vk^2})\delta^{KI}) \bar{\Delta}^{1}_{p} \\
& + (\vk \sigma_3^{KI} -(-1+\sqrt{1+\vk^2})\epsilon^{KI}) \bar{\Delta}^{3}_{p}
\Bigg] \psi_I \  k^{pq} \ \gen{P}_q  + \cdots
\end{aligned}
\ee
where one needs to use \eqref{eq:swap-lambda}.
The quantities $\bar{\Delta}^{3}_{p'},\bar{\Delta}^{1}_{p'}$ are defined by $(\Delta^{3}_{p'} \theta_K)^\dagger \check{\gamma}^0 = \bar{\theta}_K \bar{\Delta}^{3}_{p'}$ and $(\Delta^{1}_{p'} \theta_K)^\dagger \check{\gamma}^0 = \bar{\theta}_K \bar{\Delta}^{1}_{p'}$.

\subsection*{\sl Order $\theta^2$}
We need to compute the action of $\op$ and $\op^{-1}$ at order $\theta^2$ just on generators $\gen{P}_m$. Indeed the operators $\op_{(2)}$ and $\opinv_{(2)}$ acting on generators $\gen{Q}^I$ contribute only at quartic order in the Lagrangian.
First we find
\be
\begin{aligned}
\op_{(2)}({\gen{P}}_m) = 
- \frac{\vk}{2} \bar{\theta}_K \Bigg[ 
& \delta^{KI} \left(  -\bg_q \left(\bg_m +\frac{i}{4} \lambda_m^{np} \bg_{np} \right) 
+ \frac{i}{4} \lambda^{np}_{q} \bg_{np}  \bg_m \right) \\
& -\frac{1}{2} \epsilon^{KI} \left(  \bg_q \, {\lambda_m}^n \bg_n 
- {\lambda_q}^p \bg_p \bg_m  \right) 
\Bigg] \theta_I \ \eta^{qr} \gen{P}_r
+\cdots\,,
\end{aligned}
\ee
%%%%%%%%
that gives
\bea
%begin{aligned}
 -\opinv_{(0)} \circ \op_{(2)} \circ \opinv_{(0)}(e^m\gen{P}_m) = 
&-& \frac{\vk}{2} \bar{\theta}_K \ e^m {k_m}^n \ \Bigg[ 
 \delta^{KI} \left(  \bg_u \left(\bg_n +\frac{i}{4} \lambda_n^{pq} \bg_{pq} \right)
- \frac{i}{4} \lambda^{pq}_{\ u} \bg_{pq}  \bg_n \right) \nonumber \\
 & +&\frac{1}{2} \epsilon^{KI} \left(  \bg_u {\lambda_n}^p \bg_p 
- {\lambda_u}^p \bg_p \bg_n  \right) 
\Bigg] \theta_I {k}^{uv} \ \gen{P}_v 
+\cdots\,.
%\end{aligned}
\eea
%%%%%%%%%%
Also here the dots stand for contributions proportional to $\gen{J}_{mn}$ that we are omitting.
%%%%%%%%%%%%%%
The last formula that we will need is
\bea
%\begin{aligned}
&&  -\opinv_{(1)} \circ \op_{(1)} \circ \opinv_{(0)}(e^m\gen{P}_m) =  \\
&& - \frac{\vk}{4} \bar{\theta}_K \ e^m {k_m}^n \ \Bigg[ 
(-1+\sqrt{1+\vk^2}) \delta^{KJ} \bigg( \left( \bg_u -\frac{i}{2} \lambda_u^{pq}\bg_{pq}   \right) \left(\bg_n +\frac{i}{2} \lambda_n^{rs} \bg_{rs}\right) 
+ {\lambda_u}^p\bg_p   {\lambda_n}^r \bg_r \bigg) \nonumber \\
& &~~~~~+(-1+\sqrt{1+\vk^2}) \epsilon^{KJ} \bigg(-{\lambda_u}^p \bg_p \left(\bg_n +\frac{i}{2} \lambda_n^{rs} \bg_{rs}\right) 
+ \left( \bg_u -\frac{i}{2} \lambda_u^{pq}\bg_{pq}   \right) {\lambda_n}^r \bg_r \bigg) \nonumber\\
& & ~~~~~-\vk \sigma_1^{KJ} \bigg( \left( \bg_u -\frac{i}{2}\lambda_u^{pq} \bg_{pq}   \right) \left(\bg_n +\frac{i}{2} \lambda_n^{rs} \bg_{rs}\right) 
- {\lambda_u}^p\bg_p   {\lambda_n}^r \bg_r \bigg) \nonumber \\
& &~~~~~+\vk \sigma_3^{KJ} \bigg( {\lambda_u}^p\bg_p \left(\bg_n +\frac{i}{2} \lambda_n^{rs} \bg_{rs}\right) 
+ \left( \bg_u -\frac{i}{2} \lambda_u^{pq} \bg_{pq}  \right) {\lambda_n}^r \bg_r \bigg)
\Bigg] \theta_J {k}^{uv} \ \gen{P}_v 
+\cdots\,,  \nonumber 
%\end{aligned}
\eea
and it was obtained by using~\eqref{eq:swap-lambda}.

\subsection{Contribution $\lagr_{\{101\}}$ }\label{app:der-Lagr-101}
Here we show how to write $\lagr_{\{101\}}$ in the form~\eqref{eq:Lagr-101}.
It is easy to see that the insertion of $\opinv_{(0)}$ between two odd currents does not change the fact that the expression is anti-symmetric in $\a,\b$ and we have
\be\label{eq:orig-lagr-101}
\lagr_{\{101\}} = - \frac{\tilde{g}}{2} \epsilon^{\a\b} \left( -\sigma_1^{IK} + \frac{\vk}{1+\sqrt{1+\vk^2}} \delta^{IK} \right) (D^{IJ}_\a \theta_J)^\dagger \bg^0  D^{KL}_\b \theta_L .
\ee
The above contribution contains terms quadratic in $\pa \theta$, a feature that does no match the canonical form of the Lagrangian. 
These unwanted terms remain even in the limit of vanishing deformation.
They do not cause a problem however, since they are of the form $\epsilon^{\a\b} s^{IK} \pa_\a \bar{\theta}_I \pa_\b \theta_K$, where $s^{IK}$ is a symmetric tensor. Thus, although not vanishing, these terms  can be traded for a total derivative $\epsilon^{\a\b} s^{IK} \pa_\a \bar{\theta}_I \pa_\b \theta_K = \pa_\a(\epsilon^{\a\b} s^{IK}  \bar{\theta}_I \pa_\b \theta_K)$ and, therefore, they can be omitted.
Taking into account that
\be
(D^{IJ}_\a \theta_J)^\dagger \bg^0 = \delta^{IJ} \left(\pa_\a \bar{\theta}_J + \frac{1}{4} \bar{\theta}_J \omega^{mn}_\a \bg_{mn}  \right) + \frac{i}{2} \epsilon^{IJ} \bar{\theta}_J e^m_\a \bg_m ,
\ee
the contribution $\lagr_{\{101\}}$ can be rewritten as 
\be
\begin{aligned}
\lagr_{\{101\}} &= - \frac{\tilde{g}}{2}  \epsilon^{\a\b} \left( \sigma_1^{IK}- \frac{\vk}{1+\sqrt{1+\vk^2}} \delta^{IK} \right) \bar{\theta}_J D^{JI}_\a D^{KL}_\b \theta_L \\
&+\pa_\a \left(\frac{\tilde{g}}{2}  \epsilon^{\a\b} \left( \sigma_1^{IK}- \frac{\vk}{1+\sqrt{1+\vk^2}} \delta^{IK} \right) \bar{\theta}_J  D^{KL}_\b \theta_L \right).
\end{aligned}
\ee
The last term is the total derivative and we discard it. Although
na\"ively the result looks as a quadratic expression in $D^{IJ}$, this is not so, as we now demonstrate.
Let us split this expression into three terms
\be
\epsilon^{\a\b} s^{IK} \bar{\theta}_L D^{LK}_\a D^{IJ}_\b \theta_J= \text{WZ}_1 + \text{WZ}_2 + \text{WZ}_3
\ee
where a symmetric tensor $s^{IK}$ is kept unspecified. 
For each of these terms we then get
\be
\begin{aligned}
\text{WZ}_1 & \equiv \epsilon^{\a\b} s^{IK} \bar{\theta}_L \mathcal{D}^{LK}_\a \mathcal{D}^{IJ}_\b \theta_J  \\
& = -\frac{1}{4} \epsilon^{\a\b} s^{JL}  \bar{\theta}_L e^m_\a e^n_\b \bg_{m} \bg_{n}  \theta_J , \\
\text{WZ}_2 & \equiv \frac{i}{2} \epsilon^{\a\b} s^{IK} \bar{\theta}_L  \left( \epsilon^{IJ} \mathcal{D}^{LK}_\a (e^n_\b \bg_n  \theta_J) + \epsilon^{LK} e^m_\a \bg_m  \mathcal{D}^{IJ}_\b \theta_J \right) \\
& = + i \epsilon^{\a\b} s^{IK}  \epsilon^{JI} \bar{\theta}_J e^m_\a \bg_m  \mathcal{D}^{KL}_\b \theta_L, \\
\text{WZ}_3 & \equiv -\frac{1}{4} \epsilon^{\a\b} s^{IK}  \epsilon^{LK}  \epsilon^{IJ} e^m_\a e^n_\b \bar{\theta}_L \bg_m  \bg_n  \theta_J ,
\end{aligned}
\ee
where we used the fact that the covariant derivative $\mathcal{D}$ applied to the vielbein gives zero
\be
\epsilon^{\a\b} \mathcal{D}^{IJ}_\a (e^m_\b \bg_m  \theta ) = \epsilon^{\a\b} e^m_\b \bg_m  \mathcal{D}^{IJ}_\a  \theta .
\ee
The final result is
\bea
%\begin{aligned}
\lagr_{\{101\}} &=& - \frac{\tilde{g}}{2}  \epsilon^{\a\b} \bar{\theta}_L \, i \, e^m_\a \bg_m \left( \sigma_3^{LK}  D^{KJ}_\b \theta_J
-\frac{\vk}{1+\sqrt{1+\vk^2}} \ \epsilon^{LK}  \mathcal{D}^{KJ}_\b \theta_J \right) \\
\nonumber
&=& - \frac{\tilde{g}}{2}  \epsilon^{\a\b} \bar{\theta}_I \left( \sigma_3^{IJ}  
 -\frac{\vk}{1+\sqrt{1+\vk^2}} \ \epsilon^{IJ}  \right) \, i \, e^m_\a \bg_m \mathcal{D}_\b \theta_J  + \frac{\tilde{g}}{4}  \epsilon^{\a\b} \bar{\theta}_I  \sigma_1^{IJ}  e^m_\a \bg_m e^n_\b \bg_n  \theta_J .
%\end{aligned}
\eea

\subsection{Canonical Green-Schwarz form}\label{app:CGS}
Here, following the steps outlined in section \ref{sec:QFL}, we explain how to find the necessary field redefinitions that bring the original Lagrangian to the canonical form. 

We thus focus on the terms involving derivative couplings only.
For convenience we collect these terms here, and write separately the contributions with $\g^{\a\b}$ and $\epsilon^{\a\b}$
\bea
%\begin{aligned}
\nonumber
\lagr^{\g,\pa} &=& -\frac{\tilde{g}}{2}  \ \g^{\a\b} \bar{\theta}_I 
\Bigg[ 
 \frac{i}{2} (\sqrt{1+\vk^2} \delta^{IJ} - \vk \sigma_1^{IJ} ) \bg_n 
 -\frac{1}{4} (\vk \sigma_1^{IJ} -(-1+ \sqrt{1+ \vk^2}) \delta^{IJ} )  \lambda^{pq}_{n} \bg_{pq} 
\\
\label{eq:non-can-lagr-kin}
&&~~~~~~~~~ +\frac{i}{2} (\vk \sigma_3^{IJ} -(-1+ \sqrt{1+ \vk^2}) \epsilon^{IJ} ) {\lambda_{n}}^{p} \bg_{p} 
\Bigg] 
 ({k^n}_{m}+{k_{m}}^n) e^m_\a \pa_\b \theta_J,
%\end{aligned}
\eea
\bea
%\begin{aligned}
\lagr^{\epsilon,\pa} &=&  -\frac{\tilde{g}}{2}  \ \epsilon^{\a\b} \bar{\theta}_I 
\Bigg[ 
\Bigg(
-\frac{i}{2} (\sqrt{1+\vk^2} \delta^{IJ}  - \vk \sigma_1^{IJ} ) \bg_n 
  +\frac{1}{4} (\vk \sigma_1^{IJ} -(-1+ \sqrt{1+ \vk^2}) \delta^{IJ} )  \lambda^{pq}_{n} \bg_{pq} 
\nonumber \\
\label{eq:non-can-lagr-WZ}
&&\qquad\quad ~~~~~  -\frac{i}{2} (\vk \sigma_3^{IJ} -(-1+ \sqrt{1+ \vk^2}) \epsilon^{IJ} )  {\lambda_{n}}^{p} \bg_{p} 
\Bigg) ({k^n}_{m}-{k_{m}}^n) \\
&&\qquad\quad ~~~~~ +i  \left( \sigma_3^{IJ} - \frac{-1+ \sqrt{1+ \vk^2}}{\vk} \epsilon^{IJ} \right) \bg_m 
\Bigg]  e^m_\a \pa_\b \theta_J.
\nonumber
%\end{aligned}
\eea
To simplify this result, we first make the following redefinition of fermions 
\be\label{eq:red-fer-2x2-sp}
\theta_I \to \frac{\sqrt{1+\sqrt{1+\vk^2}}}{\sqrt{2}} \left(\delta^{IJ} + \frac{\vk}{1+\sqrt{1+\vk^2}} \sigma_1^{IJ} \right) \theta_J.
\ee
Then the kinetic part of the Lagrangian turns into
\bea
\nonumber
%\begin{aligned}
\lagr^{\g,\pa} \to &&\lagr^{\g,\pa} = \lagr^{\g,\pa}_{\1} + \lagr^{\g,\pa}_{\2},
\\
&&\qquad  \lagr^{\g,\pa}_{\1} = -\frac{\tilde{g}}{2}  \ \g^{\a\b} \bar{\theta}_I 
\Bigg[ 
 \frac{i}{2}  \delta^{IJ}  \bg_n 
+\frac{i}{2} \vk \sigma_3^{IJ}  {\lambda_{n}}^{p} \bg_{p} 
\Bigg] 
 ({k^n}_{m}+{k_{m}}^n) e^m_\a \pa_\b \theta_J,
 \nonumber
\\
&&\qquad \lagr^{\g,\pa}_{\2} = -\frac{\tilde{g}}{2}  \ \g^{\a\b} \bar{\theta}_I 
\Bigg[ 
 -\frac{1}{4} (\vk \sigma_1^{IJ} +(-1+ \sqrt{1+ \vk^2}) \delta^{IJ} )  \lambda^{pq}_{n} \bg_{pq} 
\\
\nonumber
&&\qquad\qquad\qquad\qquad -\frac{i}{2} (-1+ \sqrt{1+ \vk^2}) \epsilon^{IJ}  {\lambda_{n}}^{p} \bg_{p} 
\Bigg] 
 ({k^n}_{m}+{k_{m}}^n) e^m_\a \pa_\b \theta_J.
%\end{aligned}
\eea
\bea
%\begin{aligned}
\lagr^{\epsilon,\pa} \to && \lagr^{\epsilon,\pa} = \lagr^{\epsilon,\pa}_{\1} + \lagr^{\epsilon,\pa}_{\2},
\nonumber
\\
&& \lagr^{\epsilon,\pa}_{\1} =  -\frac{\tilde{g}}{2}  \ \epsilon^{\a\b} \bar{\theta}_I 
\Bigg[ 
-\Bigg(
\frac{i}{2}  \delta^{IJ}  \bg_n 
  +\frac{i}{2} \vk \sigma_3^{IJ}   {\lambda_{n}}^{p} \bg_{p} 
\Bigg) ({k^n}_{m}-{k_{m}}^n) 
 +i   \sigma_3^{IJ} \bg_m 
\Bigg]  e^m_\a \pa_\b \theta_J ,
\nonumber
\\
&&  \lagr^{\epsilon,\pa}_{\2} =  -\frac{\tilde{g}}{2}  \ \epsilon^{\a\b} \bar{\theta}_I 
\Bigg[ 
\Bigg(
  \frac{1}{4} (\vk \sigma_1^{IJ} +(-1+ \sqrt{1+ \vk^2}) \delta^{IJ} )   \lambda^{pq}_{n} \bg_{pq} 
\\
\nonumber
&&\qquad\qquad\qquad\qquad\qquad  +\frac{i}{2} (-1+ \sqrt{1+ \vk^2}) \epsilon^{IJ}   {\lambda_{n}}^{p} \bg_{p} 
\Bigg) ({k^n}_{m}-{k_{m}}^n) \\
\nonumber
&&\qquad\qquad\qquad\qquad\qquad  -i   \frac{-1+ \sqrt{1+ \vk^2}}{\vk} \epsilon^{IJ} \bg_m 
\Bigg]  e^m_\a \pa_\b \theta_J.
%\end{aligned}
\eea
Here we split each of the contributions into two parts according to their symmetry properties  (\ref{eq:sym_prop}).
Suppose now we perform a shift of bosons (\ref{eq:shift_bosons}).  
This shift will generate contribution to the \emph{fermionic} Lagrangian originating from the \emph{bosonic} one:
\bea
\lagr_{(0)} \to \lagr_{(0)} +\delta \lagr^{\g,m} +\delta \lagr^{\g,\pa}_{\2} +\delta \lagr^{\epsilon,m} +\delta \lagr^{\epsilon,\pa}_{\2}+\mathcal{O}(\theta^4)\, ,
\eea
where
{\small \be\label{eq:red-bos-lagr}
\begin{aligned}
\delta \lagr^{\g,m} &= + \tilde{g} \g^{\a\b} \left(  -  \pa_\a X^M \ \bar{\theta}_I \ \widetilde{G}_{MN} \left( \pa_\b f^N_{IJ} \right) \  \theta_J 
 - \frac{1}{2}  \partial_\alpha X^M \partial_\beta X^N \pa_P \widetilde{G}_{MN} \ \bar{\theta}_I \, f^P_{IJ} \theta_J  \right),\\
\delta \lagr^{\g,\pa}_{\2} &= + \tilde{g} \g^{\a\b} \left( - 2  \pa_\a X^M \ \bar{\theta}_I \ \widetilde{G}_{MN} f^N_{IJ} \ \pa_\b \theta_J  \right),\\
\delta \lagr^{\epsilon,m} &=+ \tilde{g} \epsilon^{\a\b}   \left( +  \pa_\a X^M \ \bar{\theta}_I \ \widetilde{B}_{MN} \left( \pa_\b f^N_{IJ} \right) \  \theta_J 
 + \frac{1}{2}   \partial_\alpha X^M \partial_\beta X^N \pa_P \widetilde{B}_{MN} \ \bar{\theta}_I \, f^P_{IJ} \theta_J \right), \\
\delta \lagr^{\epsilon,\pa}_{\2} &=+ \tilde{g} \epsilon^{\a\b}   \left( 2  \pa_\a X^M \ \bar{\theta}_I \ \widetilde{B}_{MN} f^N_{IJ} \ \pa_\b \theta_J \right).
\end{aligned}
\ee
}

\normalsize
\noindent
Here we have used $\pa \bar{\theta}_I \ f^M_{IJ} (X) \ \theta_J = + \bar{\theta}_I \ f^M_{IJ} (X) \ \pa \theta_J$, consequence of the symmetry properties of $f^M_{IJ} (X)$, and we cut the expansion at quadratic order in fermions.

Now one can see that picking up the coefficients $f^M_{IJ}(X)$ as 
\bea\label{eq:def-shift-bos-f}
%\begin{aligned}
f^M_{IJ}(X) &=&  e^{Mp} \Bigg[ \frac{1}{8}  \left( \vk \sigma_1^{IJ} + (-1+\sqrt{1+\vk^2}) \delta^{IJ} \right)   \lambda_{p}^{mn}   \bg_{mn}  \\
&&\qquad\qquad\qquad\qquad\qquad\qquad+\frac{1}{4} (-1+\sqrt{1+\vk^2}) \epsilon^{IJ}   {\lambda_p}^n  i\bg_n \Bigg], \nonumber
%\end{aligned}
\eea
we are able to completely remove the contribution $\lagr^{\g,\pa}_{\2}$ from the Lagrangian
\be
\lagr^{\g,\pa}_{\2} + \delta \lagr^{\g,\pa}_{\2} =0.
\ee
On the other hand, this shift of the bosonic coordinates is not able to completely remove  $\lagr^{\epsilon,\pa}_2$: the terms with\footnote{This statement is only true if one adds to the $B$-field entering the bosonic Lagrangian an exact form with components 
$B_{t\rho}=\tilde{g}\frac{\varkappa}{2}\frac{\rho}{1-\varkappa^2\rho^2}$, $B_{\phi r}=\tilde{g}\frac{\varkappa}{2}\frac{r}{1+\varkappa^2 r^2}$. Clearly, these will also generate new terms with no derivatives on fermions  in $\delta \lagr^{\eps, m}$ of~\eqref{eq:red-bos-lagr}. If these components are not included, cancellation of terms with $\delta^{IJ}, \sigma_1^{IJ}$ is not complete, but what is left over may be rewritten up to a total derivative as a term with no derivatives on fermions. These two ways of treating the problem are equivalent and eventually lead to the same result.} 
$\delta^{IJ}, \sigma_1^{IJ}$ are cancelled out, but the ones with $\epsilon^{IJ}$ are left over. However, in the Wess-Zumino like term we can perform integration by parts\footnote{Integration by parts of terms containing  $\g^{\a\b}$ would generate unwanted derivatives of the world-sheet metric and also second derivatives of $X^M$.} to rewrite the result such that derivatives will act only on the bosons
\be\label{eq:WZ-shift-bos-eps}
\begin{aligned}
\lagr^{\epsilon,\pa}_{\2} + \delta \lagr^{\epsilon,\delta}_{\2} = &
\frac{\tilde{g}}{2}  \epsilon^{\a\b} \bar{\theta}_I \frac{-1+\sqrt{1+\vk^2}}{\vk} \epsilon^{IJ}  \\
& e^m_\a  \left(
i  \delta^q_m - \frac{i}{2} \vk (k^n_{\ m} - {k_{m}}^{n} ) \lambda_n^{\ q}  + \frac{i}{2} \vk B_{mn} (k^{pn} + k^{np} ) \lambda_p^{\ q}
\right) \bg_q \pa_\b \theta_J \\
= &
\frac{\tilde{g}}{2}  \epsilon^{\a\b} \bar{\theta}_I \frac{-1+\sqrt{1+\vk^2}}{\vk} \epsilon^{IJ}   e^m_\a  i  \bg_m \pa_\b \theta_J \\
= &
-\frac{\tilde{g}}{4}  \epsilon^{\a\b} \bar{\theta}_I \frac{-1+\sqrt{1+\vk^2}}{\vk} \epsilon^{IJ} 
  \pa_\a X^M  \left(\pa_\b  e^m_M   \right)  i\bg_m \theta_J 
 + \text{tot. der.}
\end{aligned}
\ee
Here we also used an important identity
\be
k^p_{\ m} - {k_{m}}^{p}  - B_{mn} (k^{pn} + k^{np} )  =0\, .
\ee
After the shift of the bosonic coordinates, the only terms containing derivatives on fermions are $\lagr^{\g,\pa}_{\1}$ and $\lagr^{\epsilon,\pa}_{\1}$.
The shift will also introduce new couplings without derivatives on fermions, as indicated in~\eqref{eq:red-bos-lagr}.
After we collect everything together,  the total fermionic Lagrangian $\lagr_{(2)}\equiv \lagr^\g+\lagr^\eps$ becomes
%\begin{eqnarray}

\begin{align}
%\allowdisplaybreaks[4]
\lagr^{\g} = & \frac{\tilde{g}}{2}  \ \g^{\a\b} \bar{\theta}_I 
\Bigg[ 
- \frac{i}{2}  \delta^{IJ}  \bg_n 
 -\frac{i}{2} \vk \sigma_3^{IJ}  {\lambda_{n}}^{p} \bg_{p} 
\Bigg] 
 ({k^n}_{m}+{k_{m}}^n) e^m_\a \pa_\b \theta_J  \notag \\
& - \tilde{g} \g^{\a\b} \left(  -  \pa_\a X^M \ \bar{\theta}_I \ \widetilde{G}_{MN} \left( \pa_\b f^N_{IJ} \right) \  \theta_J 
 - \frac{1}{2}  \partial_\alpha X^M \partial_\beta X^N \pa_P \widetilde{G}_{MN} \ \bar{\theta}_I \, f^P_{IJ} \theta_J \right)  \notag \\
&+ \frac{\tilde{g}}{4}  \gamma^{\a\b} (k^p_{\ q}  +{k_{q}}^{p} )e^q_{\a} \ \bar{\theta}_I 
\Bigg[ \frac{i}{4} \delta^{IJ} \bg_p \omega^{rs}_\b\bg_{rs}   \notag \\
&\qquad  +\frac{1}{8} \left( -\vk \sigma_1^{IJ} -(-1+\sqrt{1+\vk^2} ) \delta^{IJ} \right)  \lambda_{p}^{mn}  \bg_{mn} \  \omega^{rs}_\b \bg_{rs}  \notag \\
&\qquad - \frac{1}{2}\left( (-1-2\vk^2+\sqrt{1+\vk^2})\delta^{IJ}-\vk(-1+2\sqrt{1+\vk^2}) \sigma_1^{IJ} \right) \ {\lambda_p}^n\bg_n   e^r_\b \bg_r \  \notag \\
&\qquad +\frac{i}{4} (\vk \sigma_3^{IJ} - (-1+\sqrt{1+\vk^2})\epsilon^{IJ}) \ {\lambda_p}^n \bg_n \left( \omega^{rs}_\b\bg_{rs}\right)   \notag \\
&\qquad +\frac{1}{2} (\vk \sigma_3^{IJ}+\sqrt{1+\vk^2}\epsilon^{IJ}) \bg_p   e^r_\b \bg_r   \label{eq:lagr-gamma-no-F-red} \\
&\qquad -\frac{i}{4} \left( \vk \sigma_3^{IJ} + (-1+ \sqrt{1+\vk^2})\epsilon^{IJ} \right)  \lambda_{p}^{mn}\bg_{mn} e^r_\b \bg_r
 \Bigg]  \theta_J  \notag \\
&  + \frac{\tilde{g}}{8}  \gamma^{\a\b} \vk e^v_\a e^m_\b \, {k^{u}}_v {k_m}^n \, \bar{\theta}_I\times  \notag  \\
&~~ \times \Bigg[  
2 (\sqrt{1+\vk^2}\delta^{IJ}+\vk \sigma_1^{IJ}) \left( \bg_u\left(\bg_n +\frac{i}{4} \lambda_n^{pq} \bg_{pq} \right)
- \frac{i}{4}  \lambda^{pq}_{u}\bg_{pq} \bg_n\right)  \notag \\
&\qquad  + \epsilon^{IJ} \left(  \bg_u {\lambda_n}^p \bg_p 
-   {\lambda_u}^p\bg_p \bg_n \right)   \notag \\
&\qquad+\left(-(-1+\sqrt{1+\vk^2}) \delta^{IJ} -\vk \sigma_1^{KI}\right) \left( \bg_u -\frac{i}{2} \bg_{pq}  \lambda_u^{pq} \right) \left(\bg_n +\frac{i}{2} \lambda_n^{rs} \bg_{rs}\right)  \notag \\
&\qquad + \left((1+2\vk^2-\sqrt{1+\vk^2}) \delta^{IJ} -\vk(1-2\sqrt{1+\vk^2}) \sigma_1^{IJ}\right) {\lambda_u}^p\bg_p  {\lambda_n}^r \bg_r \bigg)  \notag \\
&\qquad +\left( \vk \sigma_3^{IJ}-(-1+\sqrt{1+\vk^2}) \epsilon^{IJ} \right)  {\lambda_u}^p \bg_p\left(\bg_n +\frac{i}{2} \lambda_n^{rs} \bg_{rs}\right) \notag \\
&\qquad + \left( \vk \sigma_3^{IJ}+ (-1+\sqrt{1+\vk^2}) \epsilon^{IJ} \right)\left( \bg_u -\frac{i}{2} \bg_{pq}  \lambda_u^{pq} \right) {\lambda_n}^r \bg_r  
\Bigg] \theta_J \notag
\end{align}
%\end{eqnarray}
and
\begin{align}
\lagr^{\epsilon} = 
& - \frac{\tilde{g}}{2}  \epsilon^{\a\b} \bar{\theta}_I  \sigma_3^{IJ}  
   \, i \, e^m_\a \bg_m \pa_\b \theta_J  \notag \\
& - \frac{\tilde{g}}{2}  \ \epsilon^{\a\b} \bar{\theta}_I 
\Bigg[ 
- \frac{i}{2}  \delta^{IJ}  \bg_n 
 -\frac{i}{2} \vk \sigma_3^{IJ}  {\lambda_{n}}^{p} \bg_{p} 
\Bigg] 
 ({k^n}_{m}-{k_{m}}^n) e^m_\a \pa_\b \theta_J \notag \\
& - \tilde{g} \epsilon^{\a\b}   \left( +  \pa_\a X^M \ \bar{\theta}_I \ \widetilde{B}_{MN} \left( \pa_\b f^N_{IJ} \right) \  \theta_J 
 + \frac{1}{2}   \partial_\alpha X^M \partial_\beta X^N \pa_P \widetilde{B}_{MN} \ \bar{\theta}_I \, f^P_{IJ} \theta_J  \right) \notag\\
&-\frac{\tilde{g}}{4}  \epsilon^{\a\b} \bar{\theta}_I \frac{-1+\sqrt{1+\vk^2}}{\vk} \epsilon^{IJ} 
  \pa_\a X^M  \left(\pa_\b  e^m_M   \right)  i\bg_m \theta_J \notag \\
& - \frac{\tilde{g}}{8}  \epsilon^{\a\b} \bar{\theta}_I \left(- \sigma_3^{IJ}  
 +\frac{\vk}{1+\sqrt{1+\vk^2}} \ \epsilon^{IJ}  \right) \, i \, e^m_\a \bg_m \omega^{np}_\b \bg_{np} \theta_J  \notag\\
&+ \frac{\tilde{g}}{4}  \epsilon^{\a\b} \bar{\theta}_I  \left( \vk \delta^{IJ} +\sqrt{1+\vk^2} \sigma_1^{IJ} \right)  e^m_\a \bg_m e^n_\b \bg_n  \theta_J\notag \\
&- \frac{\tilde{g}}{4}  \epsilon^{\a\b} (k^p_{\ q} - {k_{q}}^{p} )e^q_{\a} \ \bar{\theta}_I 
\Bigg[ \frac{i}{4} \delta^{IJ} \bg_p \omega^{rs}_\b\bg_{rs}  \notag\\
&  +\frac{1}{8} \left( -\vk \sigma_1^{IJ} -(-1+\sqrt{1+\vk^2} ) \delta^{IJ} \right)  \lambda_{p}^{mn}  \bg_{mn} \  \omega^{rs}_\b \bg_{rs} \notag\\
& - \frac{1}{2}\left( (-1-2\vk^2+\sqrt{1+\vk^2})\delta^{IJ}-\vk(-1+2\sqrt{1+\vk^2}) \sigma_1^{IJ} \right) \ {\lambda_p}^n\bg_n   e^r_\b \bg_r \ \notag\\
& +\frac{i}{4} (\vk \sigma_3^{IJ} - (-1+\sqrt{1+\vk^2})\epsilon^{IJ}) \ {\lambda_p}^n \bg_n \left( \omega^{rs}_\b\bg_{rs}\right)  \notag\\
& +\frac{1}{2} (\vk \sigma_3^{IJ}+\sqrt{1+\vk^2}\epsilon^{IJ}) \bg_p   e^r_\b \bg_r  \label{eq:lagr-epsilon-no-F-red} \\
& -\frac{i}{4} \left( \vk \sigma_3^{IJ} + (-1+ \sqrt{1+\vk^2})\epsilon^{IJ} \right)  \lambda_{p}^{mn}\bg_{mn} e^r_\b \bg_r
 \Bigg]  \theta_J \notag\\
& - \frac{\tilde{g}}{8}  \epsilon^{\a\b} \vk e^v_\a e^m_\b \, {k^{u}}_v {k_m}^n \, \bar{\theta}_I\times   \notag \\
&\quad \times \Bigg[  
 2 (\sqrt{1+\vk^2}\delta^{IJ}+\vk \sigma_1^{IJ}) \left( \bg_u\left(\bg_n +\frac{i}{4} \lambda_n^{pq} \bg_{pq} \right)
- \frac{i}{4}  \lambda^{pq}_{u}\bg_{pq} \bg_n\right) \notag\\
&\qquad  + \epsilon^{IJ} \left(  \bg_u {\lambda_n}^p \bg_p 
-   {\lambda_u}^p\bg_p \bg_n \right)  \notag\\
&\qquad+\left(-(-1+\sqrt{1+\vk^2}) \delta^{IJ} -\vk \sigma_1^{KI}\right) \left( \bg_u -\frac{i}{2} \bg_{pq}  \lambda_u^{pq} \right) \left(\bg_n +\frac{i}{2} \lambda_n^{rs} \bg_{rs}\right) \notag\\
&\qquad + \left((1+2\vk^2-\sqrt{1+\vk^2}) \delta^{IJ} -\vk(1-2\sqrt{1+\vk^2}) \sigma_1^{IJ}\right) {\lambda_u}^p\bg_p  {\lambda_n}^r \bg_r \bigg) \notag\\
&\qquad +\left( \vk \sigma_3^{IJ}-(-1+\sqrt{1+\vk^2}) \epsilon^{IJ} \right)  {\lambda_u}^p \bg_p\left(\bg_n +\frac{i}{2} \lambda_n^{rs} \bg_{rs}\right) \notag\\
&\qquad + \left( \vk \sigma_3^{IJ}+ (-1+\sqrt{1+\vk^2}) \epsilon^{IJ} \right)\left( \bg_u -\frac{i}{2} \bg_{pq}  \lambda_u^{pq} \right) {\lambda_n}^r \bg_r  
\Bigg] \theta_J\, , \notag \end{align}
%\ee
where the function $f^M_{IJ}(X)$ is defined in~\eqref{eq:def-shift-bos-f}.

\medskip

In order to put the present Lagrangian (\ref{eq:lagr-gamma-no-F-red}), (\ref{eq:lagr-epsilon-no-F-red}) in the canonical form we redefine fermions as $\theta_I \to U_{IJ}\theta_J$, where the matrix $U$ acts both on the $2$-dimensional space corresponding to the labels $I,J$ and on the 
16-dimensional spinor representation space.
We will search for $U$ in the factorised form where we attribute the corresponding factors to the AdS and sphere, respectively,
\be\label{eq:red-ferm-Lor-as}
\begin{aligned}
\theta_I &\to( U_{IJ}^{\alg{a}}\otimes  U_{IJ}^{\alg{s}})\theta_J\,,
\\
\theta_{I\ul{\a}\ul{a}} &\to( U_{IJ}^{\alg{a}})_{\ul{\a}}^{\ \ul{\nu}}  (U_{IJ}^{\alg{s}})_{\ul{a}}^{\ \ul{b}}\theta_{J\ul{\nu}\ul{b}}\,.
\end{aligned}
\ee
This is not the most general redefinition, but it will serve a purpose.
Each of the matrices $U_{IJ}^{\alg{a}}$ and $U_{IJ}^{\alg{s}}$ may be expanded over independent tensors spanning the space of all $2\times 2$ matrices
\be
U_{IJ}^{\alg{a},\alg{s}}=\delta_{IJ}\, U_{\delta}^{\alg{a},\alg{s}}+\sigma_{1\, IJ}\, U_{\sigma_1}^{\alg{a},\alg{s}}+\epsilon_{IJ}\, U_{\epsilon}^{\alg{a},\alg{s}}+\sigma_{3\, IJ}\, U_{\sigma_3}^{\alg{a},\alg{s}}\,.
\ee
The objects $U_{\mu}^{\alg{a},\alg{s}}$ with $\mu=\delta,\sigma_1,\epsilon,\sigma_3$ are then $4\times 4$ matrices that may be written in the convenient basis of $4\times 4$ gamma matrices.
From the Majorana condition~\eqref{eqMajorana-cond-compact-not} we find that in order to preserve $\theta_I^\dagger \bg^0=+ \theta_I^t \, (K\otimes K)$ under the field redefinition, we have to require that
\be
\bg^0\, \Big((U_{\mu}^{\alg{a}})^\dagger \otimes (U_{\mu}^{\alg{s}})^\dagger\Big) \bg^0= -(K\otimes K) \Big((U_{\mu}^{\alg{a}})^t \otimes (U_{\mu}^{\alg{s}})^t\Big) (K\otimes K)\,.
\ee
We choose to impose the following individual conditions   $\check{\g}^0\, (U_{\mu}^{\alg{a}})^\dagger  \check{\g}^0= K (U_{\mu}^{\alg{a}})^t K$ and $(U_{\mu}^{\alg{s}})^\dagger=- K (U_{\mu}^{\alg{s}})^t K$ which are then 
solved by
\be
\la{Uas}
\begin{aligned}
& U^{\alg{a}}_\mu = f^{\alg{a}}_{\mu} \mathbf{1} + i f^p_\mu \check{\g}_p  + \frac{1}{2} f^{pq}_\mu \check{\g}_{pq} ,
\qquad
U^{\alg{s}}_\mu = f^{\alg{s}}_{\mu} \mathbf{1} - f^p_\mu  \hat{\g}_p - \frac{1}{2} f^{pq}_\mu \hat{\g}_{pq} .
\end{aligned}
\ee
The factors of $i$ are chosen here in such a way that 
the coefficients $f$ are \emph{real} functions of bosonic coordinates, in accord with  \eqref{eq:symm-prop-5dim-gamma} and~\eqref{eq:herm-conj-prop-5dim-gamma}.
For the Dirac conjugate matrices we then find
\be
\begin{aligned}
& \bar{U}^{\alg{a}}_\mu = f^{\alg{a}}_{\mu} \mathbf{1} + i f^p_\mu \check{\g}_p  - \frac{1}{2} f^{pq}_\mu \check{\g}_{pq}  ,
\qquad
\bar{U}^{\alg{s}}_\mu = f^{\alg{s}}_{\mu} \mathbf{1} - f^p_\mu  \hat{\g}_p + \frac{1}{2} f^{pq}_\mu \hat{\g}_{pq}  .
\end{aligned}
\ee
Here the coefficients $f$ are the same as in  eq.(\ref{Uas}).

A transformation we are looking for must bring the kinetic part of the Lagrangian to the canonical form, that is 
\be
\begin{aligned}
\lagr^{\g,\pa}_{\1} \to &
 -\frac{\tilde{g}}{2} \g^{\a\b} \, i  \, \bar{\theta}_I \, \delta^{IJ} \, \widetilde{e}^m_\a \bg_m \pa_\b \theta_J, \\
\lagr^{\epsilon,\pa}_{\1} \to &
 -\frac{\tilde{g}}{2} \epsilon^{\a\b} \, i  \, \bar{\theta}_I \, \sigma^{IJ}_3 \, \widetilde{e}^m_\a \bg_m \pa_\b \theta_J ,
\end{aligned}
\ee
where $\widetilde{e}^m_\a$ is the deformed vielbein given in~\eqref{eq:def-vielb-comp}.
The matrices  $U^{\alg{a}}_\mu$ and $U^{\alg{s}}_\mu$ which do this job are constructed as follows. For ${U}_{\mu}^{\alg{a}}$
we put all the coefficients $f$ to zero, except those which are chosen to be
\be
\begin{aligned}
f^{\alg{a}}_{\delta} &=
\frac{1}{2} \sqrt{\frac{\left(1+\sqrt{1-\vk ^2 \rho ^2}\right) \left(1+\sqrt{1+\vk ^2 \rho ^4 \sin ^2\zeta }\right)}{\sqrt{1-\vk ^2 \rho ^2} \sqrt{1+\vk ^2 \rho ^4 \sin
   ^2\zeta }}}, \\
f^1_\delta &=
-\frac{\vk ^2 \rho ^3 \sin \zeta }{f^{\alg{a}}_{\text{den}}}, \\
f^{04}_{\sigma_3} &=
\frac{\vk  \rho  \left(1+\sqrt{1+\vk ^2 \rho ^4 \sin ^2\zeta }\right)}{f^{\alg{a}}_{\text{den}}}, \\
f^{23}_{\sigma_3} &=
\frac{\vk  \rho ^2 \sin \zeta  \left(1+\sqrt{1-\vk ^2 \rho ^2}\right)}{f^{\alg{a}}_{\text{den}}}, \\
f^{\alg{a}}_{\text{den}} &\equiv 2 (1-\vk ^2 \rho ^2)^{\frac{1}{4}} (1+\vk ^2 \rho ^4 \sin ^2\zeta )^{\frac{1}{4}} \sqrt{1+\sqrt{1-\vk ^2 \rho ^2}}  \sqrt{1+\sqrt{1+\vk ^2 \rho ^4
   \sin ^2\zeta }}\, .
\end{aligned}
\ee
Analogously, for ${U}_{\mu}^{\alg{s}}$ all the coefficients vanish except the following ones
\be
\begin{aligned}
f^{\alg{s}}_{\delta} &=
\frac{1}{2} \sqrt{\frac{\left(1+\sqrt{1+\vk ^2 r ^2}\right) \left(1+\sqrt{1+\vk ^2 r ^4 \sin ^2\xi }\right)}{\sqrt{1+\vk ^2 r ^2} \sqrt{1+\vk ^2 r ^4 \sin
   ^2\xi }}}, \\
f^6_{\delta} &=
\frac{\vk ^2 r ^3 \sin \xi }{f^{\alg{s}}_{\text{den}}}, \\
f^{59}_{\sigma_3} &=
\frac{\vk  r  \left(1+\sqrt{1+\vk ^2 r ^4 \sin ^2\xi }\right)}{f^{\alg{s}}_{\text{den}}}, \\
f^{78}_{\sigma_3} &=
\frac{\vk  r ^2 \sin \xi  \left(1+\sqrt{1+\vk ^2 r ^2}\right)}{f^{\alg{s}}_{\text{den}}}, \\
f^{\alg{s}}_{\text{den}} &\equiv 2 (1+\vk ^2 r ^2)^{\frac{1}{4}} (1+\vk ^2 r ^4 \sin ^2\xi )^{\frac{1}{4}} \sqrt{1+\sqrt{1+\vk ^2 r ^2}}  \sqrt{1+\sqrt{1+\vk ^2 r ^4
   \sin ^2\xi }} \, .
\end{aligned}
\ee
Since the corresponding transformation involves only the tensors $\delta$ and $\sigma_3$, it acts diagonally in the 2-dimensional space, {\it i.e.} separately for each of the two Majorana-Weyl fermions.
Define
\be
U_{(1)} \equiv U_\delta+U_{\sigma_3}\,,
\qquad
U_{(2)} \equiv U_\delta-U_{\sigma_3}\,,
\implies
\theta_I\to U_{(I)} \theta_I\quad I=1,2.
\ee
These matrices satisfy 
\be\label{eq:transf-rule-gamma-ferm-rot}
\begin{aligned}
& \bar{U}_{(I)} U_{(I)}  = \gen{1},
\qquad
&& \bar{U}_{(I)} \bg_m U_{(I)} =  (\Lambda_{(I)})_m^{\ n} \bg_n ,
\\
&U_{(I)} \bar{U}_{(I)} = \gen{1},
\qquad
 &&\bar{U}_{(I)} \bg_{mn} U_{(I)}  =  (\Lambda_{(I)})_m^{\ p}  (\Lambda_{(I)})_n^{\ q} \bg_{pq} ,
\end{aligned}
\ee
where there is no summation  over $I$.
In fact, the emerging $10\times 10$ matrices  $\Lambda_{(I)}$ have very simple matrix elements

{\footnotesize
\be\label{eq:Lambda-res1}
\begin{aligned}
& (\Lambda_{(I)})_0^{\ 0} = (\Lambda_{(I)})_4^{\ 4} = \frac{1}{\sqrt{1-\vk^2 \rho^2}} , 
&& (\Lambda_{(I)})_1^{\ 1} = 1, 
&&& (\Lambda_{(I)})_2^{\ 2} = (\Lambda_{(I)})_3^{\ 3} = \frac{1}{\sqrt{1+\vk^2 \rho^4 \sin^2 \zeta}}, \\
& (\Lambda_{(I)})_5^{\ 5} =(\Lambda_{(I)})_9^{\ 9}= \frac{1}{\sqrt{1+\vk^2 r^2}}, 
&& (\Lambda_{(I)})_6^{\ 6} = 1, 
&&& (\Lambda_{(I)})_7^{\ 7} =(\Lambda_{(I)})_8^{\ 8}= \frac{1}{\sqrt{1+\vk^2 r^4 \sin^2 \xi}},
\end{aligned}
\ee
\be\label{eq:Lambda-res2}
\begin{aligned}
& (\Lambda_{(I)})_0^{\ 4} = +(\Lambda_{(I)})_4^{\ 0}= (-1)^I\, \frac{\vk \rho}{\sqrt{1-\vk^2 \rho^2}}, \quad 
&& (\Lambda_{(I)})_2^{\ 3}=-(\Lambda_{(I)})_3^{\ 2}=-(-1)^I\, \frac{\vk \rho^2 \sin \zeta}{\sqrt{1+\vk^2 \rho^4 \sin^2 \zeta}},  \\
& (\Lambda_{(I)})_5^{\ 9} = - (\Lambda_{(I)})_9^{\ 5}=(-1)^I\, \frac{\vk r}{\sqrt{1+\vk^2 r^2}}, \quad 
&& (\Lambda_{(I)})_7^{\ 8}= -(\Lambda_{(I)})_8^{\ 7}=(-1)^I\, \frac{\vk r^2 \sin \xi}{\sqrt{1+\vk^2 r^4 \sin^2 \xi}}\, .
\end{aligned}
\ee
}

\normalsize
\noindent
Remarkably, these matrices are nothing else but the matrices of 10-dimensional Lorentz transformations
\be
(\Lambda_{(I)})_m^{\ p}\ (\Lambda_{(I)})_n^{\ q}\  \eta_{pq}=\eta_{mn}\,, \qquad I=1,2\,.
\ee

To implement the redefinition of fermions~\eqref{eq:red-ferm-Lor-as} in the Lagrangian, we find it efficient to use~\eqref{eq:transf-rule-gamma-ferm-rot}. We have, for instance,
\be
\bar{\theta}_K b^m \g_m \theta_I 
\to
\bar{\theta}_K \bar{U}_{(K)} b^m \g_m U_{(I)} \theta_I 
=
\bar{\theta}_K  b^m (\Lambda_{(K)})_m^{\ n} \g_n \bar{U}_{(K)} U_{(I)} \theta_I \,,
\ee
where  the identity $U_{(K)} \bar{U}_{(K)} = \gen{1}$ was inserted.
Specifically,
\be
\begin{aligned}
\bar{\theta}_1 b^m \g_m \theta_1  &\to \bar{\theta}_1  b^m {(\Lambda_1)_m}^n \g_n  \theta_1\, ,
\\
\bar{\theta}_2 b^m \g_m \theta_1  &\to \bar{\theta}_2  b^m {(\Lambda_2)_m}^n \g_n \bar{U}_{(2)} U_{(1)} \theta_1\, .
\end{aligned}
\ee
The terms with derivatives on fermions transform (here $I$ is kept fixed)
\be
\bar{\theta}_I b^m \g_m \pa_\b \theta_I 
\to
\bar{\theta}_I  b^m {(\Lambda_{(I)})_m}^n \g_n \pa_\b \theta_I
+\bar{\theta}_I  b^m {(\Lambda_{(I)})_m}^n \g_n (\bar{U}_{(I)} \pa_\b U_{(I)}) \theta_I .
\ee
The second of these terms will contribute to the coupling to the spin connection and the $B$-field. 

Finally, to compute the resulting quantities, we need  to know how the derivatives act on  $U_{(I)}$
\be
\begin{aligned}
\bar{U}_{(I)}^{\alg{a}} {\rm d} U_{(I)}^{\alg{a}}  &=\sigma_{3II}\, \frac{\vk}{2} \left(\frac{  \rho  (2 \sin \zeta {\rm d}\rho +\rho  {\rm d}\zeta  \cos \zeta)}{1+\vk ^2 \rho ^4 \sin ^2\zeta} \check{\g}_{23}+
\frac{  {\rm d}\rho }{1- \vk ^2 \rho ^2} \check{\g}_{04} \right), \\
\bar{U}_{(I)}^{\alg{s}} {\rm d} U_{(I)}^{\alg{s}} &= \sigma_{3II}\, \frac{\vk}{2}\left( -\frac{r (2 \sin \xi  {\rm d} r+r {\rm d} \xi  \cos \xi )}{1+\kappa ^2 r^4 \sin ^2\xi } \hat{\g}_{78}
-\frac{{\rm d} r}{1+\kappa ^2 r^2} \hat{\g}_{59} \right),
\end{aligned}
\ee
and also the product of matrices $U_{(I)}$ 
\be
\begin{aligned}
\bar{U}^{\alg{a}}_{(I)} U^{\alg{a}}_{(J)}&=
\delta_{IJ}\mathbf{1}_4+
\frac{\sigma_{1IJ}(\mathbf{1}_4
-i \vk ^2 \rho ^3 \sin \zeta \, \check{\g}_1)
-\epsilon_{IJ} \vk (\rho^2 \sin \zeta \, \check{\g}_{23}+ \rho \, \check{\g}_{04})
}{\sqrt{1-\vk ^2 \rho ^2} \sqrt{1+\vk ^2 \rho ^4\sin ^2\zeta }} , 
\\
\bar{U}^{\alg{s}}_{(I)} U^{\alg{s}}_{(J)}&=
\delta_{IJ}\mathbf{1}_4+
\frac{\sigma_{1IJ}(\mathbf{1}_4
- \vk ^2 r ^3 \sin \xi \, \hat{\g}_6)
+\epsilon_{IJ}\vk ( r^2 \sin \xi \, \hat{\g}_{78}+ r \, \hat{\g}_{59})
}{\sqrt{1+\vk ^2 r ^2} \sqrt{1+\vk ^2 r ^4\sin ^2\xi }} .
\end{aligned}
\ee
As a side comment, when implementing these redefinitions it is sometimes useful to work with redefined coordinates $\rho',\zeta',r',\xi'$ given by 
\be
\rho = \vk^{-1} \sin \rho',
\qquad
\sin \zeta = \vk \frac{\sinh \zeta'}{\sin^2 \rho'},
\qquad
r = \vk^{-1} \sinh r',
\qquad
\sin \xi = \vk \frac{\sinh \xi'}{\sinh^2 r'},
\ee
as it helps to simplify some expressions.

With this last redefinition of fermions done, we obtain the Lagrangian in the canonical form (\ref{eq:lagr-quad-ferm}), (\ref{eq:deform-D-op}).

\subsection{$\kappa$-symmetry}\label{app:kappa}
Here we work out an explicit form of the $\kappa$-symmetry transformations and show that under the field redefinition found in appendix \ref{app:CGS} they reduce to the standard form. 
To start with, we rewrite the equation~\eqref{eq:eps-op-rho-kappa} in the form
\be\label{eq:kappa-var}
\op^{-1}(\alg{g}^{-1} \delta_\kappa \alg{g}) = \varrho \,,
\ee
where we also used $\varepsilon\equiv \alg{g}^{-1} \delta_\kappa \alg{g}$. Further computation  will be formally the same as the one done in section \ref{subsec:def_model}. 
We just need to perform the substitution $\pa_\a \to - \delta_{\kappa}$.
Let us express the left hand side of (\ref{eq:kappa-var}) as a linear combination of generators $\gen{P}_m$ and $\gen{Q}^I$
\be
\op^{-1}(\alg{g}^{-1} \delta_\kappa \alg{g}) = j^m_{\delta_{\kappa}} \gen{P}_m + \gen{Q}^I j_{\delta_{\kappa},I}+j^{mn}_{\delta_{\kappa}} \gen{J}_{mn}\,.
\ee
The contributions of the generators $\gen{J}_{mn}$ will not be important for us.
The coefficients $j^m_{\delta_{\kappa}} , j_{\delta_{\kappa},I}$ are the quantities that we need to compute for finding how 
$\kappa$-symmetry acts on the fields. Because $\varrho$ in the right hand side of~\eqref{eq:kappa-var} is an odd element $\varrho= \gen{Q}^I \psi_I$, we have 
\be\label{eq:kappacon}
j^m_{\delta_{\kappa}}=0, 
\qquad
j_{\delta_{\kappa},I}=\psi_I.
\ee
Expanding the above equations in powers of $\theta$, we 
actually stop at the leading order, {\it i.e.}
\be\label{eq:order-kappa-var}
\begin{aligned}
& j^m_{\delta_{\kappa}}\sim \left[\# +\mathcal{O}(\theta^2)\right]\delta_{\kappa}X+ \left[\# \theta+\mathcal{O}(\theta^3)\right]\delta_{\kappa}\theta,
\\
& j_{\delta_{\kappa},I}\sim \left[\# +\mathcal{O}(\theta^2)\right]\delta_{\kappa}\theta,
\qquad
\psi \sim \left[\# +\mathcal{O}(\theta^2)\right] \kappa,
\end{aligned}
\ee
where $\#$ stands for functions of the bosons,
in such a way that upon solving equations (\ref{eq:kappacon}) we get $\delta_{\kappa}X \sim \# \theta \kappa$ and $\delta_{\kappa}\theta \sim \# \kappa$.

Let us start computing $j^m_{\delta_{\kappa}}$. Because of the deformation, the term inside parenthesis proportional to $\gen{Q}^I$ contributes
\be
\begin{aligned}
j^m_{\delta_{\kappa}} \gen{P}_m 
&= -P^{(2)}\circ \frac{1}{\mathbf{1} - \eta R_{\alg{g}} \circ d} \left[ \left( \delta_{\kappa}X^M e^m_M + \frac{i}{2} \bar{\theta}_I \bg^m \delta_{\kappa} \theta_I + \cdots \right) \gen{P}_m -\gen{Q}^I \delta_{\kappa} \theta_I + \cdots\right]
\\
&=  -\delta_{\kappa}X^M e^m_M  {k_m}^q \ \gen{P}_q  \\
&-\frac{1}{2} \ \bar{\theta}_I  \Bigg[ \delta^{IJ} i \bg_p +
 (-\vk \sigma_1^{IJ} +(-1+\sqrt{1+\vk^2})\delta^{IJ}) \left( i\bg_p +\frac{1}{2}  \lambda^{mn}_{ p} \bg_{mn}  \right) \\
& + i \, (\vk \sigma_3^{IJ} -(-1+\sqrt{1+\vk^2})\epsilon^{IJ})  {\lambda_p}^n \bg_n
\Bigg] \delta_{\kappa} \theta_J \ k^{pq} \ \gen{P}_q + \cdots
\end{aligned}
\ee
Imposing the equation $j^m_{\delta_{\kappa}}=0$ and solving for $ \delta_{\kappa}X^M$ at leading order we get
\be
\begin{aligned}
 \delta_{\kappa}X^M =
- \frac{1}{2} \ \bar{\theta}_I e^{Mp} \Bigg[ &\delta^{IJ} i \bg_p +
 (-\vk \sigma_1^{IJ} +(-1+\sqrt{1+\vk^2})\delta^{IJ}) \left( i\bg_p +\frac{1}{2}  \lambda^{mn}_{ p} \bg_{mn}  \right) \\
& + i \, (\vk \sigma_3^{IJ} -(-1+\sqrt{1+\vk^2})\epsilon^{IJ})  {\lambda_p}^n \bg_n
\Bigg] \delta_{\kappa} \theta_J + \cdots.
\end{aligned}
\ee
The computation for $j_{\delta_{\kappa},I} $ gives simply
\be
\begin{aligned}
\gen{Q}^I j_{\delta_{\kappa},I} 
&= (P^{(1)}+P^{(3)}) \circ \frac{1}{\mathbf{1} - \eta R_{\alg{g}} \circ d} \left[  \gen{Q}^I \delta_{\kappa} \theta_I + \cdots\right]
\\
&=   \frac{1}{2}\left( (1+\sqrt{1+\vk^2})\ \delta^{IJ} -\vk \sigma_1^{IJ} \right) \gen{Q}^J \delta_{\kappa} \theta_I + \cdots.
\end{aligned}
\ee
When we compute the two projections of $\varrho$ as defined in~\eqref{eq:def-varrho-kappa-def} at leading order we can set $\theta=0$.
Then we just have
\be
\begin{aligned}
P^{(2)} \circ \op^{-1} A_\b & = P^{(2)} \circ\op^{-1} \left( e^m_\b \gen{P}_m +\cdots \right) = e_{\b m} k^{mn} \gen{P}_n,
\\
P^{(2)} \circ  \optilde^{-1} A_\b & = P^{(2)} \circ  \optilde^{-1} \left( e^m_\b \gen{P}_m +\cdots \right) = e_{\b m} k^{nm} \gen{P}_n,
\end{aligned}
\ee
where the second result can be obtained from the first one sending $\vk \to - \vk$.
Explicitly,
\be
\begin{aligned}
\varrho^{(1)} &=\frac{1}{2} (\gamma^{\a\b} - \epsilon^{\a\b})   e_{\b m} k^{mn} \left( \gen{Q}^1 \gen{P}_n + \gen{P}_n \gen{Q}^1\right)\kappa_{\a 1},\\
\varrho^{(3)} &=  \frac{1}{2} (\gamma^{\a\b} + \epsilon^{\a\b})  e_{\b m} k^{nm} \left( \gen{Q}^2 \gen{P}_n + \gen{P}_n \gen{Q}^2\right)\kappa_{\a 2} ,
\end{aligned}
\ee
A direct computation shows that
\be
\gen{Q}^I \check{\gen{P}}_m + \check{\gen{P}}_m \gen{Q}^I = -\frac{1}{2} \gen{Q}^I \check{\bg}_m,
\qquad
\gen{Q}^I \hat{\gen{P}}_m + \hat{\gen{P}}_m \gen{Q}^I = +\frac{1}{2} \gen{Q}^I \hat{\bg}_m.
\ee
We get
\be
\begin{aligned}
\varrho^{(1)} &=\gen{Q}^1 \psi_1, 
\qquad
 \psi_1=\frac{1}{4} (\gamma^{\a\b} - \epsilon^{\a\b})   \left( -e_{\b m} k^{mn} \check{\bg}_n + e_{\b m} k^{mn} \hat{\bg}_n \right)\kappa_{\a 1} \, ,\\
\varrho^{(3)} &=\gen{Q}^2 \psi_2,
\qquad
\psi_2=  \frac{1}{4} (\gamma^{\a\b} + \epsilon^{\a\b})    \left( -e_{\b m} k^{nm} \check{\bg}_n + e_{\b m} k^{nm} \hat{\bg}_n \right)\kappa_{\a 2}\, .
\end{aligned}
\ee
Finally, we solve the equation $j_{\delta_{\kappa},I}=\psi_I$, obtaining the $\kappa$-variation of fermions 
\be
\delta_{\kappa} \theta_I =  \frac{1}{1+\sqrt{1+\vk^2}}\left( (1+\sqrt{1+\vk^2})\delta^{IJ} + \vk \sigma_1^{IJ} \right) \psi_J.
\ee
Setting $\vk=0$ the formulas are simplified to 
\be\label{eq:kappa_undeformed}
\begin{aligned}
 \delta_{\kappa}X^M &= -\frac{i}{2} \ \bar{\theta}_I \delta^{IJ} e^{Mp}    \bg_p  \delta_{\kappa} \theta_J + \cdots,
\\
\delta_{\kappa} \theta_I &=  \psi_I,
\\
\psi_1&=\frac{1}{4} (\gamma^{\a\b} - \epsilon^{\a\b})  \left( -e_{\b}^m  \check{\bg}_m + e_{\b}^m \hat{\bg}_m \right)\kappa_{\a 1},
\\
\psi_2&=  \frac{1}{4} (\gamma^{\a\b} + \epsilon^{\a\b})   \left( -e_{\b}^m  \check{\bg}_m + e_{\b}^m \hat{\bg}_m \right)\kappa_{\a 2},
\end{aligned}
\ee
showing that the $\kappa$-symmetry variation is the standard as expected.
\bigskip

To put the original Lagrangian in the canonical form we performed the field redefinitions and now we have to understand how
the $\kappa$-symmetry transformations look like for the redefined fields.   
Upon rotation the variation of fermions is modified as
\be
\theta_I \to U_{IJ} \theta_J
\implies
\delta_{\kappa} \theta_I \to U_{IJ} \delta_{\kappa} \theta_J + \delta_{\kappa} U_{IJ}  \theta_J,
\ee
and since we are considering $\delta_{\kappa} \theta$ at leading order, in the following we will drop the term containing $\delta_{\kappa} U_{IJ}$.
We first redefine our fermions as in (\ref{eq:red-fer-2x2-sp})
%\be
%\theta_I \to \frac{\sqrt{1+\sqrt{1+\vk^2}}}{\sqrt{2}} \left(\delta^{IJ} + \frac{\vk}{1+\sqrt{1+\vk^2}} \sigma_1^{IJ} \right) \theta_J.
%\ee
and we get
\be
\begin{aligned}
 \delta_{\kappa}X^M &=
- \frac{1}{2} \ \bar{\theta}_I e^{Mp} \Bigg[ \delta^{IJ} i \bg_p 
- (\vk \sigma_1^{IJ} +(-1+\sqrt{1+\vk^2})\delta^{IJ}) \frac{1}{2}  \lambda^{mn}_{ p} \bg_{mn}   \\
& + i \, (\vk \sigma_3^{IJ} -(-1+\sqrt{1+\vk^2})\epsilon^{IJ})  {\lambda_p}^n \bg_n
\Bigg] \delta_{\kappa} \theta_J + \cdots,
\\
\delta_{\kappa} \theta_I &=  \sqrt{\frac{2}{1+\sqrt{1+\vk^2}}} \ \psi_I\, .
\end{aligned}
\ee
When we shift the bosons as in (\ref{eq:shift_bosons}), their variation is modified to $\delta_{\kappa}X^M \to \delta_{\kappa}X^M +2\bar{\theta}_I f^M_{IJ} \delta_{\kappa}\theta_J +\bar{\theta}_I \delta_{\kappa}f^M_{IJ} \theta_J$. Once again, since we are considering the variation at leading order, we drop the term with $\delta_{\kappa}f^M_{IJ}$. Using the explicit form of the function $f^M_{IJ}$ given in~\eqref{eq:def-shift-bos-f},
we find that after the shift of the bosons their variation becomes
\bea
%\begin{aligned}
\nonumber
 \delta_{\kappa}X^M &=& -2\bar{\theta}_I f^M_{IJ} \delta_{\kappa}\theta_J -
 \frac{1}{2} \ \bar{\theta}_I e^{Mp} \Bigg[\delta^{IJ} i \bg_p 
- (\vk \sigma_1^{IJ} +(-1+\sqrt{1+\vk^2})\delta^{IJ}) \frac{1}{2}  \lambda^{mn}_{ p} \bg_{mn}   \\
& & \qquad\qquad\qquad\qquad+ i \, (\vk \sigma_3^{IJ} -(-1+\sqrt{1+\vk^2})\epsilon^{IJ})  {\lambda_p}^n \bg_n
\Bigg] \delta_{\kappa} \theta_J + \cdots
\\
\nonumber
&=& - \frac{i}{2} \ \bar{\theta}_I e^{Mm} \left( \delta^{IJ} \bg_m 
 +  \vk \sigma_3^{IJ}  {\lambda_m}^n \bg_n
\right) \delta_{\kappa} \theta_J + \cdots.
%\end{aligned}
\eea
The shift does not affect $\delta_{\kappa} \theta_I$ at leading order.
The final result is obtained by implementing the bosonic-dependent rotation of fermions~\eqref{eq:red-ferm-Lor-as}
\be
\begin{aligned}
 \delta_{\kappa}X^M &= - \frac{i}{2} \ \bar{\theta}_I \bar{U}_{(I)} \ e^{Mm} \left( \delta^{IJ} \bg_m 
 +  \vk \sigma_3^{IJ}  {\lambda_m}^n \bg_n
\right) \ U_{(I)} \delta_{\kappa} \theta_J + \cdots \\
&= 
- \frac{i}{2} \ \bar{\theta}_I \delta^{IJ} \widetilde{e}^{Mm}  \bg_m  \delta_{\kappa} \theta_J + \cdots,
\\
\delta_{\kappa} \theta_1 &=  \sqrt{\frac{2}{1+\sqrt{1+\vk^2}}} \left(  \frac{1}{4} (\gamma^{\a\b} - \epsilon^{\a\b}) \bar{U}_{(1)}  \left( -\check{e}_{\b m} k^{mn} \check{\bg}_n + \hat{e}_{\b m} k^{mn} \hat{\bg}_n \right)\kappa_{\a 1} \right)\, ,
\\
\delta_{\kappa} \theta_2 &=  \sqrt{\frac{2}{1+\sqrt{1+\vk^2}}} \left(\frac{1}{4} (\gamma^{\a\b} + \epsilon^{\a\b})  \bar{U}_{(2)}  \left( -\check{e}_{\b m} k^{nm} \check{\bg}_n + \hat{e}_{\b m} k^{nm} \hat{\bg}_n \right)\kappa_{\a 2} \right)\, .
\end{aligned}
\ee
The variation of bosons already appears to be related to the one of fermions in the standard way.
It has actually the same form as in the undeformed case, but with the vielbein of the deformed theory.
We can  also bring the variation of fermions to the standard form if we use the fact that for both expressions
\be
\begin{aligned}
\sqrt{\frac{2}{1+\sqrt{1+\vk^2}}} \  \bar{U}_{(1)}  \left( -\check{e}_{\b m} k^{mn} \check{\bg}_n + \hat{e}_{\b m} k^{mn} \hat{\bg}_n \right)\kappa_{\a 1} = 
\left( -\widetilde{e}_{\b}^m \check{\bg}_m + \widetilde{e}_{\b}^m  \hat{\bg}_m \right) \widetilde{\kappa}_{\a 1}\, , 
\\
\sqrt{\frac{2}{1+\sqrt{1+\vk^2}}} \ \bar{U}_{(2)}  \left( -\check{e}_{\b m} k^{nm} \check{\bg}_n + \hat{e}_{\b m} k^{nm} \hat{\bg}_n \right)\kappa_{\a 2} =
\left( -\widetilde{e}_{\b}^m \check{\bg}_m + \widetilde{e}_{\b}^m  \hat{\bg}_m \right) \widetilde{\kappa}_{\a 2}\, ,
\end{aligned}
\ee
where we have inserted the identity $\mathbf{1}=U_{(I)}\bar{U}_{(I)}$ and defined 
\be\label{eq:def-kappa-tilde-k-symm}
 \widetilde{\kappa}_{\a I} \equiv \sqrt{\frac{2}{1+\sqrt{1+\vk^2}}} \ \bar{U}_{(I)} \kappa_{\a I}.
\ee
To summarise, we find the following expressions 
\be\label{eq:kappa-var-16}
\begin{aligned}
 \delta_{\kappa}X^M &= - \frac{i}{2} \ \bar{\theta}_I \delta^{IJ} \widetilde{e}^{Mm}  \bg_m  \delta_{\kappa} \theta_J + \cdots,
\\
\delta_{\kappa} \theta_I &=    \widetilde{\psi}_I,
\\
\widetilde{\psi}_1&=\frac{1}{4} (\gamma^{\a\b} - \epsilon^{\a\b})  \left( -\widetilde{e}_{\b}^m  \check{\bg}_m + \widetilde{e}_{\b}^m \hat{\bg}_m \right) \widetilde{\kappa}_{\a 1},
\\
\widetilde{\psi}_2&=  \frac{1}{4} (\gamma^{\a\b} + \epsilon^{\a\b})   \left( -\widetilde{e}_{\b}^m  \check{\bg}_m + \widetilde{e}_{\b}^m \hat{\bg}_m \right)\widetilde{\kappa}_{\a 2},
\end{aligned}
\ee
to be compared with their undeformed counterpart (\ref{eq:kappa_undeformed}). As we see, the only difference 
is an appearance of tilde in \eqref{eq:kappa-var-16} which signifies the quantities of the deformed background.

One can also write $\kappa$-variations in terms of $32$-dimensional fermions $\T$. To this end,
we introduce $32$-dimensional spinors $\widetilde{K}$ which have chirality opposite to that of  $\T$
\be
\widetilde{K} \equiv \left( \begin{array}{c} 0 \\ 1 \end{array} \right) \otimes \widetilde{\kappa}.
\ee
The variations above are then written as
\be
\begin{aligned}
 \delta_{\kappa}X^M &= - \frac{i}{2} \ \bar{\T}_I \delta^{IJ} \widetilde{e}^{Mm}  \G_m  \delta_{\kappa} \T_J + \cdots,
\\
\delta_{\kappa} \T_I &= -\frac{1}{4} (\delta^{IJ} \gamma^{\a\b} - \sigma_3^{IJ} \epsilon^{\a\b})  \widetilde{e}_{\b}^m  \G_m  \widetilde{K}_{\a J}\, .
\end{aligned}
\ee
The 10-dimensional gamma matrices $\G_m$ are defined in appendix~\ref{sec:10-dim-gamma}.
%%%%%%%%

\smallskip

Let us now look at the $\kappa$-variation of the world-sheet metric, which expression is given in~\eqref{eq:defin-kappa-var-ws-metric}.
This variation starts at first order in fermions. Then we have to compute
\be
\begin{aligned}
P^{(1)}\circ \widetilde{\op}^{-1}( A^{\b}_+ ) &= 
P^{(1)}\circ \widetilde{\op}^{\text{inv}}_{(0)} (- \gen{Q}^{I} \, D^{\b IJ}_+ \theta_J) + P^{(1)}\circ \widetilde{\op}^{\text{inv}}_{(1)} ( e^{m\b}_+\gen{P}_{m}  )+\mathcal{O}(\theta^3)\,,
\\
P^{(3)}\circ {\op}^{-1}( A^{\b}_- ) &= 
P^{(3)}\circ \opinv_{(0)} (- \gen{Q}^{I} \, D^{\b IJ}_- \theta_J) + P^{(3)}\circ \opinv_{(1)} ( e^{m\b}_-\gen{P}_{m}  )+\mathcal{O}(\theta^3)\,.
\end{aligned}
\ee
Let us start from the last line. We have
\bea
&&P^{(3)}\circ \opinv_{(0)} (- \gen{Q}^{I} \, D^{\b IJ}_- \theta_J) = -\left(\frac{1}{2} (1+\sqrt{1+\vk^2}) \, \delta^{I2}- \frac{\vk}{2} {\sigma_1}^{I2} \, \right) \gen{Q}^{2} D^{\b IJ}_- \theta_J \nonumber \\
&&P^{(3)}\circ \opinv_{(1)} ( e^{m\b}_-\gen{P}_{m}  ) = -\frac{\vk}{4} \gen{Q}^2 \  e^{m\b}_- {k_m}^n \ \Bigg[ 
 \left((1+\sqrt{1+\vk^2})\delta^{2J} -\vk \sigma_1^{2J}\right) \left(i \bg_n - \frac{1}{2} \lambda_n^{pq} \bg_{pq} \right) \nonumber \\
 && \qquad\qquad + i \left((1+\sqrt{1+\vk^2}) \epsilon^{2J} + \vk \sigma_3^{2J}\right) {\lambda_n}^p \bg_p 
\Bigg] \theta_J\, ,
\eea
where quantities with the subscript ``+'' or ``-" are defined through \eqref{defpm}.
For the first line we can use that $\widetilde{\op}^{\text{inv}}_{(0)}$ and $\opinv_{(0)}$ coincide on odd elements, while on even elements their action is equivalent to sending $\vk\to-\vk$, and we can write
\be
\widetilde{\op}^{\text{inv}}_{(0)}(\gen{Q}^I)=\opinv_{(0)}(\gen{Q}^I)\,,
\qquad
\widetilde{\op}^{\text{inv}}_{(0)}(\gen{P}_m)=k^n_{\ m} \gen{P}_n +\# \gen{J}\,,
\ee
where $k^n_{\ m}=\eta^{nn'}\eta_{mm'} k_{n'}^{\ m'}$.
On the other hand, the action of $\optilde_{(1)}$ on even elements is minus the one of $\op_{(1)}$
\be
\optilde_{(1)}(\gen{P}_m)=-\op_{(1)}(\gen{P}_m)\,.
\ee
These considerations need to be taken into account when computing the action of $\widetilde{\op}^{\text{inv}}_{(1)}$ on $\gen{P}_{m}$.
Then we find
\bea\nonumber
&&P^{(1)}\circ \widetilde{\op}^{\text{inv}}_{(0)} (- \gen{Q}^{I} \, D^{\b IJ}_+ \theta_J) = -\left(\frac{1}{2} (1+\sqrt{1+\vk^2}) \, \delta^{I1}- \frac{\vk}{2} {\sigma_1}^{I1} \, \right) \gen{Q}^{1} D^{\b IJ}_+ \theta_J \, , \\
&&P^{(1)}\circ \widetilde{\op}^{\text{inv}}_{(1)} ( e^{m\b}_+\gen{P}_{m}  ) = +\frac{\vk}{4} \gen{Q}^1 \  e^{m\b}_+ {k^n}_m \ \Bigg[ 
 \left((1+\sqrt{1+\vk^2})\delta^{1J} -\vk \sigma_1^{1J}\right) \left(i \bg_n - \frac{1}{2} \lambda_n^{pq} \bg_{pq} \right) 
 \nonumber \\
&&\qquad\qquad + i \left((1+\sqrt{1+\vk^2}) \epsilon^{1J} + \vk \sigma_3^{1J}\right) {\lambda_n}^p \bg_p 
\Bigg] \theta_J\, .
\eea
When computing the commutators in~\eqref{eq:defin-kappa-var-ws-metric}, we should care only about the contribution proportional to the identity operator, as the others yield a vanishing contribution after we multiply by $\Upsilon$ and take the supertrace.

We write the result for the variation of the world-sheet metric, after the redefinition~\eqref{eq:red-fer-2x2-sp} has been done
\bea
%\begin{aligned}
\delta_\kappa \g^{\a\b} &=&  \frac{2i\, \sqrt{2}}{\sqrt{1+\sqrt{1+\vk^2}}} \Bigg[
\bar{\kappa}^\a_{1+} \Bigg( \delta^{1J}\pa^{\b}_+  - \frac{1}{4} \delta^{1J} \omega^{\b mn}_+ \bg_{mn} +\frac{i}{2} (\sqrt{1+\vk^2}\eps^{1J}+\vk \sigma_3^{1J} ) e^{m\b}_+ \bg_m
\nonumber \\
&-&\frac{\vk}{2}  e^{m\b}_+ {k^n}_m \  \Bigg(
\delta^{1J} \left(i \bg_n - \frac{1}{2} \lambda_n^{pq} \bg_{pq} \right) 
+ i \left(\sqrt{1+\vk^2} \epsilon^{1J} + \vk \sigma_3^{1J}\right) {\lambda_n}^p \bg_p 
\Bigg)\Bigg)
\\
&& \qquad \qquad+\bar{\kappa}^\a_{2-} \Bigg( \delta^{2J}\pa^{\b}_-  - \frac{1}{4} \delta^{2J} \omega^{\b mn}_- \bg_{mn} +\frac{i}{2} (\sqrt{1+\vk^2}\eps^{2J}+\vk \sigma_3^{2J} ) e^{m\b}_- \bg_m
\nonumber \\
&+&\frac{\vk}{2}  e^{m\b}_- {k_m}^n \  \Bigg(
\delta^{2J} \left(i \bg_n - \frac{1}{2} \lambda_n^{pq} \bg_{pq} \right) 
+ i \left(\sqrt{1+\vk^2} \epsilon^{2J} + \vk \sigma_3^{2J}\right) {\lambda_n}^p \bg_p 
\Bigg)\Bigg)\Bigg] \theta_J.
%\end{aligned}
\nonumber
\eea
Here we have written the result in terms of $\bar{\kappa}=\kappa^\dagger\bg^0$.
We do not need to take into account the shift of the bosonic fields (\ref{eq:shift_bosons}), since it only matters at higher orders in fermions.
To take into account the last fermionic field redefinition and write the final form of the variation of the world-sheet metric, we split the result into ``diagonal''and ``off-diagonal''  in the labels $I,J$
\bea
%\begin{aligned}
\nonumber
\delta_\kappa \g^{\a\b}|_{\text{diag}} &=&  2i \Bigg[
\bar{\tilde{\kappa}}^\a_{1+} \Bigg(\pa^{\b}_+ +\bar{U}_{(1)}\pa^{\b}_+U_{(1)}\\
\nonumber
&&~~~~ -  \frac{1}{4} \left(\omega^{\b mn}_+ (\Lambda_{(1)})_{m}^{\ m'}(\Lambda_{(1)})_{n}^{\ n'}\bg_{m'n'} -\vk e^{m\b}_+ {k^n}_m \lambda_n^{pq}  (\Lambda_{(1)})_{p}^{\ p'}(\Lambda_{(1)})_{q}^{\ q'}\bg_{p'q'}\right) \\
\nonumber
&&~~~~ +  \frac{i \vk }{2}  e^{m\b}_+ \left((\Lambda_{(1)})_{m}^{\ m'}\bg_{m'} -{k^n}_m \  \left(
 (\Lambda_{(1)})_{n}^{\ n'}\bg_{n'} +   \vk  {\lambda_n}^p (\Lambda_{(1)})_{p}^{\ p'}\bg_{p'} 
\right)\right)
 \Bigg)\theta_1
\\
&&~~+\bar{\tilde{\kappa}}^\a_{2-} \Bigg(\pa^{\b}_- +\bar{U}_{(2)}\pa^{\b}_-U_{(2)}\\
\nonumber
&&~~~~ - \frac{1}{4} \left(\omega^{\b mn}_- (\Lambda_{(2)})_{m}^{\ m'}(\Lambda_{(2)})_{n}^{\ n'}\bg_{m'n'} +\vk e^{m\b}_- {k_m}^n \lambda_n^{pq}  (\Lambda_{(2)})_{p}^{\ p'}(\Lambda_{(2)})_{q}^{\ q'}\bg_{p'q'}\right) \\
\nonumber
&&~~~~
-\frac{i \vk }{2}  e^{m\b}_- \left((\Lambda_{(2)})_{m}^{\ m'}\bg_{m'} -{k_m}^n \  \left(
 (\Lambda_{(2)})_{n}^{\ n'}\bg_{n'} -   \vk  {\lambda_n}^p (\Lambda_{(2)})_{p}^{\ p'}\bg_{p'} 
\right)\right)
 \Bigg)\theta_2
\Bigg] ,
%\end{aligned}
\eea
\bea
%\begin{aligned}
\nonumber
\delta_\kappa \g^{\a\b}|_{\text{off-diag}} &=&   - \sqrt{1+\vk^2}\Bigg[
\bar{\tilde{\kappa}}^\a_{1+}  \bar{U}_{(1)}U_{(2)} e^{m\b}_+ \left(  (\Lambda_{(2)})_{m}^{\ m'}  \bg_{m'}
-\vk   {k^n}_m \      {\lambda_n}^p(\Lambda_{(2)})_{p}^{\ p'} \bg_{p'} \right)\theta_2
\\
&&~~-\bar{\tilde{\kappa}}^\a_{2-}  \bar{U}_{(2)}U_{(1)} e^{m\b}_-\left(  (\Lambda_{(1)})_{m}^{\ m'} \bg_{m'}
+\vk   {k_m}^n \      {\lambda_n}^p(\Lambda_{(1)})_{p}^{\ p'} \bg_{p'} \right) \theta_1
\Bigg].
%\end{aligned}
\eea
Looking at the diagonal contribution, we find that the terms containing rank-1 gamma matrices actually vanish, as they should.
The rest yields exactly the expected couplings to spin connection and $H^{(3)}$
\be
\begin{aligned}
\delta_\kappa \g^{\a\b}|_{\text{diag}}&=2i\Bigg[ 
\bar{\tilde{\kappa}}^\a_{1+} \left( \pa^{\b}_+ -\frac{1}{4} \widetilde{\omega}^{\b mn}_+ \bg_{mn} +\frac{1}{8} \tilde{e}^{m\b}_+H_{mnp} \bg^{np}\right)\theta_1\\
&\quad\quad\quad+\bar{\tilde{\kappa}}^\a_{2-} \left( \pa^{\b}_- -\frac{1}{4} \widetilde{\omega}^{\b mn}_- \bg_{mn} -\frac{1}{8} \tilde{e}^{m\b}_- H_{mnp} \bg^{np}\right)\theta_2
 \Bigg]\,.
\end{aligned}
\ee
When we consider the off-diagonal contribution we find that it gives the RR fields
\bea
%\begin{aligned}
\delta_\kappa \g^{\a\b}|_{\text{off-diag}}&=&2i\left( -\frac{1}{8} e^{\varphi} \right)\Bigg[ 
\bar{\tilde{\kappa}}^\a_{1+} \left( \bg^n F^{(1)}_n + \frac{1}{3!}\bg^{npq} F^{(3)}_{npq}+\frac{1}{2\cdot 5!} \bg^{npqrs} F^{(5)}_{npqrs}\right) \tilde{e}^{m\b}_+ \bg_m\, \theta_2
\nonumber \\
&+&\bar{\tilde{\kappa}}^\a_{2-} \left( -\bg^n F^{(1)}_n + \frac{1}{3!}\bg^{npq} F^{(3)}_{npq}-\frac{1}{2\cdot 5!} \bg^{npqrs} F^{(5)}_{npqrs}\right) \tilde{e}^{m\b}_- \bg_m\, \theta_1
 \Bigg]\,,
%\end{aligned}
\eea
where the components of the RR couplings appear to be the same as in~\eqref{eq:flat-comp-F1}-\eqref{eq:flat-comp-F3}-\eqref{eq:flat-comp-F5}.
Putting these results together, we find the standard $\kappa$-transformation also for the world-sheet metric (\ref{eq:standardkappametric}).
Rewriting of this variation in terms of 32-dimensional spinors is straightforward.

\section{Quantisation of the light-cone Hamiltonian}\label{lcHam}
%%%%%%%%%%%%%%%%%%%%%%%%%%%%%%
\subsection{Light-cone Hamiltonian}

Our starting point is the Lagrangian with 16 $\kappa$-gauge-fixed fermions $\Theta_a$ written in the form 
\bea
\mathcal{L}=&-&\frac{\tilde{g}}{2}\gamma^{\alpha\beta} \partial_\alpha X^M \partial_\beta X^N \hat{G}_{MN} +\frac{\tilde{g}}{2}\epsilon^{\alpha\beta} \partial_\alpha X^M \partial_\beta X^N \hat{B}_{MN} +\\
\nonumber
&&~~~~+i\frac{\tilde{g}}{2} \g^{\a\b}\pa_\a X^M {\Theta}_af^{ab}_M \pa_\b\Theta_b  \, -i\frac{\tilde{g}}{2}\eps^{\a\b}\pa_\a X^M  {\Theta}_aw^{ab}_M \pa_\b\Theta_b\,.\eea
Here we define the effective metric and B-field
\bea
\hat{G}_{MN}=\widetilde{G}_{MN} +G_{MN}^{(1)}\, , \qquad
\hat{B}_{MN}=\widetilde{B}_{MN}  + B_{MN}^{(1)}\,,  \eea
where $G_{MN}^{(1)}$ and $B_{MN}^{(1)}$ are quadratic in fermions. 

We write the Lagrangian as ($\eps^{\tau\sigma}=1$)
\bea
\mathscr{L}=-\frac{\tilde{g}}{2}\gamma^{\tau\tau}\hat{G}_{MN}\dot{X}^M\dot{X}^N-{\tilde{g}}(\gamma^{\tau\sigma}\hat{G}_{MN}-\hat{B}_{MN})\dot{X}^MX'^{N}-\frac{\tilde{g}}{2}\Omega_M\dot{X}^M+D\, .
\eea
Here 
\be
\Omega_M=-i\gamma^{\tau\tau}{\Theta}_af^{ab}_M \dot\Theta_b-i\gamma^{\tau\sigma}{\Theta}_af^{ab}_M \Theta'_b
+i{\Theta}_aw^{ab}_M\Theta'_b\, ,
\ee
and
\be
D=-\frac{\tg}{2}\gamma^{\sigma\sigma}\Big[\hat{G}_{MN}X'^MX'^N-i X'^M{\Theta}_af^{ab}_M \Theta_b'\Big]
+\frac{i\tg}{2}\gamma^{\sigma\tau}  X'^M{\Theta}_af^{ab}_M \dot\Theta_b+\frac{i\tg}{2}X'^M{\Theta}_aw^{ab}_M \dot\Theta_b\, .
\ee
The canonical momentum is
\bea
p_M&=&-\tg \gamma^{\tau\tau}\hat G_{MN}\dot{X}^N-\tg  \gamma^{\tau\sigma}\hat{G}_{MN}X'^N+\tg \hat{B}_{MN}X'^N-\frac{\tg}{2}\, \Omega_M\, ,\eea
and therefore
\bea
\dot{X}^M=-\frac{1}{\tg \gamma^{\tau\tau}}\hat{G}^{MN}\Bigg(p_N+\tg \gamma^{\tau\sigma}\hat{G}_{NL}X'^L-\tg \hat{B}_{NL}X'^L+\frac{\tg}{2}\, \Omega_N\Bigg)\, .
\eea
We define the Routhian 
\be
R=p_M\dot{X}^M-\mathscr{L}=-\frac{\tilde{g}}{2}\gamma^{\tau\tau}\hat{G}_{MN}\dot{X}^M\dot{X}^N-D\, ,\ee
and expressing $\dot{X}^M$ in terms of the momenta $p_M$ we then find the phase space version of the Lagrangian up to quadratic order in fermions
\bea\nonumber
\mathscr{L}=p_M\dot{X}^M-\frac{i}{2}p_M G^{MN} \Theta_a f^{ab}_N\dot{\Theta}_b\, +\frac{i}{2}\tg X'^M  \Theta_aw^{ab}_M\dot{\Theta}_b+\frac{\gamma^{\tau\sigma}}{\gamma^{\tau\tau}}C_1+\frac{1}{2\tg\gamma^{\tau\tau}}C_2
\, ,
\eea
\bea
C_1=p_MX'^M-\frac{i}{2}p_M G^{MN} \Theta_a f^{ab}_N{\Theta}'_b\, +\frac{i}{2}\tg X'^M  \Theta_aw^{ab}_M{\Theta}'_b \, ,\eea
\bea\nonumber
C_2&=&\hat{G}^{MN}p_Mp_N+\tg^2\hat{G}_{MN}X'^MX'^N-2\tg \hat{G}^{MN}p_M\hat{B}_{NK}X'^K\\
&-&i\tg^2X'^M {\Theta}_af^{ab}_M \Theta'_b \, +i\tg p_M G^{MN}{\Theta}_aw^{ab}_N\Theta'_b  \, .\eea\smallskip
The light-cone coordinates are introduced through
\bea
t&=&x_+-a x_-\, , ~~~\phi=x_++(1-a)x_-\, , \nonumber \\
p_t&=&(1-a)p_--p_+\, , ~~~p_{\phi}=p_++a p_-\,  ,
\eea
and the l.c. gauge is 
\bea
x_+=\tau\,,\quad p_+=1\,.
\eea
We write the kinetic term and the first constraint as
\bea\nonumber
\mathscr{L}_{\rm kin}=p_k\dot{x}^k-\frac{i}{2}\Theta_a f^{ab}_{(0)}\dot{\Theta}_b-\frac{i}{2}\Theta_a \Big(f^{ab}_{(1)} -\tg w^{ab}_{(1)}\Big)\dot{\Theta}_b +p_-\, ,
\eea
\bea
C_1=x'_-+p_kX'^k-\frac{i}{2}\Theta_a f^{ab}_{(0)}{\Theta}'_b-\frac{i}{2}\Theta_a \Big(f^{ab}_{(1)} -\tg w^{ab}_{(1)}\Big){\Theta}'_b \, ,
\eea
where
\bea
p_M G^{MN} f^{ab}_N= f^{ab}_{(0)} +  f^{ab}_{(1)}+\cdots \,,\quad  X'^M  w^{ab}_M = w^{ab}_{(1)}+\cdots\,.
\eea
Here $f_{(0)}$ is a constant matrix which squares to the identity, while  $f_{(1)}$ and $w_{(1)}$ are quadratic in transversal bosons.

Now one sees that to get the canonical Poisson structure up to quartic order in the fields one performs the following shift of fermions
\bea
\Theta_a\to \Theta_a - {1\ov2}\Theta_c \Big(f^{cd}_{(1)}-\tg w^{cd}_{(1)}\Big)f_{da}^{(0)} \,,
\eea
where $f_{ac}^{(0)} f^{cb}_{(0)} =\delta_a^b$, and  $f^{(0)}$ as a matrix  coincides with  $f_{(0)}$.

After this shift and up to the sixth order terms in the fields the first constraint takes the form
\bea
C_1=x'_-+p_kx'^k-\frac{i}{2}\Theta_a f^{ab}_{(0)}{\Theta}'_b \, ,
\eea
and from $C_1=0$ one finds
\bea
x_-'=-p_kx'^k+\frac{i}{2}\Theta_a f^{ab}_{(0)}{\Theta}'_b \,.
\eea
The Hamiltonian is then found by solving the second constraint for $p_-$. The quartic Hamiltonian is too complicated to be  presented here but the quadratic Hamiltonian has the same form as in the undeformed case up to some $\vk$-dependent factors. 

\subsection{Quantisation}

To quantise the model we introduce the two-index notations for the world-sheet fields which differs from the one used in the review \cite{Arutyunov:2009ga}
by the exchange of the indices 1 and 2, and $\dot{1}$ and $\dot{2}$ for all bosonic and fermionic fields: $1\leftrightarrow2$, $\dot1\leftrightarrow\dot2$. In addition the fermions should be multiplied by factors $\pm i$, to be precise
\be
\theta^{1\dot\a}\to i \theta^{2\dot\a}\,\quad \theta^{2\dot\a}\to i \theta^{1\dot\a}\,\quad 
 \eta^{\a\dot1}\to -i \eta^{\a\dot2}\,\quad \eta^{\a\dot2}\to -i \eta^{\a\dot1}\,.\quad 
\ee
This transformation is a symmetry of the undeformed model so the T-matrix would reduce to the standard one at $\vk=0$.

Rewriting the quadratic Lagrangian density in terms of the
two-index fields, one gets
 \bea\la{L2b} 
 \L_2 = P_{a\da}\dot{Y}^{a\da}  +P_{\a\dal}\dot{Z}^{\a\dal} +
i\,\eta_{\a\da}^\dagger\dot{\eta}^{\a\da}+
i\,\theta_{a\dal}^\dagger\dot{\theta}^{a\dal} - \H_2\,, \eea 
where the density of the quadratic Hamiltonian is given by 
\bal\la{dH2}
\H_2 &={1\ov 4}P_{a\da} P^{a\da} + {1\ov 4}P_{\a\dal} P^{\a\dal}+(1+\vk^2)\left(Y_{a\da} Y^{a\da} +
Y'_{a\da}Y'^{a\da}  +Z_{\a\dal}
Z^{\a\dal} +Z'_{\a\dal}Z'^{\a\dal}\right) \\ &+
\sqrt{1+\vk^2}\left(\eta_{\a\da}^\dagger\eta^{\a\da} + {1\ov
2}\eta^{\a\da}\eta'_{\a\da} -{1\ov 2}\eta^{\dagger\a\da}
\eta_{\a\da}'^\dagger + \theta_{a\dal}^\dagger\theta^{a\dal}  +
{1\ov 2}\theta^{a\dal}\theta'_{a\dal} -{1\ov
2}\theta^{\dagger a\dal} \theta_{a\dal}'^\dagger\right)\,.\eal
The fields satisfy the canonical equal-time (anti)commutation relations
\bea\nonumber
&&\hspace{-0.7cm}[\, Y^{a\da}(\s,\tau)\,  ,\, P_{b\db}(\s',\tau) \, ] = i\, \delta^a_{b}
\delta^\da_{\db}\delta(\s-\s') \,,\quad
[\, Z^{\a\dal}(\s,\tau)\,  ,\, P_{\b\dbe}(\s',\tau) \, ] = i\, \delta^\a_{\b} \delta^\dal_{\dbe}\delta(\s-\s') \,,\\\nonumber
&&\hspace{-0.7cm} \{\, \theta^{a\dal}(\s,\tau)\, ,\, \theta_{b\dbe}^\dagger(\s',\tau)\,
\}=\delta^a_{b} \delta^\dal_{\dbe}\delta(\s-\s')\,,\quad  \{\,
\eta^{\a\da}(\s,\tau)\, ,\, \eta_{\b\db}^\dagger(\s',\tau)\,
\}=\delta^\a_{\b} \delta^\da_{\db}\delta(\s-\s') \, ,~~~ \eea 
and we
 choose the following mode decompositions for the bosonic fields
\bea\la{bosrep}\begin{aligned} &~~~~Y^{a\da}(\s,\tau) = {1\ov
\sqrt{2\pi}}\int\,{\rm d}p\,{1\ov 2\sqrt{\om_p}}\left( e^{ip\s}
a^{a\da}(p,\tau) + e^{-ip\s}\eps^{ab}\eps^{\da\db}
a_{b\db}^\dagger(p,\tau)\right)\,,~~~~~~~~
\\ &~~~~P_{a\da}(\s,\tau) = {1\ov \sqrt{2\pi}}\int\,{\rm d}p\,\,i\,
\sqrt{\om_p}\left( e^{-ip\s} a_{a\da}^\dagger(p,\tau)-
e^{ip\s}\eps_{ab}\eps_{\da\db} a^{b\db}(p,\tau)\right)\,,\\
&~~~~Z^{\a\dal}(\s,\tau) = {1\ov \sqrt{2\pi}}\int\,{\rm d}p\,{1\ov
2\sqrt{\om_p}}\left( e^{ip\s} a^{\a\dal}(p,\tau) +
e^{-ip\s}\eps^{\a\b}\eps^{\dal\dbe}
a_{\b\dbe}^\dagger(p,\tau)\right)\,, \\ &~~~~P_{\a\dal}(\s,\tau) =
{1\ov \sqrt{2\pi}}\int\,{\rm d}p\,i\, \sqrt{\om_p}\left( e^{-ip\s}
a_{\a\dal}^\dagger(p,\tau)- e^{ip\s}\eps_{\a\b}\eps_{\dal\dbe}
a^{\b\dbe}(p,\tau)\right)\,,
\end{aligned}\eea and similarly for fermionic ones\footnote{Note that the mode decomposition for fermions is slightly different from the one used in the review  \cite{Arutyunov:2009ga} which in fact leads to a T-matrix which differs from the one computed in \cite{Klose:2006zd} by some signs. The mode decomposition used here gives in the undeformed case the T-matrix from \cite{Klose:2006zd}.}
\bea\la{ferrep}\begin{aligned} &~~~~~\theta^{a\dal}(\s,\tau)=
{e^{-i\pi/4}\ov \sqrt{2\pi}}\int\,{{\rm
d}p\ov\sqrt{\om_p}}\,\left( -ie^{ip\s}\,f_p\, a^{a\dal}(p,\tau)
+ie^{-ip\s}\,h_p\,\eps^{ab}\eps^{\dal\dbe}
a_{b\dbe}^\dagger(p,\tau)\right)\,,\\
 &~~~~~\eta^{\a\da}(\s,\tau)=
{e^{-i\pi/4}\ov \sqrt{2\pi}}\int\,{{\rm
d}p\ov\sqrt{\om_p}}\,\left( ie^{ip\s}\,f_p\, a^{\a\da}(p,\tau)
-ie^{-ip\s}\,h_p\,\eps^{\a\b}\eps^{\da\db}
a_{\b\db}^\dagger(p,\tau)\right) \,. ~~~~~~~~\end{aligned}\eea
Here the creation $a_{M\dot{M}}^\dagger$ and annihilation
$a^{M\dot{M}}$ operators are conjugate to each other:
$\big(a^{M\dot{M}}\big)^\dagger = a_{M\dot{M}}^\dagger$.
Then,  the
frequency $\om_p$ is given by 
\bal
\om_p = \sqrt{1+\vk^2}\sqrt{1+p^2}\, ,
\eal and the quantities
 \bal f_p &= {1+i{\nu\ov p}\ov \sqrt{1+{\nu^2\ov p^2}}}\sqrt{{\om_p+ \sqrt{1+\vk^2}\ov 2}}\,,\quad h_p =
 \sqrt{1+\vk^2}{ p\ov 2f_p}\,,\\
 & |f_p|^2-|h_p|^2= \sqrt{1+\vk^2}\,,\quad |f_p|^2+|h_p|^2=\om_p \,,
\eal
play the role of the fermion wave functions.

Omitting  the time
dependence in all the operators and total derivative terms, one finds that the
quadratic Lagrangian takes the diagonal form
 \bea\nonumber
L_2 =\int\,{\rm d}\s\, \L_2 =\int\,{\rm d}p\,
\sum_{M,\dot{M}}\left(
i\,a_{M\dot{M}}^\dagger(p)\dot{a}^{M\dot{M}}(p)  -\om_p\,
a_{M\dot{M}}^\dagger(p)a^{M\dot{M}}(p)\right)\,, \eea 
with the creation and annihilation operators satisfying
the canonical  relations
\bea\la{comrel} [\, a^{M\dot{M}}(p,\tau)\,  ,\,
a_{N\dot{N}}^\dagger(p',\tau) \, \} = \delta^M_{N}\,
\delta^{\dot{M}}_{\dot{N}}\, \delta(p-p') \, , \eea where we take
the commutator for bosons, and the anti-commutator for fermions.

\section{Equations of motion of type IIB supergravity}\label{app:IIBsugra}
In this appendix we collect the action and the equations of motion of type IIB supergravity.
The field content comprises Neveu-Schwarz--Neveu-Schwarz (NSNS) and Ramond-Ramond (RR) fields:
\begin{itemize}
\item[] \textbf{NSNS}: the metric $G_{MN}$, the dilaton $\varphi$, and the anti-symmetric two-form $B_{MN}$ with field strength $H_{MNP}$; 
\item[] \textbf{RR}: the axion $\chi$, the anti-symmetric two-form $C_{MN}$, and the anti-symmetric four-form $C_{MNPQ}$. 
\end{itemize}
The RR field strengths are defined as
\bea
&&F_{M}=\pa_{M}\chi\, , \\
&&F_{MNP}=3\pa_{[M}C_{NP]} +\chi H_{MNP}\, , \\
&&F_{MNPQR}=5\pa_{[M}C_{NPQR]}-15(B_{[MN}\pa_{P}C_{QR]}-C_{[MN}\pa_{P}B_{QR]})\, .\eea
Square brackets $[,]$ are used to denote the anti-symmetriser, for example,
\be
H_{MNP}=3\pa_{[M}B_{NP]}=\frac{3}{3!}\sum_{\pi}(-1)^{\pi}\pa_{\pi(M)}B_{\pi(N) \pi(P)}=\pa_{M}B_{NP}+\pa_{N}B_{PM}+\pa_{P}B_{MN}\, ,
\ee
where we have to sum over all permutations $\pi$ of indices $M$, $N$ and $P$, and the sign $(-1)^{\pi}$ is $+1$ for even and $-1$ for odd permutations.
The equations of motion of type IIB supergravity in the {\it string frame} may be found by first varying the action~
%\cite{Dall'Agata:1997ju,Dall'Agata:1998va}
\be
\begin{aligned}
S=\frac{1}{2\kappa^2}\int {\rm d}^{10}X \Bigg[\sqrt{-G}\Bigg(& e^{-2\varphi}\Big(R+4\pa_{M}\varphi\pa^{M}\varphi-\frac{1}{12}H_{MNP}H^{MNP}\Big)  \\
&-\frac{1}{2}\pa_{M}\chi \pa^{M}\chi -\frac{1}{12}F_{MNP}F^{MNP}-\frac{1}{4\cdot 5!}F_{MNPQR}F^{MNPQR}
\Bigg) \,   \\
&+\frac{1}{8\cdot 4!}\eps^{M_1\ldots M_{10}}C_{M_1M_2M_3M_4}\pa_{M_5}B_{M_6M_7}\pa_{M_8}C_{M_9M_{10}}\Bigg]\, ,
\end{aligned}
\ee
and after that by imposing the self-duality condition for the five-form\footnote{With this convention the flux of $F_5$ through the deformed sphere is negative.}
\be\label{eq:sel-duality-F5-curved}
F_{M_1M_2M_3M_4M_5}=+\frac{1}{5!}\sqrt{-G}\eps_{M_1\ldots M_{10}}F^{M_6M_7M_8M_9M_{10}}\,.
\ee
Here $G$ is the determinant of the metric, $R$ the Ricci scalar, and for the anti-symmetric tensor $\eps$ we choose the convention $\eps^{0\ldots 9}=1$ and $\eps_{0\ldots 9}=-1$.
%The last term in the action is the Chern-Simons term.
Let us write the equations of motion for all the fields.

\medskip

\noindent 
{\sl Equation for the dilaton $\varphi$}  
%{\small
\be\label{eq:eom-dilaton}
4\pa^{M}\varphi\pa_{M}\varphi-4\pa^{M}\pa_{M}\varphi
-4\pa_{M}G^{MN}\pa_{N}\varphi-2\pa_{M}G_{PQ}G^{PQ}\pa^{M}\varphi=R-\frac{1}{12}H_{MNP}H^{MNP}\, .
\ee
%}
Note that $\pa_{M}G_{PQ}G^{PQ} = 2\pa_{M}\log\sqrt{-G}$.

\bigskip

\noindent 
{\sl Equation for the two-form $B_{MN}$}

\be\label{eq:eom-B}
\pa_{P}\Big(\sqrt{-G}e^{-2\varphi}H^{MNP})+\sqrt{-G} F_P F^{MNP}+\frac{1}{6}\sqrt{-G}F^{MN Q RS} F_{QRS}=0
\ee
This equation has been derived by using~\eqref{eq:eom-C2} and \eqref{eq:eom-C4}.

\bigskip

\noindent 
{\sl Equation for the axion $\chi$}  

\be\label{eq:eom-axion}
\pa_{M}\Big(\sqrt{-G}\pa^{M}\chi\Big)=\frac{1}{6}\sqrt{-G}F_{MNP}H^{MNP}\, .
\ee

\bigskip

\noindent 
{\sl Equation for the two-form $C_{MN}$}

\be\label{eq:eom-C2}
\pa_{P}(\sqrt{-G}F^{MNP})-\frac{1}{6}\sqrt{-G}F^{MNQ RS}H_{QRS}=0
\ee

\bigskip

\noindent 
{\sl Equation for the four-form $C_{MNPQ}$} 
\be\label{eq:eom-C4}
\pa_{N}\left( \sqrt{-G} F^{NM_1M_2M_3M_4}\right)=-\frac{1}{36}\epsilon^{M_1\ldots M_4 M_5\ldots M_{10}}H_{M_5M_6M_7}F_{M_8M_9M_{10}}\, .
\ee

\bigskip

\noindent 
{\sl Einstein equations} 
\be\label{Einstein}
R_{MN}-\frac{1}{2}G_{MN}R=T_{MN}\, ,
\ee
where the stress tensor is
{\footnotesize
\be
\begin{aligned}
T_{MN}=G_{MN}\Bigg[2\pa^{ P}(\pa_{ P}\varphi)-2G^{PQ}\Gamma^{R}_{PQ}\pa_{R}\varphi-2\pa_{P}\varphi\pa^{P}\varphi-
\frac{1}{24}H_{PQR}H^{PQR}-\frac{1}{4}e^{2\varphi}F_{P}F^{P}-\frac{1}{24}e^{2\varphi}F_{PQR}F^{PQR}\Bigg] \\
-2\pa_{M}\pa_{N}\varphi+2\Gamma^{P}_{MN}\pa_{P}\varphi+\frac{1}{4}H_{MPQ}H_{N}^{\ PQ}+\frac{1}{2}e^{2\varphi}F_{M}F_{N}
+\frac{1}{4}e^{2\varphi}F_{MPQ}F_{N}^{\ PQ}+\frac{1}{4\cdot 4!}e^{2\varphi}F_{MPQRS}F_{N}^{\ PQRS}\, , \\
\end{aligned}
\ee
}
and the Christoffel symbol is 
\be
\Gamma^{P}_{MN}=\frac{1}{2}G^{PQ}(\pa_{M}G_{NQ}+\pa_{N}G_{MQ}-\pa_{Q}G_{MN})\, .
\ee

\end{document}